\documentclass[a4paper,prx,reprint,superscriptaddress,notitlepage,floatfix]{revtex4-2}
\usepackage[utf8]{inputenc}
\usepackage[T1]{fontenc}
\usepackage{subfiles}

\usepackage{graphicx}
\usepackage{tikz-cd}
\usetikzlibrary{calc,positioning} % Important: need this for alignment

\usepackage{amsmath}
\usepackage{amssymb}
\usepackage{stmaryrd}
\usepackage{fourier-orns}
\usepackage{commath}
\usepackage{mathrsfs}
\usepackage{braket}
\usepackage{xcolor}
\usepackage{bm}
\usepackage{csquotes}
\usepackage{graphicx}
\usepackage{booktabs}
\usepackage{enumitem}
\usepackage{siunitx}
\usepackage{microtype}

\usepackage{tikz}

\usepackage{hyperref}
\usepackage{bookmark}

\definecolor{tab_blue}{HTML}{1F77B4}
\definecolor{tab_orange}{HTML}{FF7F0E}
\definecolor{tab_green}{HTML}{2CA02C}
\definecolor{tab_red}{HTML}{D62728}
\definecolor{tab_purple}{HTML}{9467BD}
\definecolor{tab_brown}{HTML}{8C564B}
\definecolor{tab_pink}{HTML}{E377C2}
\definecolor{tab_gray}{HTML}{7F7F7F}
\definecolor{tab_olive}{HTML}{BCBD22}
\definecolor{tab_cyan}{HTML}{17BECF}

\hypersetup{%
  breaklinks=true,
  colorlinks=true,
  linkcolor=tab_blue,
  filecolor=tab_blue,
  urlcolor=tab_blue,
  citecolor=tab_blue,
}

\def\dd{\text{d}}

\def\ee{\text{e}}

\def\vec#1{\bm{#1}}

\definecolor{tab_blue}{HTML}{1F77B4}
\definecolor{tab_orange}{HTML}{FF7F0E}
\definecolor{tab_green}{HTML}{2CA02C}
\definecolor{tab_red}{HTML}{D62728}
\definecolor{tab_purple}{HTML}{9467BD}
\definecolor{tab_brown}{HTML}{8C564B}
\definecolor{tab_pink}{HTML}{E377C2}
\definecolor{tab_gray}{HTML}{7F7F7F}
\definecolor{tab_olive}{HTML}{BCBD22}
\definecolor{tab_cyan}{HTML}{17BECF}

\usepackage{dcolumn}% Align table columns on decimal point 
\usepackage{subfiles} % Best loaded last in the preamble

\def\hm{h_{\mu}}
\def\hn{h_{\nu}}
\def\hl{h_{\lambda}}

\def\tphi{\tilde{\phi}}
\def\pl{p_{\lambda}}

\def\infev{\lambda} % infinitesimal eigenvalue symbol
\def\dtev{\Lambda} % integrated eigenvalue symbol

\usepackage{amsmath}
\usepackage{amssymb}
\usepackage{dcolumn}% Align table columns on decimal point 
\usepackage{braket}

\usepackage{subfiles} % Best loaded last in the preamble
\usepackage{hyperref}

\def\dd{\text{d}}

\def\ee{\text{e}}

\def\vec#1{\mathbf{#1}}

\def\infev{\lambda} % infinitesimal eigenvalue symbol
\def\dtev{\Lambda} % integrated eigenvalue symbol

\newcommand{\response}[1]{\textcolor{black}{#1}}
\newcommand{\blue}[1]{\textcolor{black}{#1}}
\DeclareMathOperator*{\argmin}{arg\,min}

\tikzset{
  funnel/.pic={
    \newdimen\funnelWidth
    \pgfmathsetlength\funnelWidth{0.15cm}
    \newdimen\funnelHeight
    \pgfmathsetlength\funnelHeight{0.7*0.5cm}
    \newdimen\funnelRadius
    \pgfmathsetlength\funnelRadius{0.7*0.8cm}
    \pgfmathsetmacro\funnelAngle{42}
    \begin{scope}[yshift=-0.4cm,xshift=0cm]
    \path ({1/2*\funnelWidth},0) coordinate (br) -- ++(0,\funnelHeight) coordinate (mr) -- ++(90-\funnelAngle:{\funnelRadius}) coordinate (tr);
    \path ({-1/2*\funnelWidth},0) coordinate (bl) -- ++(0,\funnelHeight) coordinate (ml) -- +(90+\funnelAngle:{\funnelRadius}) coordinate (tl);
    \path[fill=tab_red!50] (br) -- (mr) -- (tr) -- (tl) -- (ml) -- (bl) -- cycle;
    \draw[tab_red!50!black,thick] (br) -- (mr) -- (tr) -- (tl) -- (ml) -- (bl) -- cycle;
    \end{scope}
  }
}

\newcommand{\funnelicon}[1][0.4]{% Default scale is 0.4
    \begin{tikzpicture}[scale=#1, transform shape, baseline=(current bounding box.south),]
        \pic {funnel};
    \end{tikzpicture}%
}

\newcommand{\funnelupicon}[1][0.4]{%
    \begin{tikzpicture}[scale=#1, transform shape, baseline=(current bounding box.south),]
        \pic[rotate=180] {funnel};
    \end{tikzpicture}%
}

\begin{document}

%\title{Information theory for data-driven identification of collective variables in dynamical systems}
\title{Information theory for dimensionality reduction in dynamical systems} %\\ with applications to physical and biological data}
%\title{Information theory for data-driven model reduction in physics and biology}

\author{Matthew S. Schmitt*}
\affiliation{University of Chicago, James Franck Institute, 929 E 57th Street, Chicago, IL 60637}
\affiliation{University of Chicago, Department of Physics, 929 E 57th Street, Chicago, IL 60637}
\author{Maciej Koch-Janusz*}
\affiliation{University of Chicago, James Franck Institute, 929 E 57th Street, Chicago, IL 60637}
\affiliation{Haiqu, Inc., 95 Third Street,  San Francisco, CA 94103, USA}
\affiliation{Department of Physics, University of Zurich, 8057 Zurich, Switzerland}
\author{Michel Fruchart}
\affiliation{University of Chicago, James Franck Institute, 929 E 57th Street, Chicago, IL 60637}
\affiliation{ESPCI, Laboratoire Gulliver, 10 rue Vauquelin, 75231 Paris cedex 05}
\author{Daniel S. Seara}
\affiliation{University of Chicago, James Franck Institute, 929 E 57th Street, Chicago, IL 60637}
\author{Michael Rust}
\affiliation{University of Chicago, Department of Molecular Genetics and Cell Biology, Chicago, IL, 60637}
\affiliation{University of Chicago, Department of Physics, 929 E 57th Street, Chicago, IL 60637}
\author{Vincenzo Vitelli}
\affiliation{University of Chicago, James Franck Institute, 929 E 57th Street, Chicago, IL 60637}
\affiliation{University of Chicago, Department of Physics, 929 E 57th Street, Chicago, IL 60637}
\affiliation{University of Chicago, Kadanoff Center for Theoretical Physics, 933 E 56th St, Chicago, IL 60637}

% At least three keywords are required at submission. Please provide three to five keywords, separated by the pipe symbol.

\begin{abstract}
The dynamics of many-body systems can often be captured in terms of only a few relevant variables. 
Mathematical and numerical approaches exist to identify these variables by exploiting a separation of time scales between slow relevant and fast irrelevant variables, but such a separation of scales is not always obvious or even available. 
In this work, we introduce an information-theoretic framework for dimensionality reduction in dynamical systems that bypasses this limitation by instead identifying relevant variables based on how predictive they are of the system's future.
To do so, we mathematically formalize the intuition that model reduction is about keeping \enquote{relevant} information while throwing away \enquote{irrelevant} information.
We characterize the solution of the resulting optimization problem and prove that it reduces to standard approaches when a separation of time scales is indeed present in the dynamics.
Importantly, we find that within this framework, the problems of identifying relevant variables and identifying their effective dynamics decouple and may be solved separately.
This makes the method tractable in practice and enables us to derive dimensionally-reduced variables from data with neural networks.
%The resulting dynamic dimensionality reduction method does not make any reference to slow and fast variables, and hence can be used in cases where these are not known a priori or do not even exist.
%which solves the intertwined objectives of identifying relevant variables and finding their latent evolution.
%We show how to make the method tractable in practice using a combination of analytical and algorithmic tools, and prove that it reduces to standard approaches when a separation of time scales is indeed present in the dynamics. 
Combined with existing equation learning methods, the procedure introduced in this work reveals the dynamical rules governing the system's evolution in a data-driven manner. 
We illustrate these tools in diverse settings including simulated chaotic systems, uncurated satellite recordings of atmospheric fluid flows, and experimental videos of cyanobacteria colonies in which we discover an emergent synchronization order parameter.
\end{abstract}
%A key result of our work is that, within this information-theoretic framework, the identification of the latent variables and of their effective dynamics can be performed separately (Sec. II B 4). 

\maketitle

\section{Introduction}
The exhaustive description of biological or physical systems is usually impractical due to the sheer volume of information involved. 
As an example, the fluids in a steam engine
may be described by a \mbox{$10^{27}$--dimensional} state vector containing the positions and momenta of every particle in the engine.
Yet, for most practical purposes, the working of this engine can be effectively described using only a small number of thermodynamic variables like pressure and temperature, from which most of the observables we care about can be reconstructed.
Similar dimensionality reductions can be achieved for dynamical systems spanning the natural sciences.
%systems ranging from diffusing particles to biochemical molecules and complex networks.

Standard approaches to dynamical reduction rely on a separation of timescales between \emph{slow variables} (that we usually care about) and \emph{fast variables} (that we usually want to remove from the effective description).
These approaches, which can be carried out analytically or numerically, include multiple-scale analysis~\cite{Pavliotis2010}, adiabatic elimination~\cite{Kuramoto1984,Haken1983,Kuramoto2019,Kuramoto1989,Mori1998},
center manifold reduction \cite{Crawford1991,Kuramoto2019},
spectral submanifolds~\cite{Haller2016,Cenedese2022,Jain2021,Liu2024}, renormalization methods~\cite{Chen1996,kunihiro2022geometrical}, diffusion maps \cite{Nadler2005,coifman2006_diffusionmaps}, or projection-operator (Mori-Zwanzig) methods~\cite{Mori1965,Zwanzig1961,Hijon2010}.
These approaches can be augmented by data-driven techniques, some of which have been proposed to occur in the brain \cite{abbasi_chklovskii_2016_cells_do_pca,pughe-sanford_neurons_2025}, ranging from simple statistical methods such as principal component analysis (PCA) to more sophisticated methods like dynamical mode decompositions \cite{rowley2009_dmd,schmid_2010_dmd,ModernKoopman,romeo_learning_2021,supekar_learning_2023} and neural-network based methods for latent variable learning \cite{ModernKoopman,brunton2020_review,Otto2021,Vlachas2022,Brunton2025,Rydzewski2023,Meila2024,sarra_marquardt_2021_prl,dillavou_murphy2025comparing}. % todo distribute citations on the simple/NN boxes as needed

However, phenomena that lack a clear timescale separation put these approaches in jeopardy. 
This lack of scale separation occurs in various situations ranging from climate \cite{Lucarini2023} to ecosystems \cite{Ferretti2025} and disordered systems \cite{Kurchan2023,Franz2022}, mainly because many overlapping timescales corresponding to different kinds of interconnected processes are at play.
In this situation, we are back to square one: in addition to a scheme that performs model reduction, we need to identify the \enquote{latent variables} that are best suited to predict the future of the whole system.
Worse, we must solve these two intertwined problems at the same time. 

The goal of this work is: 
(i) to formulate precisely the joint optimization problem that should produce the optimal latent variables and their effective dynamics and
(ii) to show that it reduces to the expected result when a separation of timescales is present; and
(iii) to implement numerically the optimization procedure in the general case.
In order to do so, we develop an information-theoretic framework for collective variable identification (Sec.~\ref{sec1:formalism}).
Very much like MP3 compression retains information that matters most to the human ear \cite{Jayant1993}, our reduced description of the system keeps information that matters most to predict the future \blue{\cite{creutzig2008,Lipshutz_2020_slowfeatures}}.
Formalizing this intuitive statement requires some care in order to avoid trivial results that manual approaches usually avoid through physical intuition. 

A key result of our work is that, within this information-theoretic framework, the identification of the latent variables and of their effective dynamics can be performed separately (Sec.~\ref{info_separates_concerns}).
The corresponding optimization problem can then be related to a lossy compression problem known as the information bottleneck \cite{tishby_IB,bialek2012biophysics,creutzig2008} (Sec.~\ref{soft_sizes}). 
Once the latent variables are identified, existing methods can readily yield the effective dynamics \cite{brunton_sindy,Kondrashov2015} (examples are carried out in detail in Sec.~\ref{sec4:characterizing_latent_variables}).
% TODO other things? reviews? 

We show analytically how and under what conditions the standard operator-theoretic formalism of dynamical systems, which underlies most methods of model reduction, naturally emerges from the optimal compression solution (Sec.~\ref{sec2:opt_enc}). 
In particular, we show analytically that our method yields the dominant eigenfunction of the evolution (transfer) operator in limit of high compression under the assumption of Markovian dynamics.
As a consequence, our framework reproduces the expected result when a separation of timescales is present, namely that the latent variables are the slow variables. 
In the context of deep learning, this result also sheds light on the workings of %variational autoencoders and variational IB 
the variational autoencoder-like neural networks that solve the variational information bottleneck problem \cite{kingma2022vae,Diederik2019,alemi2016_dvib}. % TODO as discussed in XXX 

%MI to measure non-Markovianity {quantifying_non_markovianity}
In addition, our information-theoretic framework provides criteria for when to stop increasing the complexity of the latent description, by directly measuring how informative relevant variables are of their own future and of the future of the entire system (Sec.~\ref{sec:whentostop}). 
We also discuss how to quantify and possibly reduce the non-Markovianity of latent variables (Sec.~\ref{quantifying_non_markovianity}).
Finally, we illustrate how to combine our approach with equation learning algorithms to produce a fully-fledged model discovery pipeline yielding both the latent variables and their effective dynamics (Sec.~\ref{sec:sinegordon}).

Section \ref{sec5:applications} showcases our framework in uncontrolled situations: satellite movies of atmospheric flows downloaded from YouTube (Sec.~\ref{sec:dvib_fluids}) and experimental measurements of experimental microscopy videos of cyanobacteria colonies (Sec.~\ref{sec:cyanobac}).
In these situations, multiple timescales that are not necessarily well-separated are present (like circadian rhythms and cell growth in cyanobacteria experiments, Sec.~\ref{sec:cyanobac}), the Markovian hypothesis used in Sec.~\ref{sec2:opt_enc} is not guaranteed to hold, and the data is high-dimensional, precluding naive optimization procedures. 
Nevertheless, our procedure produces meaningful interpretable latent variables that can be related to the underlying physics or biology.

\section{Model discovery as compression}
\label{sec1:formalism}

\subsection{Dimensionality reduction for dynamics}

Consider a system whose state is described by a (possibly random) variable $X_t$, which might correspond to anything from the position of a single particle to an image of a fluid flow or the fluorescent molecules in a living system (Fig.~\ref{fig_mr}a,b).
The full state can be high dimensional, with a number of dimensions equal to the number of particles in a gas, or the number of pixels in an image. 
The system's dynamics are described by an evolution operator $\mathcal{U}_{\Delta t}$ that produces the future state of the system $X_{t+\Delta t} = \mathcal{U}_{\Delta t}(X_t)$ as a function of the current state $X_t$, represented graphically as:
\begin{equation}
    \begin{tikzpicture}[baseline=(current  bounding  box.center)]
    \node (Xt) {$X_t$};
    \node[right=3.75cm of Xt] (Xtdt) {$X_{t+\Delta t}$};
    \draw[->] (Xt) -- (Xtdt);
    \path (Xt) -- (Xtdt) node[midway,above] {physical dynamics $\mathcal{U}_{\Delta t}$};
    \path (Xt) -- (Xtdt) node[midway,below] {\includegraphics[width=0.8cm]{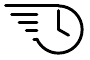}};
    \label{tikz:right_X_evolve}
\end{tikzpicture}
\end{equation}
Even when $X_t$ is very high dimensional, in many cases it turns out that these dynamics are effectively low-dimensional.
This means that we can compress $X_t$ into a low-dimensional ``latent'' variable $H_t$ using an encoding function $h$:
\begin{equation}
    \begin{tikzpicture}[baseline=(current  bounding  box.center)]
    \node (Xt) {$X_t$};
    \node[below=1.25cm of Xt] (Ht) {$H_t$};
    \node[right=-0.1cm of Ht] {$= h(X_t)$};
    \draw[->] (Xt) -- (Ht);
    \path (Xt) -- (Ht) node[midway,right] {encoder $h$};
    \path (Xt) -- (Ht) node[midway,left=0.6cm,coordinate] (funnel pos) {};
    \pic at (funnel pos) {funnel};
    %node[midway,left] {\includegraphics[width=0.8cm]{media/icons/icon_funnel.pdf}};
    \label{tikz:down_encode}
\end{tikzpicture}
\end{equation}
where we have represented the encoder $h$ with a funnel to convey the idea that it should compress information.
%which are low-dimensional and very informative of the future state of the system. 
The dynamics of the latent variable are described by an operator $\mathcal{V}_{\Delta t}$ that acts on the reduced state $H_{t+\Delta t}=\mathcal{V}_{\Delta t}(H_t)$:
\begin{equation}
    \begin{tikzpicture}[baseline=(current  bounding  box.center)]
    \node (Ht) {$H_t$};
    \node[right=3.5cm of Ht] (Htdt) {$H_{t+\Delta t}$};
    \draw[->] (Ht) -- (Htdt);
    \path (Ht) -- (Htdt) node[midway,above] {latent dynamics $\mathcal{V}_{\Delta t}$};
    \path (Ht) -- (Htdt) node[midway,below] {\includegraphics[width=0.8cm]{media/evo_icon.pdf}};
    \label{tikz:right_H_evolve}
\end{tikzpicture}
\end{equation}
We can then attempt to recover the future state of the full system by inverting the encoding with a decoding function $\tilde{h}^{-1}$ acting as
\begin{equation}
\begin{tikzpicture}[baseline=(current  bounding  box.center)]
    \node (Xtdt) {$\tilde{X}_{t+\Delta t}$};
    \node[right=-0.1cm of Xtdt] {$\equiv \tilde{h}^{-1}(H_{t+\Delta t})$};
    \node[below=1.25cm of Xtdt] (Htdt) {$H_{t+\Delta t}$};
    \draw[<-] (Xtdt) -- (Htdt);
    \path (Xtdt) -- (Htdt) node[midway,right] {decoder $\tilde{h}^{-1}$};
    \path (Xtdt) -- (Htdt) node[midway,left=0.6cm,coordinate] (funnel pos) {};
    \pic[rotate=180] at (funnel pos) {funnel};
    %\path (Xt) -- (Ht) node[midway,left] {\includegraphics[width=0.8cm]{media/icons/icon_funnel_up.pdf}};
    \label{tikz:up_decode}
\end{tikzpicture}
\end{equation}
where we have represented the decoder as a inverted funnel.
In general, information is lost during the encoding procedure and as a result the predicted state and the true state will differ. 
Consequently, there is finite distance 
\begin{equation}
    d = \lVert  X_{t+\Delta t} - \tilde{h}^{-1}(H_{t + \Delta t}) \rVert
    \label{distance_d}
\end{equation}
that we want to make as small as possible.
In other words, we want to minimize the non-commutativity of the diagram
\begin{equation}
    \begin{tikzpicture}[baseline=(current  bounding  box.center)]
    \node (Xt) {$X_t$};
    \node[right=2cm of Xt] (Xtdt) {$X_{t+\Delta t}$};
    \draw[->] (Xt) -- (Xtdt);
    \path (Xt) -- (Xtdt) node[midway,above] {$\mathcal{U}_{\Delta t}$};
    \path (Xt) -- (Xtdt) node[midway,below] {\includegraphics[width=0.6cm]{media/evo_icon.pdf}};

    \node[below=2.85cm of Xt] (Ht) {$H_t$};
    \draw[->] (Xt) -- (Ht);
    \path (Xt) -- (Ht) node[midway,right] {$h$};
    \path (Xt) -- (Ht) node[midway,left=0.6cm,coordinate] (funnel pos) {};
    \pic at (funnel pos) {funnel};

    \node[right=3.5cm of Ht] (Htdt) {$H_{t+\Delta t}$};
    \draw[->] (Ht) -- (Htdt);
    \path (Ht) -- (Htdt) node[midway,above] {${\mathcal{V}}_{\Delta t}$};
    \path (Ht) -- (Htdt) node[midway,below] {\includegraphics[width=0.6cm]{media/evo_icon.pdf}};

    \node[above=1.6cm of Htdt] (bad) {$\tilde{X}_{t+\Delta t}$};
    \draw[->] (Htdt) -- (bad);
    \path (Htdt) -- (bad) node[midway, left=0.6cm, coordinate] (funnel pos) {};
    \pic[rotate=180] at (funnel pos) {funnel};
    \path (Htdt) -- (bad) node[midway,right] {$\tilde{h}^{-1}$};

    \coordinate (d_start) at ($(Xtdt.center)+(45:0.55cm)$);
    \coordinate (d_end) at ($(bad.center)+(45:0.55cm)$);
    \coordinate (d_mid) at ($(d_start)!0.5!(d_end)$);

    \draw[tab_red, |<->|, line width=0.8pt] (d_end) -- (d_start) node[midway,inner sep=0.1cm,tab_red,fill=white] (d) {$d$};

    \node[tab_red, align=center] (ncom) at ($(Ht)!0.5!(Htdt)+(0,2.5cm)$) {noncommutativity};
    \draw[->,>=latex,tab_red] (ncom.north) to[out=40,in=-150] (d);
    
\end{tikzpicture}
\label{eq:noncommutative_diagram}
\end{equation}

However, minimizing $d$ is not our only goal. If it were so, we could simply set $h$ and $\tilde{h}^{-1}$ to the identity and $\mathcal{V}=\mathcal{U}$ to perfectly reproduce the original dynamics.
Instead, we wish to simultaneously minimize $d$ while ensuring that the size of the latent variable $H_t$ (e.g. its dimension) is as small as possible. 
%All in all, this suggests that we should optimize over the encoders $h$ and latent evolutions $\mathcal{V}$ to get
%\begin{equation}
%     \{h_{\text{opt}}, \mathcal{S}_{\text{opt}}\} \equiv \arg\min_{h, %\mathcal{S}, h^{-1}} \dim H_t +  \beta{\color{red}\, d}
%\end{equation}
%in which $\beta$ is a tradeoff parameter.
To find a good latent representation of the state and its dynamics, we encode these requirements in an optimization problem
\begin{align}
    (h_*, \mathcal{V}_*, \tilde{h}^{-1}_*)
    = \argmin_{\funnelicon[0.4], {\includegraphics[width=0.5cm]{media/evo_icon.pdf}}, \,\funnelupicon[0.4]} {\text{size}(H_t)} +  \beta\, { d},
    \label{eq:objective_symbolic}
\end{align}
so that the optimal encoder, latent evolution, and decoder (decorated with a star) minimize the objective function (loss function) $\mathcal{L} = \text{size}(H_t) +  \beta\, {d}$, where $\beta$ determines the trade off between predictive accuracy and dimensionality reduction.

The optimization problem Eq.~\eqref{eq:objective_symbolic} is daunting because it requires simultaneously optimizing over three complex objects.  
This can be realized in certain settings \cite{lusch2018,takeishi2017_neurips}, but it is too difficult in general.
It turns out that this difficulty is drastically alleviated by a careful implementation of the idea that we now sketch. Up to now, we have glossed over two key points. First, the encoder is in general not invertible, which naturally leads to a probabilistic decoder. Second, the distance function used to compute $d$ must be chosen to be consistent with this probabilistic picture. 
As we shall see, a careful choice of distance dramatically simplifies the optimization problem.

\subsection{Information theory for dimensionality reduction}
% HERE: probabilistic encoding etc. information theory

\subsubsection{Compression requires noninvertible encoders}
\label{crne}

As $X_t$ is a very-high dimensional variable while $H_t$ is (hopefully) a low-dimensional one, the encoding $h$ will generally be non-invertible. 
The non-invertibility of the encoder is the essence of the compression: it is a feature, not a bug. 
The requirement that $H_t$ does not contain \enquote{too much} information is contained in the quantity $\text{size}(H_t)$ that we try to make as small as possible in Eq.~\eqref{eq:objective_symbolic}.

The precise meaning of $\text{size}(H)$ depends on the nature of the variable $H$. When $H$ lives in a discrete space (for instance, if it is a finite chain of bits), then the cardinality of the set of possible states is a reasonable measure of $\text{size}(H)$. 
When $H$ is a continuous variable (say $H \in \mathbb{R}^n$), the cardinality of the set of possible $H$ is infinite, and is therefore not a good measurement of the available size. 
In principle, one can encode an arbitrarily large amount of information in just a single real number (e.g. in all the decimals).
However, the dimension of a space is preserved by continuous bijections \cite{sagan_space-filling_1994}. 
This suggests that in this case, the dimension of (the space where lives) $H_t$ is an reasonable measure of $\text{size}(H)$, provided that we impose that the encoder is continuous.
The addition of noise into $h$ also limits how much information about $X_t$ may be stored in $H_t$  \cite{strouse2016deterministicIB,slonimphd}. In this case, the size of $H_t$ may be captured by e.g. the entropy of the distribution $p(H_t)$.

\subsubsection{Noninvertible encoders require probabilistic decoders}

The \enquote{inverse} of a non-invertible encoder could produce any of the possibly many states $X_t$ that map to a given value of $H_t$. The decoder $\tilde{h}^{-1}$ is therefore defined probabilistically via a conditional distribution: instead of obtaining $\tilde{X}_{t+\Delta t}$ as $\tilde{h}^{-1}(H_{t+\Delta t})$, it is a random variable with distribution $p(\tilde{X}_{t+\Delta t}|H_{t+\Delta t})$ and we can think of $\tilde{h}^{-1}(H_{t+\Delta t})$ as a sample from this distribution. 
Hence, Eq.~\eqref{tikz:up_decode} can more precisely be represented as
\begin{equation}
\begin{tikzpicture}[baseline=(current  bounding  box.center)]
    \node{\includegraphics{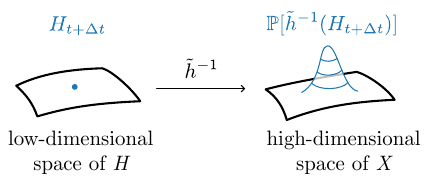}};
\end{tikzpicture}
\end{equation}
This is analogous to a procedure producing a thermodynamic ensemble compatible with the state variables. 
For instance, the canonical ensemble associates to every macrostate $H_t=(N, V, T)$ defined by a number of particles, volume, and temperature a distribution $p_{\text{can}}(\xi|N,V,T)$ over the set of microstates $\xi$ from which the decoded state is drawn, namely, $\xi = \tilde{h}^{-1}(H_t)$ follows the distribution $p_{\text{can}}(\xi | N,V,T)$.

\subsubsection{Noise makes dynamics fathomable}
\label{noise_helps}

The necessity of a probabilistic decoder leads us to embrace the probabilistic nature of all quantities involved: indeed, $X_t$, $H_t$, and so on can all be treated on an equal footing as random variables \footnote{From now on, we denote random variables by uppercase $X_t$ and their values by lowercase $x_t$.}, which allows us to handle stochastic (noisy) evolutions.
This turns out to be a key element in our analysis.
The idea that randomness regularizes dynamics has emerged in the study of many-body dynamical systems, where chaos provides a justification for the introduction of statistical ensembles \cite{Eckmann1985,Gaspard2010}.
This idea is key in reconciling deterministic dynamics with thermodynamic and hydrodynamic approaches. 
Indeed, noise (in the form of internal chaotic dynamics and/or external perturbation) tends to induce dynamics that are effectively low-dimensional on long time scales \cite{Cvitanovic2012,froyland2005,gaspard1993}.
From a mathematical perspective, this can be understood in terms of the spectrum of the transfer operator associated to the physical dynamics $\mathcal{U}_{\Delta t}$, which may, in certain cases, be split into a finite-dimensional part corresponding to the slow dynamics, plus a (possibly infinite-dimensional and complicated) part that can be neglected on long time scales because it describes the fast dynamics.
This is illustrated in Fig.~\ref{fig_mr}d, see Sec.~\ref{sec2:opt_enc} for technical details.
This idea has forcefully been highlighted in Ref.~\cite{Cvitanovic2012}, where it is argued that noise limits how much we can resolve states in a chaotic limit, suggesting the existence of a finite-dimensional description whose dimension is limited by the dynamics rather than our observational abilities. 
Mathematically, this translates into the existence of an optimal partition of phase space associated to a finite-dimensional probabilistic description~\cite{Froyland2001,froyland2005,Cvitanovic2012}.
Our information-theoretic perspective allows us to extend these ideas to the case of continuous variables, as we discuss in detail in Sec.~\ref{sec:whentostop}.

\subsubsection{Information theoretic distance separates concerns}
\label{info_separates_concerns}

The distance $d$ in Eq.~\eqref{distance_d} should be compatible with the probabilistic nature of the decoder, and of the other quantities.
In particular, $d$ should compare the statistics of decoded states $\tilde{X}_{t+\Delta t}$ with those of the true $X_{t+\Delta t}$, rather than comparing individual samples.
To do so, we measure how predictive the latent variable $H_{t+\Delta t}$ is of the future state $X_{t+\Delta t}$ in terms of the mutual information between these variables (SI~Sections~S1~and~S2) by setting
\begin{equation}
    d = - I\left([\tilde{h}^{-1}\circ \mathcal{V}_{\Delta t}\circ h](X_t), X_{t+\Delta t}\right).
\end{equation}
As shown in Appendix~\ref{main_apx:MI_from_DKL}, mutual information emerges naturally when taking into account the probabilistic nature of the decoder.
Notice the sign on the right-hand side of the equation: to minimize the distance, we should maximize the mutual information. 

The mutual information is not a true metric function, which in turn allows us to abstract away from the geometry of the underlying space. Instead of measuring the similarity of variables in terms of how close they are, it is measured in terms of how predictive they are of each other in a system-independent way.
This has an important consequence: as a result of the data processing inequality, we can show that our objective function then simplifies to one depending only on $h$. 
More precisely, we prove in Appendix~\ref{main_apx:information_processing_inequality} that the optimal encoder is given by
\begin{align}
    h_*=\argmin_{\vcenter{\hbox{\funnelicon[0.4]}}} 
    \quad\text{size}( H_t)
    -  \beta I\big(H_t, X_{t+\Delta t}\big)
    \label{eq:modelreduction_objective_simple}
\end{align}
%The steps to this result are shown in detail in \blue{Appendix~\ref{main_apx:information_processing_inequality}}.
This simplification is dramatic: instead of simultaneously learning the encoding function \funnelicon[0.4], the latent evolution operator \includegraphics[width=0.5cm]{media/evo_icon.pdf}, and the decoder \funnelupicon[0.4], we only need to care about the encoding.

In particular, we can completely forget about the latent evolution at this stage.
In the language of computer science, this implements a separation of concerns \cite{Dijkstra1982} between the choice of optimal latent variables and the discovery of an optimal latent evolution.
Consequently, we view model discovery in the rest of this paper as a task of finding the optimal reduced variables with which to represent our system.
The evolution rule $\mathcal{V}$ and decoder $\tilde{h}^{-1}$ are \emph{induced} by the choice of $h$, and we may view their identification as a secondary, downstream task to be performed after the proper variables have been found.

\subsubsection{Continuous information-theoretic sizes help optimization}
\label{soft_sizes}

In Sec.~\ref{crne}, we have focused on situations where the $\text{size}(H_t)$ is a discrete quantity, like a number of bits or a dimension. 
However, it is not practical to have discrete quantities in optimization problems.
Using an information-theoretic constraint, we can limit the effective size available to $H_t$ without enforcing a hard constraint.
To do so, we replace in the cost function $\text{size}(H_t)$ by the term $I(H_t, X_t)$ that explicitly forces the reduced representation to lose information about the full state. 
The objective found in this way 
\begin{align}
   \mathcal{L}_{\text{IB}}=I(H_t,X_t) - \beta I(H_t,X_{t+\Delta t}),
    \label{ib_obj}
\end{align}
is known as the information bottleneck (IB) \cite{tishby_IB,bialek2012biophysics}.
It has been used (along with related mathematical \cite{Slonim2000,strouse2016deterministicIB,piran_dual_2020} and computational~\cite{abdelaleem2023deep,alemi2016_dvib,abdelaleem2025} extensions) in a variety of contexts, from document clustering~\cite{Slonim2000} to analysis of neural networks~\cite{shwartz-ziv_opening_2017} to biological signaling networks \cite{bauer_trading_2021,bauer_information_2023} to, most relevant for us, dynamical systems~\cite{creutzig2008,still_information_2014,still_2010,sachdeva2021,Palmer2015}. 

We allow the encoder to be a possibly stochastic function, meaning that values of $h(X_t)$ are drawn from a conditional probability distribution $\funnelicon[0.3]=p(H_t|X_t)$.
The optimal encoder is the solution to our optimization problem
\begin{align}
    \label{optenc}
    p^*_{H_t|X_t} %(h_t|x_t) = 
    = \argmin_{\vcenter{\hbox{\funnelicon[0.4]}}} \;\mathcal{L}_{\text{IB}}.
\end{align}

\subsubsection{Distilling the encoder yields relevant variables}
\begin{figure*}[tp]
    \includegraphics[width=1.0\textwidth]{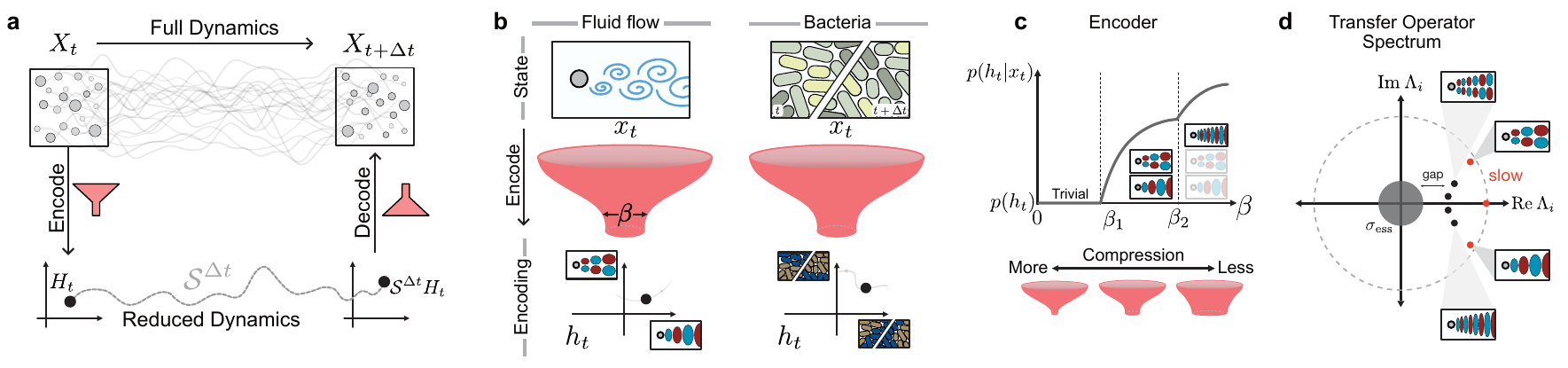}
    \caption{\label{fig_mr}\textbf{\blue{Data-driven identification of interpretable dynamical variables via the information bottleneck.}}
    (a) A reduced description of a system requires both an encoding procedure that isolates relevant variables, and a set of dynamical equations that evolve the relevant variables in time.
    In our information theoretic framework these two steps decouple (\blue{Appendix~\ref{main_apx:model_reduction}}).
    (b) The information bottleneck compresses high-dimensional state variables $x_t$, into simpler encoding variables $h_t$ with a controllable trade-off between the degree of compression and the predictive power about the system's future.
    With deep neural networks, the encoding can be computed directly from data of observed fluid flows (left) or biological datasets, such as fluorescently labeled bacteria colonies (right). 
    In general, the state of the variable $x_t$ may comprise time-lagged variables of the intensity field, $x_t=\{I_t, I_{t+\Delta t}\}$ (right).
    The amount of compression is determined by the ``width'' of the bottleneck $\beta$ (see Eq.~\eqref{ib_obj}).
    The resulting compressed, or encoded, variables $h_t$ represent collective variables most predictive of the system's future. %that evolve slowly in time.
    (c) Schematic evolution of the encoder $p(h_t|x_t)$ for varying compression strength $\beta$. For low $\beta$ (high compression), the encoder is trivial and forgets everything about the input $x_t$. After the first IB transition at $\beta_1$, the encoder becomes non-trivial by gaining some dependence on $x_t$; some features of the input are able to pass through the bottleneck.
    At subsequent IB transitions, additional features are learned.
    (d) The point spectrum of the transfer operator contains several slowly decaying modes (red). We show that the most predictive variables that IB systematically extracts correspond to the slowest eigenfunctions of the transfer operator, associated to eigenvalues $\Lambda_i$ with $|\Lambda_i|\approx 1$.
    In fluid flows, the slowest-decaying eigenfunctions typically represent large-scale coherent patterns of the flow field, while faster-decaying eigenfunctions correspond to variations over shorter length scales. In general the system may exhibit an essential spectrum, however for noisy systems this is small (see main text).}
\end{figure*}

The form of the optimal encoder depends strongly on the value of the trade-off parameter $\beta$. 
For small $\beta$ the compression term dominates and the optimal encoder is trivial, losing all information about the system.
For intermediate $\beta$ the compression term does not allow $X_t$ to be completely represented by $H_t$, so features of $X_t$ must compete to pass through to the encoding variable (Fig.~\ref{fig_mr}b).
These selected features are reflected in an encoder that provides the optimal trade-off between retained information and predictability \cite{PhysRevE.79.041925} (\blue{Appendix~\ref{main_apx:IB_in_general}}).
In the optimal encoder, information about $X_t$ has been removed both by reducing the dimensionality, as well as by injecting noise into $h_t$.
Increasing $\beta$ can impact both of these, either by increasing the dimensionality to learn new features, or by reducing noise to better identify the features that have already been found. 

\blue{We obtain the most relevant dimensionally-reduced variables with a two-step procedure. First, we minimize the IB objective to find the optimal encoder $p^*(h_t|x_t)$.}
In order to best identify the features given a fixed set of components, in the second step we modify the IB encoder by discarding the part attributable to noise alone to recover a deterministic function $h(X_t)$. This effectively increases the coupling of the reduced variables to the already-learned features. 
This \enquote{distillation} procedure is detailed in Sec.~\ref{si_sec:distillation}.
We shall see that this procedure isolates deterministic, interpretable reduced variables in both discrete and continuous settings.

%Our goal is to connect the dynamical properties of the system to the features learned by the encoder. %We will then use this understanding to construct numerical methods applicable to experimental data. 

%Any system in the real world is subject to measurement noise or random fluctuations or both. This induces uncertainty in our knowledge of the system's true state.
\section{The optimal encoder in terms of the transfer operator}
\label{sec2:opt_enc}
%In any realistic experimental setting, the presence of noise or uncertainty means we cannot predict precisely the future state of a system but instead only a likely distribution of future states. Our prediction of the state at $\Delta t$ in the future is then represented mathematically as $p(x_{t+\Delta t}|x_t)$, the probability of observing state $x_{t+\Delta t}$ given the current state $x_t$.

\subsection{General results}
%In an autonomous
For Markovian (``memoryless'') dynamics, the conditional probability distribution $p(x_{t+\Delta t}|x_t) \equiv p_{X_{t+\Delta t}|X_t}(x_{t+\Delta t}|x_t)$ of observing state $x_{t+\Delta t}$ at any $\Delta t$ in the future given the current state $x_t$ completely characterizes the dynamics of the system, and determines how probability distributions evolve in time.
The evolution of probability distributions can then be cast as
\begin{equation}
p_{X_{t+\Delta t}} = U^{\Delta t}[p_{X_{t}}],
\end{equation}
the action on probability distributions $p_{X_{t}}$ of the linear transfer operator $U^{\Delta t}$ defined by
\begin{equation}
    \label{Uop_int}
    U^{\Delta t}[p_{X_{t}}](x)
    \equiv
    \int p(x|x') p_{X_t}(x') \dd x'
\end{equation}
The transfer operator $U^{\Delta t}$ can be decomposed as
\begin{equation}
    \label{spectral_decomposition}
    U^{\Delta t} = \sum_{n} \ket{\rho_n} \ee^{\infev_n \Delta t}  \bra{\phi_n} + U_{\text{ess}}
\end{equation}
where (using notations standard in quantum mechanics, explained in SI) $\ket{\rho_n}$ denote right eigenfunctions with eigenvalue $\dtev_n \equiv \ee^{\infev_n \Delta t}$ and $\bra{\phi_n}$ are the corresponding left eigenfunctions. 
The $\infev_n$ are the eigenvalues of the infinitesimal generator of $U^{\Delta t}$, known as the Fokker-Planck operator (Fig.~\ref{fig_mr}d). 
The operator $U_{\text{ess}}$ corresponds to the so-called essential spectrum, and we assume that it can be neglected. 
This is usually possible when the system is subjected to even a small amount of noise, or when some amount of uncertainty is present in the measurements ~\cite{gaspard1995,froyland2013}.
The eigenfunctions $\bra{\phi_n}$ in Eq.~\eqref{spectral_decomposition} are in some sense ``natural'' features of the dynamics, as they evolve independently in time.

Our key observation is that the optimal IB encoder in Eq.~\eqref{optenc}, which ``filters'' out the relevant features of $X_t$, can be expressed in terms of the eigenvalues $\infev_n$ and left eigenfunctions $\phi_n$ of (the generator of) $U$, 
\begin{equation}
    \label{optimal_encoding_spectral}
    \!\!
    p_{\beta}^*(h_t|x_t) = \frac{p_{\beta}^*(h_t)}{\mathcal{N}(x_t)} \, \exp\left[ \beta \sum_{n} \ee^{\infev_n \Delta t} \phi_n(x_t) f_n(h_t) \right]
\end{equation}
where $f_n(h_t)$ are factors that do not depend on $x_t$. For an outline of the mathematical steps leading to this see \blue{Appendix~\ref{main_apx:IB_and_eigenfunctions} and SI Section~\ref{si_sec:derivations}}. 
The combined effect of the factors in the exponent determines what features the encoder learns about the state $X_t$.
In general, there may be a large number of non-zero factors $f_n$ so that the learned features are difficult to interpret in terms of individual eigenfunctions $\phi_n$.
However, things become simple in the limit of small $\beta$, or high compression.
When $\beta$ is small the encoder is trivial: $p(h_t|x_t)=p(h_t)$ so the value $h_t$ is assigned at random with no regard to the state of the system. No feature has been learned, and all factors $f_n$ are equal to zero.
As $\beta$ is increased, the encoder undergoes a series of transitions at $\beta=\beta_1< \beta_2<\beta_3...$ where new features are allowed to pass through the bottleneck (Fig.~\ref{fig_mr}c) \cite{gedeon2002,wu2020,chechik2005,ngampruetikorn2021schwab}. 
The first transition happens at a finite value of $\beta_1$ when the first, most predictive, feature is learned.

\textbf{Theorem} --
\textit{
Consider a transfer operator $U$ with infinitesimal generator $\mathcal{L}_{U}$. Assume that $\mathcal{L}_{U}$ has a discrete spectrum with eigenvalues satisfying $0=\text{Re }\infev_0>\text{Re }\infev_1>\text{Re }\infev_2\gg\text{Re }\infev_3...$. 
Then, for $\beta$ just above the first IB transition $\beta_1$ the optimal encoder is given approximately by}
\begin{equation}
    p^*_\beta(h|x) = \frac{1}{\mathcal{N}(x)} p^*_\beta(h) \, \exp\left(\beta \, \ee^{\infev_1 \Delta t} \phi_1(x) f_1(h)\right)
    \label{encoder_mostinformativefeature}
\end{equation}
\textit{with corrections due to the second eigenfunction given by $f_2(h) \approx f_1(h) \ee^{-\Gamma\Delta t} + \mathcal{O}(\ee^{-2\Gamma \Delta t})$ where $\Gamma=\infev_1-\infev_2>0$ denotes the spectral gap.
For equilibrium systems exhibiting no probability fluxes in the steady state, this result is exact.}

This result shows that in the limit of high compression, the encoder's dependence on $x_t$ is given by the first left eigenfunction $\phi_1(x_t)$, which is the slowest-varying function of the state under dynamics given by $U$.
Therefore, Eq.~\eqref{encoder_mostinformativefeature} makes precise the intuitive statement that slow features are the most relevant for predicting the future.
Our analytical result is valid for arbitrary, including non-Gaussian, variables.

\textbf{Proof} -- We sketch here the main steps; see Appendix~\ref{main_apx:IB_and_eigenfunctions} and SI Section~\ref{si_sec:derivations} for a more detailed treatment.
To show Eq.~\eqref{encoder_mostinformativefeature} we perform a perturbative analysis of the full optimal encoder Eq.~\eqref{optimal_encoding_spectral}. Before the transition, the coefficients $f_n(h)$ ($n>0$) are zero. 
The instability at the first IB transition at $\beta=\beta_1$ corresponds to the emergence of negative eigenvalues in the Hessian of the IB loss which may be computed explicitly (Appendix~\ref{main_apx:IB_and_eigenfunctions} and SI Section S4). The values of $f_i$ can be read off from the eigenvector corresponding to the first negative eigenvalue, which show dependence only on $f_1$. The transition is continuous, so that after the transition the encoder depends only on $\phi_1(x)$ to the desired order. We verify our theorem numerically in the SI.

We have assumed that the systems we consider are (or can be made) Markovian; in the SI we provide a detailed discussion of how to measure the memory (Section~\ref{si_sec:information_theory_for_dynamics} and SI Fig.~\ref{si_fig:lorenz_memory}) and how it changes our below results (Section \ref{si_sec:expansion_limitations} and SI Fig.~\ref{si_fig:transients_exact_IB_pseudospectr}).
One notable assumption underpinning this result, in addition to Markovianity, is that the system is in a statistical steady state, so that $x_t$ is sampled from the steady state distribution. When this assumption is violated, eigenfunctions of the transfer operator may cease to carry information about the system's long-time dynamics and instead its pseudospectrum may become relevant \cite{trefethen2005spectra} (SI Section~\ref{si_sec:expansion_limitations}).

We further observe numerically that this picture holds true more generally: at successive IB transitions, the learned features correspond to successive modes of the transfer operator (SI Figs.~\ref{si_fig:doublewell_stability} and \ref{si_fig:triplewell_stability}). %(Fig.~\ref{fig:doublewell_IB}{} and SI~Fig.~S3). 
%even for relatively small spectral gaps, IB learns an encoder which depends only on the first eigenfunction.
This is reminiscent of the exact results known for the specific case of Gaussian IB,
where the encoder learns successive eigenvectors of a matrix related to the covariance of the joint $X_t,X_{t+\Delta t}$ distribution at each IB transition \cite{chechik2005,creutzig2008}.

Due to the smallness of $f_1$ at the transition, the optimal IB encoder couples only weakly to the eigenfunction $\phi_1$, so that the dynamics of $h_t$ will be dominated by noise. Given the maximal number of available bits in memory, or variables used in an analytical model, the IB solution will thus not make the best use of the resources (IB is optimal with respect to the amount of information retained, but not the size of its representation \cite{strouse2016deterministicIB}). However, we can use the optimal IB encoding as a filter to identify $\phi_1$, and then ``distill'' it by discarding the noisy part of the encoding, effectively increasing the coupling between $h_t$ and $\phi_1(x_t)$ while keeping the number of components fixed. In practice, this is done by taking the most likely value for the reduced state, $h_t=\arg\max_h p(h|x_t)$ (see SI Section~\ref{si_sec:distillation}). 
Similar procedures are performed in other probabilistic dimensionality reduction approaches, \emph{e.g.}~probabilistic principal component analysis \cite{bishop2006pattern}.
There one finds a stochastic mapping from the full state $x$ to the reduced representation $h$, and then discards the noise to obtain the deterministic reduction.

Together this lays the foundation of an operational prescription for model \blue{discovery} which is built on IB for identifying relevant, slow, features.
As we show later, this insight can be leveraged to systematically learn these slow variables directly from data with neural networks \cite{alemi2016_dvib}, which in turn may be used as inputs to an equation learning pipeline.

\subsection{Information decay and the spectrum of the transfer operator: an example}
\label{sec:infodecay}

To develop intuition for information in a dynamical system, we turn to the simple example of a Brownian particle in a double-well potential. 
This may represent, for example, a molecule with a single degree of freedom that transitions between two metastable configurations \cite{chu2007}.
In the overdamped limit, the state of the particle is completely determined by its position $X_t\in \mathbb{R}$, with dynamics given by the Langevin equation
\begin{subequations}
\label{eq:pitchfork_dynamics}
\begin{align}
    \dot x_t &= -\partial_x V(x_t) + \sigma \eta_t.
    \\
    V(x) &= \frac{1}{4}(\mu - x^2)^2
\end{align}
\end{subequations}
Here, $\eta_t$ is unit-variance white noise, $\sigma$ controls its strength, and $\mu$ controls the shape of the potential $V(x)$.

\begin{figure}[tp]
    \centering
    \includegraphics[width=0.45\textwidth]{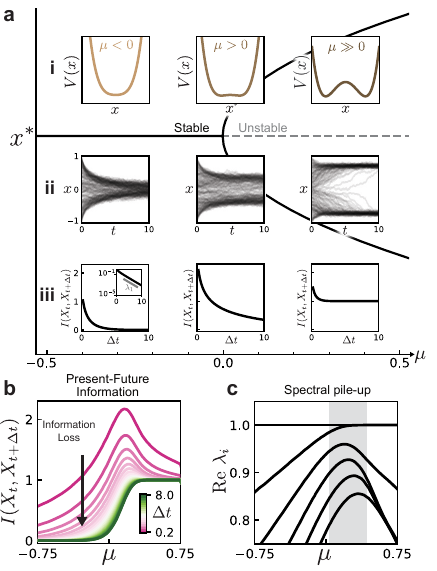}
    \caption{\textbf{Information and the spectrum of the transfer operator.}
    (a) Fixed point (FP) diagram of the dynamics given by Eq.~\eqref{eq:pitchfork_dynamics} for zero noise. 
    There is a bifurcation at $\mu=0$ where the stable FP at $x=0$ becomes unstable and two new stable FPs appear at $\pm\sqrt{\mu}$.
    (a,i) The corresponding potential $V(x)$ for varying $\mu$, with the emergence of a double-well structure for $\mu>0$.
    (a,ii) Dynamics of the system Eq.~\eqref{eq:pitchfork_dynamics} for varying values of $\mu$ corresponding to the potentials above, with uniformly-distributed initial conditions. 
    (a,iii) Loss of information between the initial condition and the future state. Inset shows scaling given by the first eigenvalue of the transfer operator. 
    (b) Spectrum of the transfer operator $U$, showing a pile-up of eigenvalues for $\mu\gtrsim 0$. These are related to the eigenvalues of its infinitesimal generator by $\Lambda_i = \ee^{\lambda_i\Delta t}$.
    (c) Mutual information between the present and future state for varying time delay $\Delta t$ and bifurcation parameter $\mu$.
    }
    \label{fig:doublewell_infoloss}
\end{figure}

The deterministic dynamical system undergoes a bifurcation at $\mu=0$ (Fig.~\ref{fig:doublewell_infoloss}a). 
%For $\mu<0$, the potential features a single well, which then becomes two for positive $\mu$ (Fig.~\ref{fig:doublewell_infoloss}a). For negative $\mu$, the trajectories all converge to a fixed point at $x=0$, while for $\mu>0$ they fluctuate around the fixed points at $x=\pm \sqrt{\mu}$. 
%When $\mu$ is small but positive, the particle may occasionally transition from one potential well to the other because of thermal fluctuations.
To quantify the amount of information about the future state $X_{t+\Delta t}$ contained in the initial state $X_t$ we compute their mutual information. 
The dynamics of $X_t$ are Markovian, so that for any sequence of times $t_0<t_1<t_2$, $p(X_{t_2}|X_{t_1}, X_{t_0})=p(X_{t_2}|X_{t_1})$. 
From the data processing inequality, one has \cite{cover2012elements}
\begin{align*}
    I(X_{t_2},X_{t_0})\leq I(X_{t_1},X_{t_0}),
\end{align*}
which implies that information can only decrease in time.

What governs the rate at which information decays?
Here we can already see the role of the spectrum of the dynamics' transfer operator.
By exploiting the spectral expansion of the conditional distribution $p(x_{t+\Delta t}|x_t)$ one finds that for long times the information decays as
\begin{align}
\!\!
    I(X_t,X_{t+\Delta t}) = e^{2\infev_1\Delta t}\langle\phi_1^2\rangle\langle\rho_1^2/\rho_0^2\rangle + \mathcal{O}(e^{2\infev_2\Delta t})
\end{align}
where expectations are taken over the steady state distribution (SI Section~\ref{si_sec:information_theory_for_dynamics}).
Asymptotically, the information decay is set by the value of $\infev_1$, the rate of decay of the slowest-varying function $\phi_1(x)$ under the dynamics of $U$.
In the limit of infinite time, for any value of $\mu$ even weak noise will cause the mutual information to become zero as there is a non-zero (perhaps exponentially small) probability of hopping between the wells \cite{gaspard1995}.

The loss of information in time depends on the bifurcation parameter $\mu$ as summarized in Fig.~\ref{fig:doublewell_infoloss}b. Note the peak in $I(X_{t},X_{t+\Delta t})$ for small, positive $\mu$. 
This corresponds to dynamics where observation of $X_t$ strongly informs the future state; recall that the mutual information is maximized when the conditional entropy $\mathcal{H}(X_{t+\Delta t}|X_t)\approx 0$ (\blue{SI Section~\ref{si_sec:background}}).
In contrast, for large positive or negative $\mu$, $X_t$ is not as informative of $X_{t+\Delta t}$ even for small times: the initial state is quickly forgotten as the particle approaches the bottom of the single (for $\mu<0$) or double (for $\mu>0$) well.

This phenomenon is reminiscent of critical slowing down, which occurs in the noise-free system as $\mu$ passes through the bifurcation at $\mu=0$. 
For the deterministic dynamics, the slowing down is reflected in the spectrum as a ``pile up'' of eigenvalues to form a continuous spectrum \cite{gaspard1995}.
In the presence of noise, although the continuous spectrum becomes discrete \cite{gaspard1995, froyland2013} there is still a pile-up of eigenvalues characterized by several eigenvalues becoming close to 1 (Fig.~\ref{fig:doublewell_infoloss}c).
This pile-up gives rise to the information peak seen in Fig.~\ref{fig:doublewell_infoloss}b.
The peak is not solely due to the closing spectral gap $\lambda_1-\lambda_2$, but is also impacted by the subdominant eigenvalues which accumulate at $\mu\approx 0.2$ (SI Fig.~\ref{si_fig:gapfig}).

\begin{figure*}[tp]
    \includegraphics[width=\textwidth]{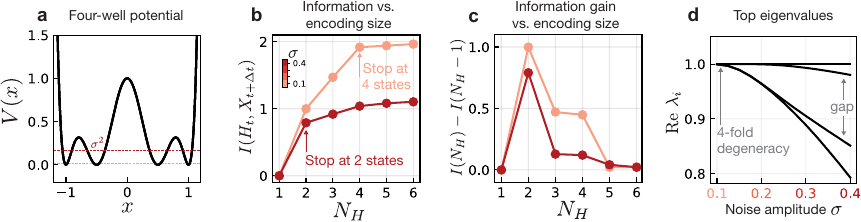}
    \caption{\label{fig:4well}\textbf{``Knowing when to stop''.} 
    The spectral properties of the transfer operator determine the necessary complexity (i.e. ``when to stop'' \cite{Cvitanovic2012}) of the reduced model, which is reflected in mutual information.
    (a) Four-well potential in which a Brownian particle fluctuates. The magnitude $\sigma$ of the fluctuating noise is related to an energy scale $E_{\sigma}= \sigma^2$.
    (b) Information contained in the encoding variable $H_t$ about the future state $X_{t+\Delta t}$ for varying levels of noise and alphabet sizes $N_H$.
    (c) Information gain achieved by increasing the alphabet size by a single variable. This is the discrete derivative of the curve in (b).
    (d) Spectrum of the transfer operator for changing values of noise amplitude.
    }
\end{figure*}

\section{Characterizing latent variables}%{Discrete latent space}
\label{sec4:characterizing_latent_variables}

In the next two sections, we focus on the case of a discrete latent space (like in Sec.~\ref{sec:infodecay}) in order to precisely characterize and demonstrate numerically two features of the latent variables: (i) the size of the latent space effectively satisfies a law of diminishing returns, in the sense that the information gain per additional extra state quickly decreases after some point, and (ii) the dynamics of latent variables are in general not Markovian, even if the underlying dynamics are Markovian; this non-Markovianity can be quantified and in some cases reduced.

\subsection{Knowing when to stop}
\label{sec:whentostop}
For discrete encoding variables $h$, the information bottleneck partitions state space and reduces the dynamics on $x$ to a discrete dynamics on $h$.
Such reductions of complex systems to symbolic sequences via partitioning of state space have attracted attention for more than half a century in both theoretical and data-driven contexts~\cite{ulam1960,li1976_ulam,Crutchfield1983,berman2016flies,ahamed2021,costa2023,murphy2023optimized,hathcock_dillavou_stochastic_2025}.
Several works have approached this partition problem from a dynamical systems perspective, linking optimal partitions to eigenfunctions of the (adjoint) transfer operator \cite{Cvitanovic2012,froyland2005}.
In this setting, a central question is ``when to stop'' \cite{Cvitanovic2012,gaspard1993,ahamed2021,costa2023}: 
how many states does $h$ need in order to capture statistical properties of the original dynamics? 

We consider this question by finding the optimal IB encoder in the limit of low compression, $\beta \gg 1$, but fixed encoding capacity $N_H$ (where $H_t\in\{0,...,N_H-1\}$), i.e. the encoder is only restricted by the number of symbols it can use. An analogous setup was used in the context of renormalization group (RG) transformations in \cite{Goekmen2021,PhysRevE.104.064106,Goekmen2023}, which results in effective model reduction due to the ``sloppiness'', or irrelevance, of certain system variables \cite{machta2013_science,transtrum2015_sloppiness_review}.
In this regime, the encoder learned by IB is deterministic; we are learning an optimal hard partition of state space. This can be seen by noting that $I(H_t;X_{t+\Delta t})=\mathcal{H}(H_t)-\mathcal{H}(H_t|X_{t+\Delta t})$ is maximized when the latter term is zero, which happens when $x_t$ unambiguously determines $h_t$, i.e. when $p(h_t|x_t)\in \{0,1\}$ for all $x_t$. The optimal encoder is computed using the Blahut-Arimoto algorithm (Ref.~\cite{tishby_IB,cover2012elements} and SI Section~\ref{si_sec:background}).

Consider a fluctuating Brownian particle as in the double well above, where now each of the wells is split into two smaller wells, giving a total of four potential minima (Fig.~\ref{fig:4well}a).
As the system is in steady state, the variance of the fluctuations defines an energy scale $E_{\sigma}=\sigma^2=2k_\text{B} T$. 
For small $E_\sigma$, the system rarely transitions between the four potential minima. In this case, knowledge of the initial minimum is very informative of the future state of the particle.
In contrast, for large fluctuations the particle can spontaneously jump between shallow minima in each large well, so that the system immediately forgets about the precise potential minimum it was in. Information about the shallow minima has been ``washed out'', and only the information about the larger double-well structure remains.

To see this reflected in the information, we again consider an encoding of the initial state into a discrete variable $H_t\in\{0,...,N_H-1\}$.
In both the small and large noise scenarios, a variable with $N_H=2$ encodes approximately one bit of information (Fig.~\ref{fig:4well}b), corresponding to an $H_t$ which distinguishes the two large wells for $x\lessgtr 0$. For large noise this is essentially all the information that can be learned; increasing the capacity of the encoding variable beyond this provides only marginally more information about the future state (Fig.~\ref{fig:4well}c).
In the small noise case, the information between the encoding and the future state continues to increase to approximately two bits at $N_H=4$, after which it plateaus. The encoding has learned to distinguish each of the four potential wells.
These observations are reflected in the transfer operator spectrum shown in Fig.~\ref{fig:4well}d. 
For small noise, the eigenvalue $\lambda=1$ is nearly four-fold degenerate, indicating the existence of four regions that can evolve independently under $U$, giving rise to four steady state distributions satisfying $U\rho \simeq \rho$. These regions correspond to the potential minima. Hops between the separate minima are exceedingly rare, so that the dynamics essentially take place in the four minima independently.
With larger $\sigma$ the degeneracy is lifted, resulting in one dominant subleading eigenvalue followed by a gap. The corresponding eigenfunction is one which is positive (negative) on the right (left) side of the large potential barrier at $x=0$: the only relevant piece of information is which of the large wells the initial condition is contained in, and all other information is lost exponentially quickly.

In many cases, our goal is not only to find a reduced representation $H_t$ that is predictive of the future state, but one which may be evolved forward for long times as a surrogate for the full system. 
This is a practical but much more stringent requirement of our reduced state, and it may result in a different answer to the question of ``when to stop'' which we explore in the following sections.

\subsection{Quantifying non-Markovianity}
\label{quantifying_non_markovianity}

This section is dedicated to the Markovianity of the latent variables (or the lack thereof), assuming an underlying Markovian dynamics. In practice, this assumption may not hold, for instance when experimental datasets only contain a partial description of the systems. In Sec.~\ref{sec:cyanobac}, we discuss this situation as well as potential remedial strategies.

In general the dynamics of the reduced variable will be non-Markovian, meaning that its evolution will be determined not only by its current state, but also its past. 
With a fixed encoder $p(h_t|x_t)$, transition probabilities $p(h_{t+1}|h_t)$ are induced by decoding $h_t$ to $x_t$, evolving $x_t$ forward in time to $x_{t+1}$, and then encoding $x_{t+1}$ to $h_{t+1}$ (here we assume discrete time dynamics with $\Delta t=1$ for simplicity). 
Information theory provides a natural quantification of the (non-)Markovianity of the dynamics as measured by the conditional mutual information (SI Section~\ref{si_sec:information_theory_for_dynamics}):
\begin{equation}
    \text{NM}(H_t) \equiv I(H_{t-1},H_{t+1}|H_t).
\end{equation}
This quantity tells us how much additional information the past state $H_{t-1}$ contains about the future $H_{t+1}$, on top of what is contained in the present state $H_t$ (see alternative approaches in Refs.~\cite{bialeknemenmen2001predictiveinfo,tchernookov2013,costa2023}). 

\begin{figure}[tp]
    \centering
    \includegraphics[width=0.45\textwidth]{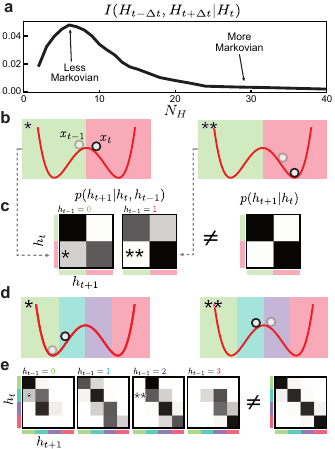}
    \caption{\textbf{Learned collective variables are generically non-Markovian}
    (a) Conditional mutual information $I(H_{t-\Delta t}, H_{t+\Delta t}|H_t)$ as a function of $N_H$.
    (b) Two possible scenarios (left and right) with the same value of $h_t$, but different $h_{t-1}$. Here, the circles represent the value of the full state $x_t$ while the colored regions represent the value of the reduced state $h_t\in\{1,2\}$ which are mapped to for each $x_t$.
    (c) Illustration of the transition probability which takes into account the previous state $h_{t-1}$ (left), compared to the Markovian transition probability obtained by marginalizing out $h_{t-1}$ (right).
    Black means the transition is likely, white means it is unlikely.
    In a Markovian system, these transition probabilities would coincide.
    (d-e) Same as in (h), but for $N_H=4$ states. 
    }
    \label{fig:doublewell_markov}
\end{figure}

To illustrate this, we again consider a particle in a double well, which we compress into a discrete variable $H_t\in\{0,...,N_H-1\}$.
Because the full system is Markovian by design, we expect that for large $N_H$ the conditional mutual information $I(H_{t-1},H_{t+1}|H_t)$ should go to zero as $H_t$ begins to more closely approximate the full state $X_t$ (Fig.~\ref{fig:doublewell_markov}a). 
Interestingly, the approach to Markovianity is highly non-monotonic, and the system is more non-Markovian with $N_H=4$ than with $N_H=2$.

The role of memory is illustrated in the two cases shown in Fig.~\ref{fig:doublewell_markov}b. 
In both cases, the current state is in the right well, so that $h_{t}=1$. However, more precise information about the state (and hence its future) is revealed by knowing the system's past. 
If $h_{t-1}=0$ (so the state was in the left well), it is likely that the full state is near the boundary between wells, so that $h_{t+1}=0$ is very likely. However, $h_{t+1}=0$ is very \emph{unlikely} when $h_{t-1}=1$, which happens when the state is deep in the right well.
This behavior can be quantified by computing the transfer probabilities $p(h_{t+1}|h_{t},h_{t-1})$ (Fig.~\ref{fig:doublewell_markov}c). These differ significantly from $p(h_{t+1}|h_{t})$, which would not be the case for Markovian dynamics.
The primary contribution to the non-Markovian dynamics occurs when $(h_{t-1},h_t)=(0,1)$ or $(1,0)$, which occurs when the state is near the boundary between the wells. 
As $N_H$ increases, the likelihood that the state is near the boundary of two regions with different $h_t$ labels will increase, which will increase the non-Markovianity of the system (Fig.~\ref{fig:doublewell_markov}d-e). However, at the same time the different $h_t$ regions will shrink in size so that $h_{t-1}$ will no longer provide any meaningful information about the system's location $x_t$. 
The trade-off between these two effects leads to the observed non-monotonic behavior of $I(H_{t-1},H_{t+1}|H_t)$.

\subsection{Learning effective dynamics of latent variables}
\label{sec:sinegordon}
In practice, latent variable identification is often only the first step of a procedure that finds the variables and their dynamics (Section~\ref{sec1:formalism}).
We now illustrate how our framework forms the basis for a complete model discovery pipeline by using sparse equation learning to extract relevant variable dynamics \cite{brunton_sindy,Kaptanoglu2022_pysindy}.
To learn continuous latent variables, we solve the IB optimization problem using an approximate variational objective introduced in Ref.~\cite{alemi2016_dvib} that can be solved with neural networks and which we refer to as variational IB. This approach is discussed in more detail in the following Section~\ref{sec5:applications} and in SI Section~\ref{si_sec:dvib}.

\begin{figure*}
    \centering
    \includegraphics[width=0.9\textwidth]{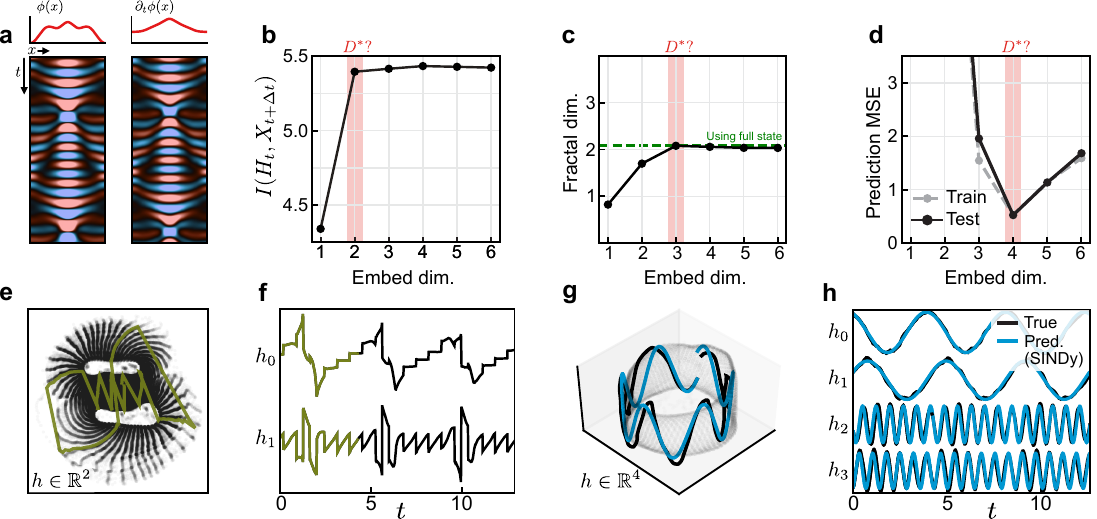}
    \caption{
    \textbf{``When to stop'' depends on what you want.}
    (a) Kymographs of the sine-Gordon system, discussed in the text. 
    Left shows the field $\phi(x)$ while the right shows time-derivative of the field.
    Initial conditions are shown above as curves.
    (b) Information content contained in the encoding for varying dimensionality of the embedding space. 
    While the information appears to plateau at $D^*=2$, this does not necessarily mean a ``proper'' embedding dimension has been reached. 
    (c) Estimated fractal dimension of the embedded dynamics for varying embedding dimension. The full fractal dimension of $\approx 2$ is only attained with an encoding dimension $D^*=3$.
    (d) Error of the predictions for the latent variable evolution produced by the dynamical system learned by SINDy. The error reaches a minimum at $D^*=4$. 
    (e) For $d=2$, the encoding manages to project the full torus into two dimensions by cutting and unfolding it (left). This introduces discontinuous jumps in the evolution of $h$.
    (f) Time evolution of both components of $h$, with highlighted part corresponding to the trajectory highlighted in (e).
    (g) Trajectory for a four-dimensional latent variable, visualized by projected onto the first three principal components. True trajectory shown in black, trajectory predicted by SINDy shown in blue with the equations given in the main text.
    (h) Predicted trajectory for all the degrees of freedom of $h$.
    }
    \label{fig:sinegordon}
\end{figure*}

We consider a system described by an underdamped, driven, one-dimensional PDE known as the sine-Gordon equation \cite{bishop_sinegordon} (Fig.~\ref{fig:sinegordon}a; see SI Section~\ref{si_sec:simulation_parameters}):
\begin{align}
    \partial_t^2\phi - \partial_x^2\phi +\sin\phi 
    =\varepsilon[-\alpha \partial_t\phi + \Gamma\cos(\omega t)].
\end{align}
For a particular choice of driving frequency $\omega$ and strength $\varepsilon\Gamma$, this system undergoes quasiperiodic dynamics, and hence lives on a toroidal manifold.

\subsubsection{When to stop?}
\label{when_to_stop_cont}

To identify the reduced variables for the system, we must make a choice about the dimensionality of $H_t$. 
As discussed above, the mutual information between $H_t$ and $X_{t+\Delta t}$ provides an indicator of ``when to stop'' (Fig.~\ref{fig:sinegordon}b). This metric plateaus already at $\dim H_t=2$.
The fractal dimension of the latent manifold, however, does not plateau until $\dim H_t=3$ (Fig.~\ref{fig:sinegordon}c). Interestingly, when using the encoding to learn a deterministic equation for the evolution of $H_t$ using SINDy, the prediction error is only minimized at $\dim H_t=4$ (Fig.~\ref{fig:sinegordon}d).

Each of these indicators of ``when to stop'' tells us something different about the system.
The encoding learned for $\dim H_t=2$ is shown in Figure~\ref{fig:sinegordon}e, with a portion of the trajectory highlighted.
Although the full dynamics takes place on a torus, the encoding neural network is able to ``cut'' the torus to embed it in two dimensions. The dynamics of the latent variables are completely deterministic: at any point on the latent manifold it is known where the system will evolve next, so that the information about the future is complete. 
However, the dynamics will be highly irregular, featuring large jumps in latent space (Fig.~\ref{fig:sinegordon}f).

The ``fraying'' visible at the boundary of the torus embedded in two dimensions leads to a fractal dimension which is less than two, although this difference may disappear in the limit of infinite data.
At three dimensions, the latent manifold is two dimensions everywhere (Fig.~\ref{fig:sinegordon}c).
However, while the torus may be embedded in three dimensions, it does not admit simple, linear dynamics until $\dim H_t=4$. 

\subsubsection{Symbolic equation learning}
% what about the things of Haller?

In dimension $\dim H_t=4$, SINDy is able to identify the correct linear equations of motion 
\begin{align}
    \dot h_0&=-\omega_1h_1 &\, \dot h_2&=\omega_2 h_3
    \nonumber
    \\
    \dot h_1&=\omega_1h_0 &\, \dot h_3&=-\omega_2 h_2        
    \label{eq:sinegordon_eq}
\end{align}
with $\omega_1=1.5$ and $\omega_2=8.7$ (Fig.~\ref{fig:sinegordon}g-h).

The above example illustrates the subtle difference between a model being the simplest in the sense of providing a minimal predictive set of variables, and the simplest in terms of the governing equations. The latter may benefit from the inclusion of informationally redundant variables. The IB objective, being purely information based, is indifferent to the geometric or analytic properties of the encodings. Such model properties, often desirable, as \emph{e.g.}~smoothness or linearity must be introduced as additional constraints.

% but what if data is non markovian ? see tests later and section xxx

\section{Applications}%{Continuous latent states}
\label{sec5:applications}
%\section{Variational IB for data-driven discovery of slow variables}
\subsection{von K{\'a}rm{\'a}n streets: variable discovery from YouTube videos}
\label{sec:dvib_fluids}

IB finds a reduced state variable by optimizing an information theoretic-objective that makes no reference to physics or dynamics. This suggests it may be used for the discovery of slow variables in situations where one lacks physical intuition.
In the regime of small $\beta$, or high compression, features of the state $x_t$ are forced to compete to make it through the bottleneck $h_t$, which allows us to extract relevant features of the system.

These variables coincide with left eigenfunctions of the transfer operator, as discussed in Section~\ref{sec2:opt_enc}.
We verify this theoretical result numerically in cases where the IB objective Eq.~\eqref{ib_obj} can be solved exactly by using an iterative scheme known as the Blahut-Arimoto algorithm \cite{tishby_IB,cover2012elements} (SI Section~\ref{si_sec:background} and \ref{si_sec:derivations}).
By taking the logarithm of the encoder as it begins to deviate from the trivial uniform encoder above $\beta_1$ we can confirm that it depends on $x$ only through $\phi_1(x)$ (SI Fig.~\ref{si_fig:doublewell_stability}a-d).
This can be independently verified by studying the stability of the trivial encoder with respect to perturbations in the parameters $f_n(h_t)$ in Eq.~\eqref{optimal_encoding_spectral}, where we see that the encoder becomes unstable to perturbations in $f_1$, as predicted by our theoretical results (\blue{SI Fig.~\ref{si_fig:doublewell_stability}e-i}).

The utility of exact IB for variable discovery is limited because it requires knowledge of the exact conditional distribution $p(x_{t+\Delta t}|x_t)$ which is difficult to estimate in many real-world scenarios.
Alternatively, as in the previous section, variational IB may be used to solve the IB objective function directly from data. 
This is achieved by introducing a tractable Ansatz for the form of $p(h_t|x_t)$; we show in SI Section~\ref{si_sec:derivations} that this does not change the encoder's dependence on transfer operator eigenfunctions.

\begin{figure}[tp]
    \centering
    \includegraphics[width=0.48\textwidth]{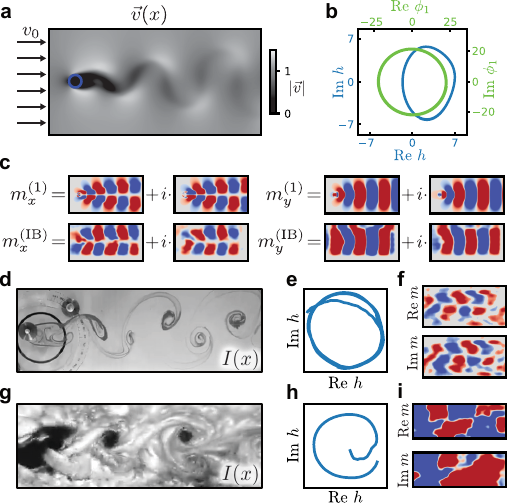}
    \caption{\textbf{Variational IB neural networks learn transfer-operator eigenfunctions in high-dimensional fluid flows.}
    (a) Fluid flow past a disk with uniform in-flow velocity $v_0$ exhibits vortex shedding behind the object in a so-called von K{\'a}rm{\'a}n street. The state of the system is given by a spatially varying two-component vector field $\vec{v}(\vec{x})$.
    (b) Dynamics in latent space (blue) traverse a nearly circular trajectory (see Supplementary Movie 1). 
    For comparison we show the evolution of the mode amplitudes obtained by projecting the velocity field onto the first DMD mode (green).
    (c) Comparison of the first Koopman mode obtained from DMD ($\vec{m}^{(1)}$) and from VIB ($\vec{m}^{(\text{IB})}$). Koopman modes from VIB are computed as gradients of the latent encoding variables as described in the main text.
    Red corresponds to positive values and blue to negative; 
    the magnitudes of the modes are not directly comparable.
    (d) Intensity field $I(x)$ from a video of water flowing past an obstacle; video from Ref.~\cite{vonkarman_youtube}.
    (e) Trajectory of the system in latent space. 
    (f) Gradients of the latent encodings with respect to the input field $I(x)$.
    (g-i) Same as (d-f), but using satellite images of clouds near the island of Guadalupe as our state $I(x)$.
    }
    \label{fig:cylinderflow_sim}
\end{figure}

Our core result linking the behavior of the learned encoder with transfer operator eigenfunctions remains valid even for variational IB applied to high-dimensional systems, which we show by considering a simulated data of fluid flow past a disk \cite{tencer2020tailored}.
The state of the system is given by a two-dimensional velocity field $\vec{v}(\vec{x}) \in \mathbb{R}^{2\times N_{\text{pixels}}}$, where $N_{\text{pixels}}\simeq \mathcal{O}(10^5)$ (Fig.~\ref{fig:cylinderflow_sim}a).
%From the full fields $x_t = [v_x, v_y]$ we learn a two-dimensional encoding $h_t\in \mathbb{R}^2$. 
Fluid flows in from the left boundary with a constant velocity $v_0\hat{e}_x$ past a disk of unit diameter. 
At Reynolds number $\text{Re}\gtrsim 150$, the fluid undergoes periodic vortex shedding behind the disk, forming what is known as a von K\'arm\'an street.

The eigenfunctions for this system are in general complex, and come in conjugate pairs: $\phi_2(x) = \phi^{*}_1(x)$. In this situation any linear combination of $\phi_1$ and $\phi_2$ will decay at the same rate, and hence we expect to learn some arbitrary combination of the two dominant eigenfunctions, or equivalently a combination of the real and imaginary parts of $\phi_1$. 
We therefore take a two-dimensional encoding variable $[h_0, h_1]$, so that it can represent the full complex eigenfunction.

Our learned latent variables are oscillatory with the correct frequency as shown in Fig. \ref{fig:cylinderflow_sim}b.
However, this alone does not indicate whether we are learning the correct eigenfunctions. 
Because the system is well approximated by linear dynamics, eigenfunctions of the adjoint transfer operator are linear functions of the state variable, and may hence be represented as $\phi_n[\vec{v}] = \langle \vec{v}(\vec{x}), \vec{m}^{(n)}(\vec{x})\rangle$, where angled brackets denote an average over space and $\vec{m}^{(n)}(\vec{x})$ are so-called Koopman modes.
The true $\vec{m}^{(n)}$ (and hence the eigenfunctions) can be thus computed via dynamic mode decomposition (DMD) \cite{rowley2009_dmd,schmid_2010_dmd} as described in the SI.
For the functions $h[\vec{v}]$ learned using variational IB, we can infer the corresponding $\vec{m}^{(\text{IB})}$ using gradients of the neural network with respect to the input field.
Fig.~\ref{fig:cylinderflow_sim}c shows these inferred modes $\vec{m}^{(\text{IB})}$ compared to the true mode $\vec{m}^{(1)}$ (\blue{SI Section~\ref{si_sec:vib_fluid}}).
This shows that variational IB not only recovers the essential oscillatory nature of the dynamics, but does so by learning the correct slowly varying functions of the state variable given by the adjoint transfer operator eigenfunctions.

%\section{Relevant variable identification in laboratory-generated and atmospheric flows}
%\label{sec:exptmovies}

Our framework provides interpretability for the learned latent variables even in messy real-world fluid flow datasets scraped directly from videos on Youtube \cite{vonkarman_youtube, noaa_youtube} (Supplementary Movie 1).
The first shows a von K\'arm\'an street which forms as water passes by a cylindrical obstacle at Reynolds number 171, with flow visualized by a dye injected at the site of the obstacle \cite{vonkarman_youtube}.
We take a background-subtracted grayscale image of the flow field as our input (Fig.~\ref{fig:cylinderflow_sim}d) and task VIB with learning a two-dimensional latent variable as above.  
Also here, variational IB learns oscillatory dynamics of the latent variables (Fig.~\ref{fig:cylinderflow_sim}e).
We visualize the function learned by the encoder by considering gradients of the latent variables, which show the same structure as those obtained for the $x$ component of the simulated data (Fig.~\ref{fig:cylinderflow_sim}f). 
This is expected, as the $x$-component of the velocity field has similar glide reflection symmetry as the intensity image.

Finally, we apply variational IB to a von K\'arm\'an street arising due to air flow around Guadalupe Island, which was imaged by a National Oceanic and Atmospheric Administration (NOAA) satellite \cite{noaa_youtube} (Fig.~\ref{fig:cylinderflow_sim}g). 
Although the video consists of only 62 frames and less than one full oscillation, the variational IB neural network learns latent variables which capture this oscillation and have the expected dependence on the input variables (Fig.~\ref{fig:cylinderflow_sim}h-i). In general the performance of VIB in the low-data limit is susceptible to overfitting \cite{abdelaleem2025_MI_est}; here this effect is mitigated by the low effective dimensionality of the system.
As in the first experimental example, the gradients of the encoding variables show the glide symmetry of $m_x$ due to the symmetry of the intensity pattern in Fig.~\ref{fig:cylinderflow_sim}g.

\subsection{Cyanobacteria: latent variable discovery in experimental data}
\label{sec:cyanobac}

\begin{figure*}[htp]
    \centering
    \includegraphics[width=0.95\textwidth]{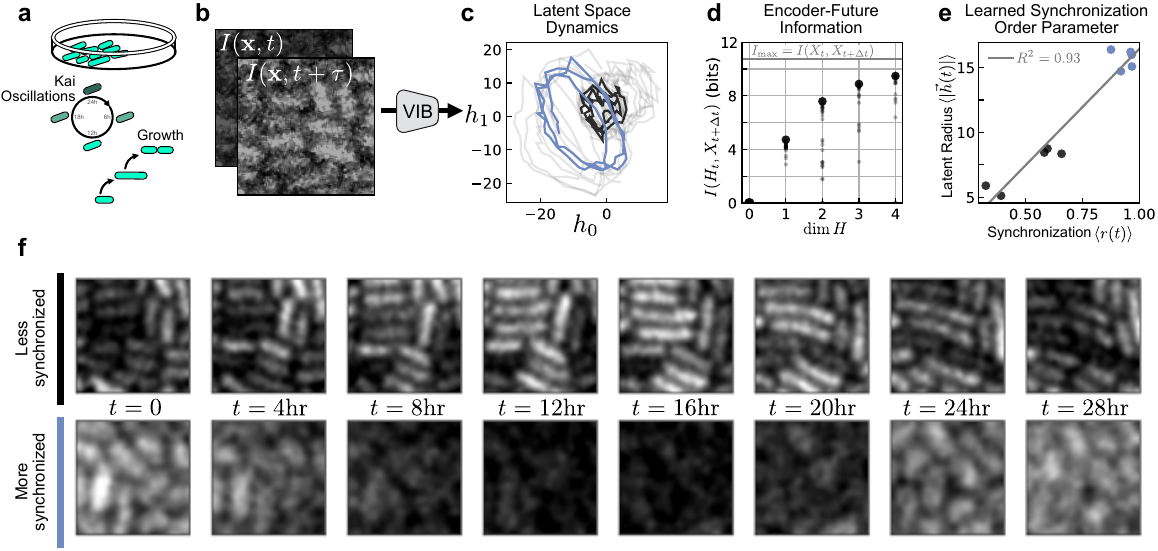}
    \caption{\textbf{Discovering slow collective variables in cyanobacteria populations} 
    (a) Fluorescent images of cyanobacteria colonies labeled with EYFP driven by the \textit{kaiBC} promoter, allowing the visualization of Kai protein transcription. 
    Colonies are imaged as they undergo cell growth and Kai expression oscillations associated with the circadian cycle.
    (b) The ``state'' used for variational IB is the time-lagged intensity field with lag time $\tau$ (see SI Fig.~S13 for details).
    (c) Variational IB embeddings of time-lagged images into two dimensions (see Supplementary Movie 2). Every line corresponds to one colony's evolution in time. Note the apparent oscillations of different radii. One large-radius (blue) and one small-radius (black) trajectory are highlighted.
    (d) Mutual information between the future state and the encoding, given by $I(X_{t+\Delta t}, H_t)$, for varying dimension of the latent space. 
    Each small point represents one training instance of the variational IB model, while the large point shows the maximum estimated value. 
    $I_{\max}=I(X_t,X_{t+\Delta t})$ is the mutual information estimated for the full dynamics. 
    (e) Mean radius versus synchronization parameter for each colony (SI Section S9). VIB identifies clusters of cells characterized by high (blue) or low (black) synchronization.
    (f) Sample time series of weakly (black) and highly (blue) synchronized colonies. We apply a Gaussian blur to better visualize the bacteria boundaries.}
    \label{fig:bacteria}
\end{figure*}

Variational IB may be used as an aid for collective variable discovery in situations where physical intuition may not be a useful guide -- collective behavior of biological organisms  (Supplementary Movie 2).
Here, we ask what the most predictive variables are for predicting the evolution of populations of cyanobacteria (\textit{Synechococcus elongatus}). The dynamics of the colonies are driven by several factors: growth and division of individual bacteria, translational motion of groups of bacteria as they are pushed by their neighbors, as well as the circadian oscillations within each bacterium (Fig.~\ref{fig:bacteria}a).
These oscillations are controlled by three Kai proteins ~\cite{Ishiura1998} and depend in particular on the ratios of the copy number of these proteins which can be tuned experimentally \cite{ChewRust2018}. %chew_ref_16?,chew_ref_17?}.

We were provided with videos of ten cyanobacteria colonies that were grown under different conditions.% that impact their dynamics. However, as a test of our method, we were blinded to these conditions until we had performed our analysis.
The videos are sequences of fluorescent images, taken once per hour, which show the clock state of each individual bacteria visualized with a fluorescent marker EYFP driven by the \textit{kaiBC} promoter. 
Here, we focus on collective variables which are predictive of the state of the interior of the colony and not the growth in area of the colonies. We therefore crop the images to the interiors of each colony (SI Fig.~\ref{sifig:cyanobac}) to isolate bacterial motion and fluorescence oscillations inside each colony .

Our input to the variational IB neural network are these cropped images augmented with a time-lagged image of the same region (Fig.~\ref{fig:bacteria}b-c).
The purpose of this time lag is to make the dynamics Markovian: because the intensity field oscillates, an observation at a single time point does not allow to determine whether the intensity is increasing or decreasing. In principle more time-lags may be necessary, and their number could be estimated from the behavior of the mutual information or the transition entropy (Ref.~\cite{costa2023} and SI Section~\ref{si_sec:information_theory_for_dynamics}). As these quantities are difficult to measure from images with limited computational resources, we keep only one lag: the state is given by pairs $X_t = \{I(\mathbf{x}, t), I(\mathbf{x}, t+\tau)\}$, where $\tau$ is the lag duration. Here we take $\tau=3$ hr and a prediction horizon $\Delta t=8$ hr, but find that different $\Delta t$ or $\tau$ does not change our results (SI Fig.~\ref{sifig:cyanobac}).

% \clearpage
% \newpage

With variational IB we compress the state $X_t$ into a latent variable $h$ of variable dimension. We train the neural network on the entire dataset of all ten colonies simultaneously.
The dynamics in latent space undergo clear oscillations, indicating that the relevant variables encode primarily the intensity fluctuations rather than, for example, the spatial locations of the bacteria.
Notably, the trajectories are essentially two-dimensional, even when the encoding space is higher dimensional.
This is reflected in the information retained about the future state, $I(X_{t+\Delta t}, H_t)$: increasing the dimension of the embedding space beyond two only marginally increases $I(X_{t+\Delta t}, H_t)$; this tells us ``when to stop'' (Fig.~\ref{fig:bacteria}d and SI Fig.~\ref{sifig:cyanobac_pca}). %We independently verify this by using principal component analysis to characterize the geometry of embedded trajectories, and find that even in higher dimensions the trajectories occupy a two dimensional subspace. 
In the following, we thus restrict our focus to $\dim H=2$.

We noticed that there were notable differences in the radius of latent space oscillations from colony to colony, two of which are highlighted in Figure~\ref{fig:bacteria}c. 
To understand this difference, we examined the original microscopy time series corresponding to both large and small latent radius (Fig.~\ref{fig:bacteria}e-f) and found that while the large-radius sample showed clear, nearly uniform oscillations in intensity, the small-radius samples appeared much more heterogeneous. 

To quantify this we consider each pixel to be an independent oscillator, akin to a spatial Kuramoto model~\cite{Ren1998, Hong2005, Laing2011}, and compute a global synchronization order parameter $r(t)$ (SI Section~\ref{si_sec:bacteria}). 
For each colony we calculate the time-averaged synchronization $\langle r(t)\rangle_t$ and find that two clusters emerge corresponding to high and low synchronization (Fig.~\ref{fig:bacteria}e).
These clusters are precisely those representing trajectories of large and small latent radius, suggesting that variational IB learns to encode the synchronization of the colony in the latent variable radius. 
As a check, we perform IB on a simulated locally-coupled Kuramoto model as a system sharing many features of the experimental system. Here we also learn an encoding in which the latent radius corresponds to the synchronization order parameter (SI Fig.~\ref{sifig:kuramoto}).

In Appendix~\ref{main_apx:VIB_comparison} and Fig.~\ref{sifig:comparison} we compare the performance of variational IB to several other model reduction methods and find that IB delivers more interpretable and well-behaved features. This is likely due to the fact that many standard methods for data-driven model reduction rely on assumptions about the dynamics which may not be appropriate in the case at hand, such as linearity.
%(in the case of DMD) or diffusive dynamics (in the case of diffusion maps \cite{coifman2006_diffusionmaps}). 
Even among deep learning methods free of such assumptions, such as time-lagged autoencoders, the variables learned by IB appear more interpretable. 
This increased interpretability is likely due to the compression term which effectively regularizes the latent space by encouraging the network to learn slow transfer operator eigenfunctions.
While there are many specific variants of DMD \cite{williams2015,Chen2012_dmdvariants,kutz2015multiresolution,Wynn2013_optimaldmd,colbrook_mpedmd} or autoencoders for dynamics \cite{takeishi2017_neurips,lusch2018,bakarji2023,wehmeyer2018_timelaggedautoencoder} that may outperform variational IB in some cases, we find that in this real-world example it yields the smoothest and most interpretable latent variables without tailored preprocessing (Appendix~\ref{main_apx:VIB_comparison} and Fig.~\ref{sifig:comparison}).

By using variational IB we could reduce a complex system with multiple dynamical components -- cell growth, division, and gene expression fluctuations -- into a low dimensional form that retains only the most relevant information for the future. In addition to the insight that the dynamics are dominated by oscillations in two dimensions, the latent variables clearly distinguished trajectories into two groups that were not apparent \textit{a priori}.
We were provided this data as a ``blind'' test with no knowledge of the underlying system.
After we performed our analysis, it was revealed to us that these bacterial colonies have been engineered to control the translational efficiency of the Kai proteins by varying theophylline concentration~\cite{ChewRust2018}. 
The synchronization order parameter discovered by variational IB corresponds to differing experimental concentrations of theophylline, which is in agreement with the findings in Ref.~\cite{ChewRust2018}.
IB can thus serve as a way to connect experimental control parameters to effective changes in dynamics. 

%The circadian clock, ubiquitous in both eukaryotes and prokaryotes, is a 24-hour biochemical oscillation that allows organisms to anticipate the day-night cycle~\cite{Harmer2001}.
%As with any clock, its utility lies in its ability to predict the future~\cite{Woelfle2004}.
%Here we address the question -- how does the accuracy of individual circadian oscillations manifest in collective behavior?

\section{Conclusion}
We have introduced an information-theoretic framework for dimensionality reduction in dynamical systems where relevant latent variables are identified based on how predictive they are about the future state of the system.
%When a separation of timescales exists, our approach reduces to standard schemes in the limit of high compression; it also goes beyond this situation and remains well-defined even when no formal separation of timescale is available.
%In addition to our main results on the discovery of optimal latent variables, 
In addition to formulating the core compression problem and characterizing its solution, our information-theoretic framework provides diagnostic tools to directly assess the goodness of the dimensionality reduction. 
In particular, we provide guidelines to decide when to stop increasing the size of the latent space given an objective, and to assess how (non-)Markovian the resulting latent dynamics are.

Our latent variable discovery procedure can be implemented exactly for low-dimensional discrete systems, or approximately with deep neural networks that can be deployed on continuous real-world data.
Importantly, our mathematical characterization of the solution the optimization problem allows us to interpret the discovered latent variables in terms of spectral properties of the transfer operator.
With cyanobacteria experiments as a case study, we have shown how variational IB can discover latent variables directly from experimental image data without significant preprocessing.
We also showed how to combine these tools with existing equation-learning methods such as SINDY to produce a fully-fledged data-driven model discovery pipeline.
In the SI we showcase further applications, including to chaotic systems, where they may yield computationally efficient methods to calculate chaotic invariants such as fractal dimensions directly from experimental data.
%We have shown how an information-theoretic approach facilitates dimensionality reduction for dynamical systems. 
%In the framework we have developed, the problem of learning optimal latent variables is shown to decouple from the problem of learning the latent evolution operator. 
%This simplification opens up the dimensionality reduction procedure to both analytical study and efficient numerical implementations.
%In particular, we can show that the reduced variables may be connected to the spectrum of the transfer operator. 
%In the limit of high compression, these variables are precisely the slow variables identified by approaches that use a separation of time scales as a starting point.

%This result allows us to interpret the latent variables learned by variational IB, a neural-network implementation of the information bottleneck that discovers latent variables directly from data.
%We use these networks to demonstrate how a purely information-theoretic objective function leads to the discovery of physically-relevant latent variables. Combined with equation-learning methods such as SINDY we can even discover their governing equations.
%In the SI we showcase further applications, including to chaotic systems, where they may yield computationally efficient methods to calculate \emph{e.g.}~fractal dimensions.

Going forward, we expect our framework to find uses in domains ranging from climate sciences where an overarching goal is to decouple as much as possible long-term climate from short-term weather despite the fact that many time scales are entangled \cite{Ghil2020,Lucarini2023} to systems biology where, despite their apparent complexity, systems ranging from single cells or cell collectives \cite{heimberg_low_2016,li_measuring_2024,achar2022francois,dubnov_identifying_2024,avraham-davidi_spatially_2025,lee_topological_2025} to entire organisms like plants \cite{plants_meroz_2024,bastien_kinematics_meroz_2016} and animals~\cite{stephens08worms,berman2016flies,seyboldt2022francois} can often be captured by low-dimensional descriptions which are difficult to identify by biological or physical considerations alone.

\section*{Acknowledgments}
The authors would like to thank Z. Ringel, M. Han and D.E. G{\"o}kmen for helpful discussions.
%M.K.-J. would like to thank Zohar Ringel for discussions.
M.S.S. acknowledges support from a MRSEC-funded Graduate Research Fellowship (DMR-2011854).
M.K.-J. gratefully acknowledges financial support from the European Union’s Horizon 2020 programme under Marie Sklodowska-Curie Grant Agreement No. 896004 (COMPLEX ML).
D.S.S. acknowledges support from a MRSEC-funded Kadanoff–Rice fellowship and the University of Chicago Materials Research Science and Engineering Center (DMR-2011854).
V.V. acknowledges support from the Simons Foundation, the Complex Dynamics and Systems Program of the Army Research Office under grant W911NF19-1-0268, the National Science Foundation under grant DMR-2118415 and the University of Chicago Materials Research Science and Engineering Center, which is
funded by the National Science Foundation under award
no. DMR-2011854.
All authors acknowledge support from the UChicago Research Computing Center which provided the computing resources for this work.

\widetext

\appendix
%\section{Appe}
%If it reduces our uncertainty a lot, these variables will have high mutual information, and if it doesn't reduce the uncertainty at all then the variables must be independent. %An equivalent formulation of mutual information is the distance (measured as a Kullback-Leibler divergence) between the joint distribution and a distribution of independent $X$ and $Y$, \eqref{meth:midef_dkl}.

\section{Mutual information as a measure of prediction error}
\label{main_apx:MI_from_DKL}
Here we show that mutual information maximization emerges naturally as a measure of non-commutativity of the diagram Eq.~\eqref{eq:noncommutative_diagram} for probabilistic encoder/decoders.
Let us measure the distance between the true distribution of future states and the decoded distribution of future states as 
\begin{align}
    d_{x,h'} = D_{\text{KL}}\left[p(X_{t+\Delta t}|X_{t}=x)\Big\lVert p(X_{t+\Delta t}|H_{t+\Delta t}=h') \right]
\end{align}
This compares how well the latent future state $h'$ predicts the distribution of $X_{t+\Delta t}$ compared to what would be obtained by evolving the true initial state $x$ forward in time. Note that $h'$ follows the distribution $p(h'|x)=\int \dd h_t\,p(h'|h_t)p(h_t|x)$ which results from first encoding $X_t$ to $H_t$ and then evolving $H_t$ to $H_{t+\Delta t}$. 

The above Kullback-Leibler divergence compares the distributions for one particular pair of values $x$ and $h'$. However, we are generally interested in the average distance over all possible values:
\begin{align}
    d =\mathbb{E}_{p(X_t,H_{t+\Delta t})}\left[d_{x,h'}|X_{t}=x,H_{t+\Delta t}=h'\right].
\end{align}
This quantity can be written in terms of mutual information:
\begin{align}
    d&=\mathbb{E}_{p(X_t,H_{t+\Delta t},X_{t+\Delta t})}\left[\log \frac{p(X_{t+\Delta t}|X_t)}{p(X_{t+\Delta t}|H_{t+\Delta t})}\right]
    %\\
    %&=\int\dd x\dd h'\dd x'\, p(h',x)\,p(x'|x)\log \frac{p(x'|x)}{p(x')}\frac{p(x')}{p(x'|h')}
    \\
    &=\mathbb{E}_{p(X_t,X_{t+\Delta t})}\left[\log \frac{p(X_{t+\Delta t}|X_t)}{p(X_{t+\Delta t})}\right] - \mathbb{E}_{p(X_{t+\Delta t},H_{t+\Delta t})}\left[\log \frac{p(X_{t+\Delta t}|H_{t+\Delta t})}{p(X_{t+\Delta t})}\right]
    \\
    &=I(X_{t},X_{t+\Delta t})-I(H_{t+\Delta t},X_{t+\Delta t})
\end{align}
To write the first line, we used the definition of the Kullback-Leibler divergence and the fact that variables follow the Markov chain $H_{t+\Delta t}-X_t-X_{t+\Delta t}$, which allowed us to write the expectation over the full joint distribution. To get the middle line we first multiplied the argument of the logarithm with $1=p(X_{t+\Delta t})/p(X_{t+\Delta t})$ and then integrated over $H_{t+\Delta t}$ in the first term and $X_t$ in the second. 
The final result shows that minimizing $d$ is equivalent to maximizing the mutual information $I(H_t,X_{t+\Delta t})$ because the first term is fixed.

The above derivation does not incorporate the fact that we use a stochastic decoder $\tilde{h}^{-1}$ which follows the distribution $q(X_{t+\Delta t}|H_{t+\Delta t})$. Instead, we implicitly assumed that we could obtain the predicted distribution $p(X_{t+\Delta t}|H_{t+\Delta t})$ by performing the decoding $H_{t+\Delta t}\rightarrow H_t\rightarrow X_t\rightarrow X_{t+\Delta t}$ using Bayes' rule. However, in some cases we may want to use a decoder which may be imperfect (for example because it is more computationally tractable).
The distance $d$ resulting from using our imperfect decoder $q$ can be derived similarly to the above
\begin{align}
    d&=\mathbb{E}_{p(X_t,H_{t+\Delta t})}\left[D_{\text{KL}}\left[ p(X_{t+\Delta t}|X_t)\Big\lVert q(X_{t+\Delta t}|H_{t+\Delta t})\right]\right]
    \\
    &=\mathbb{E}_{p(X_t,H_{t+\Delta t},X_{t+\Delta t})}\left[\log \frac{p(X_{t+\Delta t}|X_t)}{q(X_{t+\Delta t}|H_{t+\Delta t})}\right]
    \\
    %&=
    %\mathbb{E}_{p(X_t,X_{t+\Delta t},H_{t+\Delta t})}\left[\log\frac{\tilde{p}_{X_{t+\Delta t}|H_{t+\Delta t}}}{ p_{X_{t+\Delta t}|X_{t}}}\right]
    %\\
    %&=
    %\mathbb{E}_{p(\dots)}
    %\left[\log\frac{\tilde{p}_{X_{t+\Delta t}|H_{t+\Delta t}}}{ p_{X_{t+\Delta t}|H_{t+\Delta t}}}\right]+
    %\mathbb{E}_{p(\dots)}\left[\log\frac{p_{X_{t+\Delta t}|H_{t+\Delta t}}}{ p_{X_{t+\Delta t}|X_{t}}}\right]
    %\\
    &=
    I(X_{t},X_{t+\Delta t}) - I(H_{t+\Delta t},X_{t+\Delta t})
    +
    \mathbb{E}_{p(H_{t+\Delta t})}
    \left[D_{\text{KL}}\left[q(X_{t+\Delta t}|H_{t+\Delta t})\Big\lVert p(X_{t+\Delta t}|H_{t+\Delta t})\right]\right]
    \\
    &\equiv
    I(X_{t},X_{t+\Delta t}) - I(H_{t+\Delta t},X_{t+\Delta t})
    +
    \delta_{\text{dec}}[h,\tilde{h}^{-1}]
\end{align}
This distance is now given by the mutual information as before, plus a term $\delta_{\text{dec}}$ that measures the quality of our decoder. We retain its dependence on the (stochastic) encoder $h$ and decoder $\tilde{h}^{-1}$ for the next section.
This term is greater than or equal to zero, and is minimized when the decoder step gives the same result as decoding $H_{t+\Delta t}\rightarrow H_t\rightarrow X_t\rightarrow X_{t+\Delta t}$ with Bayes' rule. 
It is therefore \emph{known} what decoder minimizes $\delta_{\text{dec}}$ (and hence which minimizes $d$). This is in contrast to the situation for the encoder $h$, whose optimal solution is unknown \textit{a priori}.

\section{Separating concerns with the data processing inequality}
\label{main_apx:information_processing_inequality}

Here we show how to obtain Eq.~\eqref{eq:modelreduction_objective_simple} in the main text.
The starting point is Eq.~\eqref{eq:objective_symbolic}, represented here explicitly
\begin{align}
    h, \mathcal{V}_{\Delta t}, \tilde{h}^{-1}
    &=\argmin_{h, \mathcal{V}_{\Delta t}, \tilde{h}^{-1}}  \dim h(X_t)
    + \beta\, d\left((\tilde{h}^{-1}\circ \mathcal{V}_{\Delta t}\circ h)(X_t)), X_{t+\Delta t}\right)
    \\
    &=\argmin_{h, \mathcal{V}_{\Delta t}, \tilde{h}^{-1}} \mathcal{L}(h, \mathcal{V}_{\Delta t}, \tilde{h}^{-1}).
\end{align}
Here we write $h, \mathcal{V}_{\Delta t}, \tilde{h}^{-1}$ and their composition as functions. These functions may be stochastic, which means that the their value is sampled from a distribution, e.g. $H_{t+\Delta t}=(\mathcal{V}_{\Delta t}\circ h)(X_t)$ is drawn from the distribution 
\begin{align*}
    p(h_{t+\Delta t}|x_t)=\int\dd h_t\,\underbrace{p(h_{t+\Delta t}|h_t)}_{\text{\enquote{$\mathcal{V}_{\Delta t}$}}}\underbrace{p(h_{t}|x_t)}_{\text{\enquote{$h$}}}.
\end{align*}

There are two ways to introduce a distance function which compares the statistics of the decoded future state and the true future state. The first is simply to set $d=-I\left((\tilde{h}^{-1}\circ \mathcal{V}_{\Delta t}\circ h)(X_t)), X_{t+\Delta t}\right)$. The second is to compare the distributions of decoded future states and true future states my measuring the Kullback-Leibler divergence between them, as we do in Appendix~\ref{main_apx:MI_from_DKL}. Both of these perspectives lead to the same result. For now, we focus on the first approach and comment on the second approach after.

By replacing minimization of the distance $d$ with maximization of the mutual information, we have
\begin{align}
    \mathcal{L}_\text{MI}(h, \mathcal{V}_{\Delta t}, \tilde{h}^{-1})&=\dim h(X_t)
    - \beta\, I\left((\tilde{h}^{-1}\circ \mathcal{V}_{\Delta t}\circ h)(X_t)), X_{t+\Delta t}\right).
\end{align}
The variables in the above equation form the Markov chain
\begin{align*}
    X_{t+\Delta t}
    - X_{t}
    - h(X_t)
    - \mathcal{V}_{\Delta t}(h(X_t)) 
   - \tilde{h}^{-1}(\mathcal{V}_{\Delta t}(h(X_t) ))
\end{align*}
where $X\,\,\text{---}\,\,Y\,\,\text{---}\,\,Z$ means that $Z$ is conditionally independent of $X$, given $Y$: $p(z,x|y)=p(z|y)p(x|y)$.
As a consequence of the data processing inequality, for any transformation $Z=g(Y)$ one has
\begin{align*}
    I(X,g(Y))\leq I(X,Y).
\end{align*}
where equality is obtained if $g$ is a bijection \cite{cover2012elements}.
Consequently, we have
\begin{align*}
    I(\tilde{h}^{-1}(\mathcal{V}_{\Delta t}(h(X_t) )), X_{t+\Delta t})
    \leq
    I(\mathcal{V}_{\Delta t}(h(X_t) ), X_{t+\Delta t})
    \leq
    I(h(X_t) , X_{t+\Delta t}),
\end{align*}
where the inequalities are maximized if $\tilde{h}^{-1}$ and $\mathcal{V}_{\Delta t}$ are both bijective functions. Notably, $\mathcal{V}_{\Delta t}=\text{identity}$ is a solution. However, this should not be taken to mean that $H_t$ has no dynamics, as it will still have dynamics induced by the evolution of $X_t$.  
Consequently, 
\begin{align}
    \min_{h, \mathcal{V}_{\Delta t}, \tilde{h}^{-1}}  \mathcal{L}_\text{MI}(h, \mathcal{V}_{\Delta t}, \tilde{h}^{-1})
    &=
    \min_{h, \mathcal{V}_{\Delta t}, \tilde{h}^{-1}}  \dim h(X_t)
    - \beta\, I\left((\tilde{h}^{-1}\circ \mathcal{V}_{\Delta t}\circ h)(X_t)), X_{t+\Delta t}\right)
    \\
    &=
    \min_{h}\left(\dim h(X_t)
    - \beta\, \min_{\mathcal{V}_{\Delta t}, \tilde{h}^{-1}}  I\left((\tilde{h}^{-1}\circ \mathcal{V}_{\Delta t}\circ h)(X_t)), X_{t+\Delta t}\right)\right)
    \\
    &=
    \min_{h}\left(\dim h(X_t)
    - \beta\, I\left(h(X_t)), X_{t+\Delta t}\right)\right).
\end{align}
This shows how the only quantity we need to learn is the encoder $h$. The  latent evolution $\mathcal{V}_{\Delta t}$ and decoder $\tilde{h}^{-1}$ which minimize $\mathcal{L}_{\text{MI}}$ are induced by the encoding. While it may be difficult to numerically approximate these functions, they are at least known in principle, which is not the case for $h$.

We now illustrate how measuring the distance $d$ as a Kullback-Leibler divergence between the true distribution of future states and the predicted one leads to the same result. Plugging the distance $d$ derived in Appendix~\ref{main_apx:MI_from_DKL} into the full objective yields
\begin{align}
    \mathcal{L}'_\text{MI}(h, \mathcal{V}_{\Delta t}, \tilde{h}^{-1})&=\dim h(X_t)
    - \beta\, I\left((\mathcal{V}_{\Delta t}\circ h)(X_t), X_{t+\Delta t}\right)+\beta\delta_{\text{dec}}[h,\tilde{h}^{-1}].
\end{align}
Because of the same Markov chain property used above, we have
\begin{align*}
    I(\mathcal{V}_{\Delta t}(h(X_t) ), X_{t+\Delta t})
    &\leq
    I(h(X_t) , X_{t+\Delta t}),
\end{align*}
Consequently, 
\begin{align}
    \min_{h, \mathcal{V}_{\Delta t}, \tilde{h}^{-1}}  \mathcal{L}'_\text{MI}(h, \mathcal{V}_{\Delta t}, \tilde{h}^{-1})
    &=
    \min_{h, \mathcal{V}_{\Delta t}, \tilde{h}^{-1}}  \dim h(X_t)
    - \beta\, I\left((\mathcal{V}_{\Delta t}\circ h)(X_t), X_{t+\Delta t}\right)+\beta\delta_{\text{dec}}[h,\tilde{h}^{-1}]
    \\
    &=
    \min_{h}\left(\dim h(X_t)
    - \beta\, \min_{\mathcal{V}_{\Delta t}}   I\left((\mathcal{V}_{\Delta t}\circ h)(X_t), X_{t+\Delta t}\right)+\beta\min_{ \tilde{h}^{-1}}\delta_{\text{dec}}[h,\tilde{h}^{-1}]\right)
    \\
    &=
    \min_{h}\left(\dim h(X_t)
    - \beta\, I\left(h(X_t)), X_{t+\Delta t}\right)\right).
\end{align}
To go from the second to last line to the last line, we used the data processing inequality to minimize the second term. For the last term $\delta_{\text{dec}}$, we used the fact that for any encoder $h$, the optimal decoder (which yields $\delta_{\text{dec}}=0$) is known (see Appendix~\ref{main_apx:MI_from_DKL}). In particular, it is
\begin{align}
    p(x_{t+\Delta t}|h_{t+\Delta t})=\int \dd h_t\dd x_t p(x_{t+\Delta t}|x_t)p(x_t|h_{t})p(h_t|h_{t+\Delta t})
\end{align}
which is obtained by inverting the stochastic dynamics and the encoding using Bayes' rule. 
Therefore, the only unknown quantity is $h$: all the others are induced by a choice of $h$.

\section{A primer on the information bottleneck}
\label{main_apx:IB_in_general}
The information bottleneck was originally formulated in Ref. \cite{tishby_IB} as a rate distortion problem. Rate distortion theory describes how to maximally compress a signal (i.e. minimize its communication rate) such that it remains minimally distorted, which, however, presupposes the knowledge of a distortion function specifying which features need to be preserved \cite{cover2012elements}. 
In Ref. \cite{tishby_IB}, Tishby et al. introduce a variant of the problem where the \textit{a priori} unknown distortion function is not used, but rather one seeks to ensure that the compression retains information about an auxiliary variable $Y$ correlated with the signal. As the correlations with $Y$ define implicitly the relevant features of the signal to be preserved, $Y$ is called the relevance variable.
%In Ref. \cite{tishby_IB}, the Tishby et al. describe a variant of the problem where distortions of the original signal aren't penalized, but rather one seeks to ensure that the compressed signal retains information about an auxiliary variable $Y$. The information retained in the compression is \emph{relevant} information about $Y$ and hence $Y$ is called a relevance variable.
Concretely, we call $X$ the source signal, and $H$ denotes the compressed signal.
The random variables form a Markov chain $H - X - Y$, meaning that $H$ and $Y$ are conditionally independent given $X$:
    \begin{align*}
        p(y|h, x)&=p(y|x)p(h,x)
        \\
        p(h|x, y)&=p(h|x)p(y,x)
    \end{align*}
As noted in the main text, the IB optimization objective is given by the Lagrangian
\begin{align}
\label{apxibloss}
    \mathcal{L}_{\text{IB}}[p(h|x)] &= I(X,H) - \beta I(Y,H). 
\end{align}
To enforce normalization of the $p(h|x)$ one introduces a Lagrange multiplier $\lambda(x)$ so that the full optimization function is
\begin{equation}
\begin{split}
    \mathcal{L}_{\text{IB}}[p(h|x)]
    &= \sum_{x,h}p(h|x)p(x)\log\frac{p(h|x)}{p(h)} 
    - \beta\sum_{y,h}p(y,h)\log\frac{p(y,h)}{p(h)p(y)} 
    +  \sum_x \lambda(x)\left(1-\sum_h p(h|x)\right).
\end{split}
\end{equation}
The encoder which optimizes this objective can be solved for exactly. 
Using the following functional derivatives, 
    \begin{align*}
    	\frac{\delta }{\delta p(h|x)} p(h') &= \delta(h-h')  p(x)
	\\
	\frac{\delta }{\delta p(h|x)} p(h',y') &=  \delta(h-h') p(x,y'),
\end{align*}
one can compute the derivative of the Lagrangian.
By evaluating the derivative and setting to zero, one finds 
\begin{align*}
    \log p(h|x) = \log p(h) - \beta\sum_y p(y|x) \log \frac{1}{p(y|h)} - \lambda(x)
\end{align*}
which can be rearranged to give the optimal encoder 
\begin{align}
    p(h|x) = \frac{p(h)}{N(x)}\exp\left[ -\beta D_{KL}(p(y|x) \Vert p(y|h) )\right].
    \label{apx:IB_opt_encoding_dkl}
\end{align}
By absorbing terms which only depend on $h$ and $x$ into $p(h)$ and $N(x)$, respectively, the encoder can be expressed as
\begin{align}
p(h|x) = \frac{p(h)}{N(x)}\exp\left[ \beta \int dy\, p(y|x) \log p(y|h) \right].
\label{ib_solution}
\end{align}

When $\beta < 1$, it follows from the data processing inequality $I(X,H)\geq I(Y,H)$ that Eq.~\eqref{apxibloss} is minimized by a trivial encoder $p(h|x) = p(h)$.
In this case, $\mathcal{L}_{\text{IB}}=0$ and no information passes through the bottleneck.
As $\beta$ is increased, more information is allowed through the bottleneck until the relevant variables begin to gain a dependence on the state $x$. This occurs suddenly for a certain value of $\beta=\beta_{1}>1$ at which the encoder becomes non-uniform and $I(H,X)$ becomes non-zero. This is referred to as an IB transition; for increasing $\beta$, there may be a sequence of transitions at $\beta_2,\beta_3,...$ etc.

%\subsection*{The information bottleneck}
%The information bottleneck \cite{tishby_IB} is an example of a rate-distortion problem which seeks to find an optimal compression which minimizes some distortion measure with the original signal \cite{cover2012elements}. 
%Concretely, we call $X$ the source signal, and let $H$ denote the compressed signal.
%In IB, rather than using an \textit{a priori} unknown distortion function, one seeks to ensure that the compression retains information about an additional relevance variable $Y$.
%As noted in the main text, the IB optimization objective is given by the Lagrangian
%\begin{align}
%    \label{apxibloss}
%    \mathcal{L}_{\text{IB}}[p(h|x)] &= I(X,H) - \beta I(Y,H), 
%\end{align}
%where in our case the source signal $X$ is the state of the system $X_t$ at time $t$, and the relevance variable is the state of the system $X_{t+\Delta t}$ at a future time $t+\Delta t$.
%The encoder which optimizes this objective can be solved for exactly and is given by \cite{tishby_IB}
%\begin{align}
%p(h|x) = \frac{p(h)}{N(x)}\exp\left[ -\beta D_{KL}(p(y|x) \Vert %p(y|h) )\right].
%\label{methods:opt_encoding}
%\end{align}

\section{The optimal encoder in the limit of high compression}
\label{main_apx:IB_and_eigenfunctions}
To understand the form of the optimal encoder in terms of transfer operator eigenfunctions, we first rewrite Eq.~\eqref{ib_solution} in terms of the transition probabilities,
\begin{align}
    p(h_t|x_t) = \frac{\tilde{p}(h_t)}{\tilde{N}(x_t)}\exp\left[ \beta \int \dd x_{t+\Delta t} p(x_{t+\Delta t}|x_t) \log p(x_{t+\Delta t}|h_t) )\right].
    \label{methods:opt_encoding_time}
\end{align}
where we have absorbed terms in the exponent which depend only on $h_t$ or $x_t$ into the normalization factors.
In the above equation, we replace the transition probability with the spectral decomposition
\begin{align}
    p(x_{t+\Delta t}|x_t) = \sum_n \ee^{\lambda_n \Delta t}\rho_n(x_{t+\Delta t})\phi_n(x_t).
\end{align}
From this Eq.~\eqref{optimal_encoding_spectral} of the main text immediately follows, where 
\begin{align}
    f_n(h_t) = \int \dd x_{t+\Delta t} \rho_n(x_{t+\Delta t}) \log p(x_{t+\Delta t}|h_t), 
\end{align}
which may be interpreted as a sort of cross entropy ($\rho_n$ is generally not a probability distribution) between each right eigenfunction and the decoding of $h_t$ into the future state $x_{t+\Delta t}$.

To study the behavior of the encoder in the limit of high compression, we consider a transfer operator $U$ with infinitesimal generator $\mathcal{L}_{U}$. For $\mathcal{L}_{U}$ with a discrete spectrum with eigenvalues satisfying $0=\infev_0>\infev_1>\infev_2\gg\infev_3...$ and for $\beta$ just above the first IB transition $\beta_1$, we  show that the optimal encoder is given approximately by
\begin{equation}
    p^*_\beta(h|x) = \frac{1}{\mathcal{N}(x)} p^*_\beta(h) \, \exp\left(\beta \, \ee^{\infev_1 \Delta t} \phi_1(x) f_1(h)\right)
    \label{encoder_mostinformativefeature_methods}
\end{equation}
with corrections due to the second eigenfunction given by $f_2(h) \approx f_1(h) \ee^{-\Gamma\Delta t} + \mathcal{O}(\ee^{-2\Gamma \Delta t})$ where $\Gamma=\infev_1-\infev_2>0$ denotes the spectral gap. 
Here we assume a finite alphabet of size $N_H$, i.e. $h_{\mu}$ with $\mu\in\{1,...,N_H\}$, but in the SI we consider the case where $p(h|x)$ is Gaussian which is relevant for variational IB.
To prove the above, we compute the Hessian of the IB Lagrangian
\begin{align*}
    H_{(\mu,n),(\nu, m)}&=\frac{\partial^2}{\partial f_n(h_{\mu})\partial f_m(h_{\nu})}\mathcal{L}_{\text{IB}}
    \\
    &= 
    \left(\langle \tphi_n \tphi_m \rangle 
    - \beta \left\langle \langle\tphi_n \rangle_{p(\cdot|y)}\langle\tphi_m \rangle_{p(\cdot|y)} \right\rangle_y 
    - (1-\beta)\langle \tphi_n\rangle\langle \tphi_m\rangle\right)
    \left(\delta_{\mu\nu}p(\hm) - p(\hn)p(\hm)\right)
    %\label{eq:hessian}
\end{align*}
%where we index $H$ with a multi-index $(\mu,n)$.
where $\tilde{\phi}_n(x)=\ee^{\infev_n \Delta t}\phi_n(x)$.
We see that $H$ has the form of a Kronecker (tensor) product
\begin{align*}
    H_{(\mu,n),(\nu,m)} &= A^{\beta}_{nm}\otimes G_{\mu\nu}.
\end{align*}
Eigenvalues of $H$ are given by products of eigenvalues of $A^{\beta}$ and $G$, while eigenvectors are given by the tensor product of eigenvectors of $A^{\beta}$ and $G$. 
Note that $G$, which is nothing other than the covariance matrix for a multinomial distribution, does not depend on $\beta$. Therefore we expect the behavior at the transition to be governed by $A^{\beta}$.

We are concerned with the sign of the eigenvalues of $H$. A negative eigenvalue indicates that $\mathcal{L}_{\text{IB}}$ is unstable to a perturbation in $f$, which means the loss can be lowered by changing $f$ away from the trivial encoder at $f_n=0$. Because the eigenvalues of a tensor product of matrices are products of the eigenvalues of the component matrices, the eigenvalues of $H$ change sign when those of $A^{\beta}$ change.

For equilibrium systems, i.e. those with no steady state flux, the matrix is given by
\begin{align}
    A_{nm}=\delta_{nm}\ee^{2\lambda_n\Delta t}\left(1 - \beta \ee^{2\lambda_n\Delta t}\right)
    \label{eq:stability_mat_equilib}
\end{align}
except for the $n=m=0$ term, which is $A_{00}=0$.
The eigenvalues are given directly by these diagonal elements. For small $\beta$ these are all positive, and become unstable one after the other at 
\begin{align*}
    \beta=\ee^{-2\lambda_1\Delta t}, \ee^{-2\lambda_2\Delta t},...
\end{align*}
For non-equilibrium systems, we consider the 2$\times$2 block of $A$ to study the effect of corrections due to the second transfer operator eigenvector:
\begin{align*}
    A&=
    \begin{pmatrix}
        A_{2\times 2} & 0 \\
        0 & 0 
    \end{pmatrix}
     +\mathcal{O}(\ee^{\infev_3\Delta t})
\end{align*}
The stability of $A$ is given to the desired order by the stability of $A_{2\times 2}$. As we show in the SI Section~\ref{si_sec:derivations}, the first eigenvector of $A$ to change sign takes the form $(f_1,0)$ with a correction of size $\mathcal{O}\left(\langle\phi_1\phi_2\rangle\ee^{-\Gamma\Delta t}\right)$, which is controlled both by the size of the spectral gap ($\Gamma>0$) and the amount of overlap between $\phi_1$ and $\phi_2$.

The critical values of $\beta$ do not depend on the dimensionality of the latent variable $N_H$ \cite{gedeon_mathematical_2012,parker_bifurcation_2004}. Therefore, because we are interested in only the first perturbation from the uniform encoder, we may consider the case $N_H=2$.
As we show in the SI for both this discrete case and the case of a Gaussian encoder, the transition is continuous, so that the encoder depends only on $f_1$ after the transition, albeit weakly.

\section{Variational IB compared to other dimensionality reduction techniques}
\label{main_apx:VIB_comparison}
\begin{figure}[htp]
    \centering
    \includegraphics[width=\textwidth]{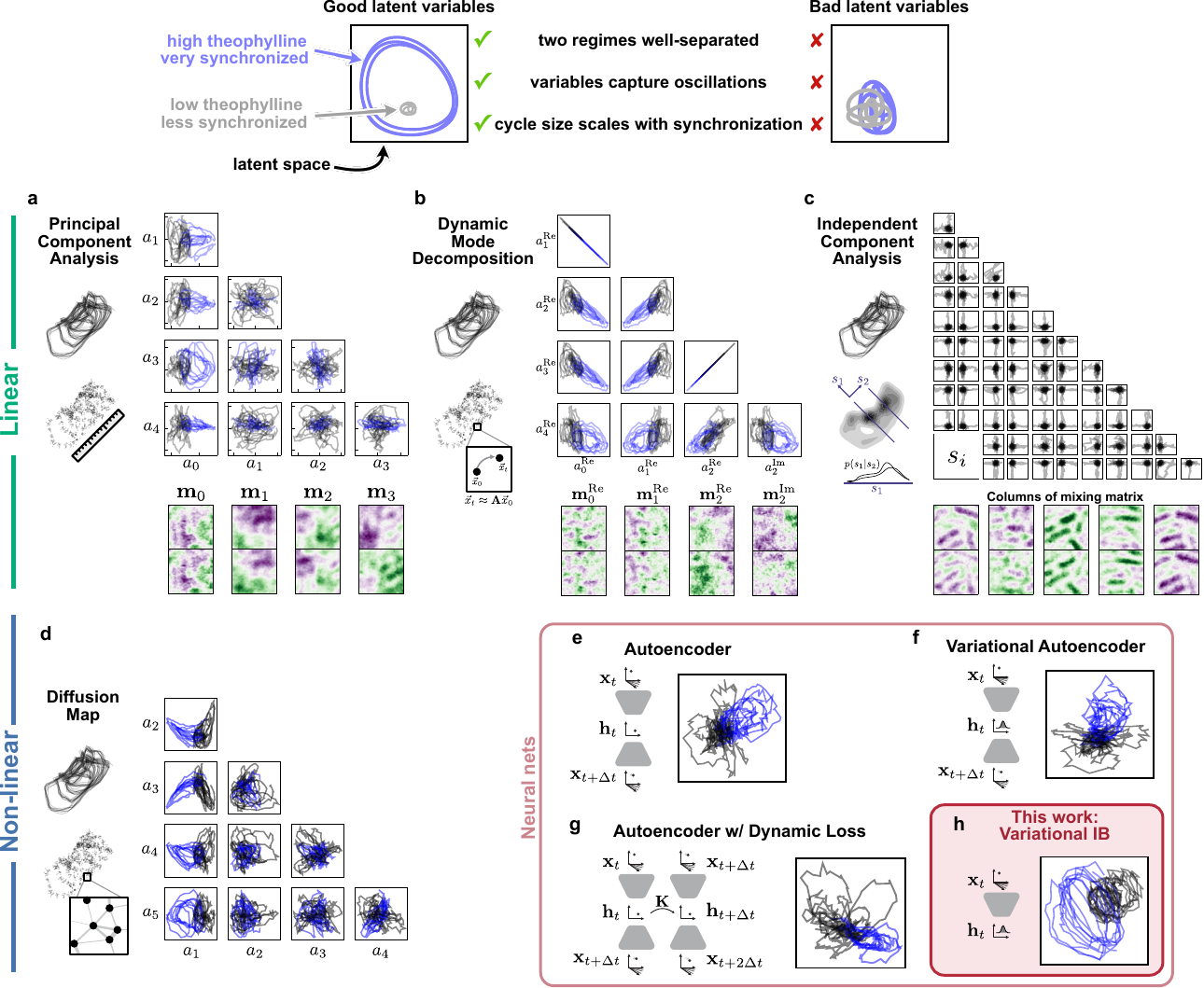}
    \caption{\textbf{Comparison of various dimensionality reduction techniques on cyanobacteria dynamics.}
    %Top row illustrates linear methods, bottom row shows non-linear methods. 
    We compare various dimensionality reduction techniques in the case study of cyanobacteria dynamics. 
    As illustrated on the top row, in this case a good dimensional reduction should produce trajectories in latent space that separate well the two regimes of low and high synchronization, that encode the (noisy) cyclic nature of the dynamics, and if possible that encode the amount of synchronization in the size (radius) of the cycle.
    For each method, we plot the trajectories of all cyanobacteria colonies when projected onto the dominant modes (whose definition depends on the particular method). Blue trajectories show the highly synchronized colonies corresponding to high theophylline concentration, black/gray trajectories show the unsynchronized colonies from low theophylline conditions.
    Unlike the first four methods, deep neural networks (bottom right) directly produce a two-dimensional latent variable and do not provide access to subleading modes.
    (a) Principal component analysis reduces the dimensionality of data by finding a projection onto the directions of most variance (sketch). 
    (Top) Projection of the cyanobacteria trajectories onto several pairs of principal components, $a_i(t)=\mathbf{x}(t)\cdot\mathbf{m}_i$. 
    %Each curve corresponds to the trajectory of one single colony evolving under one of four theophylline conditions; there are 10 trajectories in total, coming from 4 experimental conditions.
    (Bottom) The first four principal components. Here, the state is composed of two time-lagged images of the cyanobacteria colony; the top half of $\mathbf{m}$ corresponds to the earlier image, the bottom half corresponds to the lagged image.
    (b) Dynamic mode decomposition is performed on pairs of data points and aims to find a linear evolution operator $\mathbf{A}$ that links them (sketch).
    (Right) Trajectories of the projection of the state onto several eigenvectors of $\mathbf{A}$. 
    The eigenvectors themselves are shown below (the imaginary parts of the first two modes are uniformly zero).
    (c) Independent component analysis seeks a linear projection of the data onto statistically independent components $\mathbf{s}_i$. In other words, the distribution $p(s_i | s_j)$ is independent of $s_j$ (sketch).
    (Right, top) Trajectories of the projection of the state onto several independent components. 
    (Bottom) columns of the mixing matrix; the components $s_i$ describe how the weighting of these columns evolves in time.
    (d) (Left) Diffusion maps build a graph of the observed state variables with edges weights determined by their distance from one another. (Right) Projection of the trajectories onto the first non-trivial eigenfunctions (because $a_0=$const) of the inferred transfer operator on the graph. 
    (e) A (time-lagged) autoencoder is a deep neural network architecture which encodes the state $\mathbf{x}_t$ into a low-dimensional latent space. From the encoding one then tries to reconstruct the future state. 
    (Right) Trajectories of the cyanobacterial colonies when encoded into a two-dimensional latent space.
    (f) A variational autoencoder is similar to a standard autoencoder, where one instead learns a distribution over possible encodings, $p(\mathbf{h}_t|\mathbf{x}_t)$.
    (Right) Trajectories of the cyanobacterial colonies when encoded into a two-dimensional latent space.
    (g) Implementation of the network in Ref.~\cite{takeishi2017_neurips} which encodes the state using a time-lagged autoencoder with an additional loss term that aims to enforce linearity of the embedded dynamics, $\mathbf{h}_{t+\Delta t}\approx \mathbf{K}\mathbf{h}_{t}$.
    (h) Latent trajectories produced by variational IB (VIB).}
    \label{sifig:comparison}
\end{figure}

%Variational IB (VIB) is by no means the only numerical method for performing data-driven model reduction.
Here we provide a brief overview of the benefits and shortcomings of variational IB (VIB) with respect to other methods, illustrated on the cyanobacteria dataset in Fig.~\ref{sifig:comparison}. An extended discussion can be found in the SI.
%Some defining features of VIB are that it is non-linear, it is lossy, and it does not require any choice of basis functions.
%There are several classes of other dimensionality reduction techniques. 

\subsection{Linear methods}

One class of methods is based on linear projections, such as principal component analysis (PCA), dynamic mode decomposition (DMD) \cite{rowley2009_dmd,schmid_2010_dmd}, or (time-lagged) independent component analysis (TICA) \cite{ica_hyvarinen2001} (which is equivalent to DMD \cite{Klus2018}). 
These methods can be extended to take into account non-linearity by introducing a library of non-linear terms on which one then applies the above methods, such as in kernel PCA \cite{schoelkopf1998_kernelpca} or extended DMD (eDMD) \cite{williams2015}. 
These methods have the advantages, relative to VIB, that their optimization (even for the extended algorithms) relies only on linear projections which are fast and interpretable.
However, the success of these methods depends on the choice of an appropriate library of functions so that the projection onto this space is closed under the dynamics. Choosing an appropriate library is not always possible \cite{Kaiser_2021,page2019}.

\subsection{Nonlinear methods}

A second category of non-linear dimensionality reduction techniques are graph-based or similarity-based methods, which typically assume that the data is distributed on a low-dimensional manifold embedded in a higher-dimensional space \cite{tsne,mcinnes2020umap}. 
One prominent example is diffusion maps \cite{coifman2006_diffusionmaps}, which starts from a set of data snapshots and, assuming the system evolves diffusively on short times, constructs an approximate transition matrix from which one can compute eigenfunctions to parameterize the data manifold. 
The assumption of diffusive dynamics can be violated when data is not sampled sufficiently frequently. This likely explains our finding that VIB produced more well-behaved low-dimensional embeddings on the cyanobacteria dataset (Fig.~\ref{sifig:comparison}). 
VIB has the additional advantage, relative to this and similar methods, that it explicitly takes dynamics into account without the strong assumptions required by diffusion maps.
%While they have been effectively used in equilibrium systems, we found diffusion maps to perform poorly on the cyanobacteria data. We expect this is because they violate the assumption of diffusivity on small times: the time resolution of our images is fairly large, so that the dynamics undergo large jumps.

\subsection{Deep neural network based methods}

Finally, deep neural networks can be used for model reduction through encoder-decoder architectures that attempt to reconstruct the data from a low-dimensional latent space; VIB falls into this class of methods.
Some standard neural network architectures from this class include autoencoders (AEs) and variational autoencoders (VAEs). 
For dynamical systems in particular, extensions to these methods have been proposed which impose constraints on the latent dynamics, such as linearity \cite{takeishi2017_neurips,lusch2018,bakarji2023,wehmeyer2018_timelaggedautoencoder}.
Autoencoders often produce poorly-behaved latent spaces that distribute the latent variables on a narrow manifold with sharp features, see for example \cite{wehmeyer2018_timelaggedautoencoder}. By regularizing the latent embedding to encourage smoothness, variational autoencoders can remedy some of these issues. We note that the VIB loss is very similar to a VAE loss with the contrastive InfoNCE loss replacing the reconstruction loss, so we expect that for many problems these should perform similarly. 
Other dynamically-constrained architectures such as in \cite{takeishi2017_neurips,lusch2018,bakarji2023} work well for deterministic systems but it is unclear what effect stochasticity has on their performance. In our examples we have seen that VIB works well on noisy data.

\subsection{Summary}

When investigating a new system, we suggest to start by attempting to perform dimensionality reduction with linear methods such as PCA or DMD because they are fast, straightforward to implement, and easy to interpret.
In situations where linear techniques are not sufficient, VIB may be preferable to other methods because it is guaranteed to find dynamically relevant variables (in contrast to diffusion maps, t-SNE, AEs, VAEs, etc.) and it does not require that one performs the carefully tailored preprocessing steps that are required by eDMD or kernel PCA, or other variants of DMD \cite{Chen2012_dmdvariants,kutz2015multiresolution,Wynn2013_optimaldmd,colbrook_mpedmd}. Additionally, it works well even when the dynamics are highly stochastic, as we have illustrated with the cyanobacteria dataset.

\bibliography{biblio}

\clearpage
\newpage

\widetext
\begin{center}
\textbf{\large Supplementary Information}
\end{center}
%%%%%%%%%% Merge with supplemental materials %%%%%%%%%%
%%%%%%%%%% Prefix a "S" to all equations, figures, tables and reset the counter %%%%%%%%%%
\setcounter{equation}{0}
\setcounter{section}{0}
\setcounter{figure}{0}
\setcounter{table}{0}
\makeatletter
\@removefromreset{equation}{section} 
\renewcommand{\theequation}{S\arabic{equation}}
\renewcommand{\thefigure}{S\arabic{figure}}
\renewcommand{\thesection}{S\arabic{section}}
\renewcommand{\thesubsection}{\thesection.\arabic{subsection}} % Defines the S4.6 format
\renewcommand{\p@subsection}{} % Clears the default RevTeX space prefix
\renewcommand\appendixname{}
%%%%%%%%%% Prefix a "S" to all equations, figures, tables and reset the counter %%%%%%%%%%

\section{Background material}
\label{si_sec:background}
\subsection*{Mutual information and entropy}
Let $X$ be a random variable which takes values $x$ that are observed with probability $p(x)$. 
The entropy of this distribution measures the predictability of the outcome of a measurement of $X$ and is given by \cite{cover2012elements}
\begin{align*}
    \mathcal{H}(X) = -\int \dd x\, p(x)\log p(x).
\end{align*}
High entropy means high uncertainty.
The conditional entropy $\mathcal{H}(X|Y)$ between two variables is the average entropy of $p(x|y)$, where the average is taken over $p(y)$:
\begin{align}
    \mathcal{H}(X|Y)=-\int\dd y\,p(y)\int\dd x\,p(x|y)\log p(x|y).
    \nonumber
\end{align}
The mutual information between $X$ and $Y$ corresponds to the (average) reduction of uncertainty about $X$, given knowledge of $Y$:
\begin{align}
    I(X,Y) 
    &= \mathcal{H}(X) - \mathcal{H}(X|Y) 
    \nonumber
    \\
    &= D_{\text{KL}}(p(x,y)\lVert p(x)p(y)) 
    \nonumber
    \\
    &= \int\dd x\dd y\,p(x,y)\log\frac{p(x,y)}{p(x)p(y)}.
\end{align}
The three formulations of the mutual information above are equivalent and useful in different settings. For example, formulation as a Kullback-Leibler divergence immediately tells us that mutual information is non-negative.

\subsection*{Transfer operators}
\label{transfer_operators}
The transfer operator of a Markovian dynamical system describes the evolution of probability distributions
\begin{align}
    p_{X_{t+\Delta t}}(x) = U^{\Delta t}[p_{X_t}](x)
    =\int dx' p(X_{t+\Delta t}=x|X_t=x')p(X_t=x').
\end{align}
The operator may be decomposed as 
\begin{equation}
    U^{\Delta t} = \sum_{n} \ket{\rho_n} \ee^{\infev_n \Delta t}  \bra{\phi_n} + U_{\text{ess}}.
\end{equation}
For noisy dynamical systems the operator $U^{\Delta t}$ is nearly compact, meaning that its essential spectral radius will be small \cite{gaspard1995,froyland2013,Cvitanovic2012}. We therefore neglect it in the following.

Above we adopted notation standard from quantum mechanics: $\ket{\rho_n}$ is treated as a vector in the function space on which $U^{\Delta t}$ acts (for example $L^1$), while $\bra{\phi_n}$ is a vector in the dual space (for example $L^\infty$).
The right eigenfunctions can be thought of as densities, and $\rho_0$ is the steady state distribution and satisfies $U^{\Delta t}\rho_0=\rho_0$.
These vectors act on each other via
\begin{align}
    \bra{\phi_n}\rho_m\rangle\equiv\langle \phi_n,\rho_m\rangle=\int \dd x\, \phi_n(x)\rho_m(x)
\end{align}
and they satisfy the biorthogonality condition
\begin{align*}
    \langle\phi_n,\rho_m\rangle=\delta_{nm}.
\end{align*}
One may use these functions to decompose the conditional probability distribution as
\begin{align}
    p(x_{t+\Delta t}|x_t)=\sum_n e^{\lambda_n\Delta t}\rho_n(x_{t+\Delta t})\phi_n(x_t).
\end{align}

In this work we consider dynamical systems given by a Langevin equation
\begin{align}
    \dot{\vec{x}} = \vec{f}(\vec{x}) +\sqrt{2D(\vec{x})}\vec{\xi}(t)
    \label{apxeq:langevin}
\end{align}
where the first term is the deterministic part of the dynamics, and the second term corresponds to noise, where
\begin{align*}
    \langle \xi_i(t)\xi_j(t')\rangle = C_{ij} \delta(t-t').
\end{align*}
This framework captures purely deterministic dynamics, which are obtained by setting $D=0$. 
For dynamics given by a Langevin equation, probability distributions evolve according to the corresponding Fokker-Planck equation
\begin{align*}
    [\mathcal{L}\rho](\vec{x})=\partial_t \rho(\vec{x}) = 
    -\partial_{i}(f_i(\vec{x})\rho(\vec{x})) + \partial_i\partial_j(D(\vec{x})C_{ij}\rho(\vec{x})).
\end{align*}
Here we recognize $\mathcal{L}$ as the infinitesimal generator of the transfer operator $U$: evolution for a time $t$ is given by $U=\ee^{t\mathcal{L}}$.
The adjoint operator is given by the so-called backward Kolmogorov equation, and it describes the evolution of functions $\phi$. 
\begin{align}
    \mathcal{L}^\dagger \phi(x) = f_i(\vec{x})\partial_{i}(\phi(\vec{x})) + D(\vec{x})C_{ij}\partial_i\partial_j(\phi(\vec{x})).
\end{align}
For $D=0$, $\mathcal{L}$ is the generator of the so-called Perron-Frobenius operator, while its adjoint is the {Koopman} operator \cite{ModernKoopman}.
%In general, if $\mathcal{L}$ acts on $L^{p}$, then $\mathcal{L}^{\dagger}$ acts on % also: Chapman-Kolmogorov eqn (according to Erwin Frey)

\subsection*{Solving IB exactly and Ulam approximations of the transfer operator}
\label{apx:ulam}

\begin{figure}[tp]
    \centering
    \includegraphics[width=0.45\textwidth]{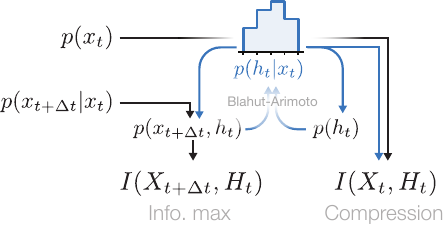}
    \caption{\label{fig:exactIB}\textbf{Exact IB.} 
    Exact IB finds an optimal encoder using an iterative Blahut-Arimoto algorithm, described in the main text, for a rate-distortion problem with Kullback-Leibler distortion \cite{tishby_IB}. This process requires access to (an estimate of) the transfer matrix and the steady state distribution.  %This procedure can be understood as the Blahut-Arimoto algorithm for a rate-distortion problem with Kullback-Leibler distortion.
    Once an optimal encoder has been found, relevant quantities such as mutual information can be computed.
    }
\end{figure}

The optimal IB encoder Eq.~\eqref{apx:IB_opt_encoding_dkl} can be found using the Blahut-Arimoto (BA) algorithm \cite{cover2012elements}. As described in detail in Refs.~\cite{tishby_IB,slonimphd} and sketched in Fig.~\ref{fig:exactIB}, the algorithm is an iterative procedure, where the encoder at iteration $k+1$ is updated according to
\begin{align*}
\begin{cases}
    p_{k+1}(h|x_t) = \frac{p_k(h)}{N_{k+1}(x)}\exp(-\beta D_{\text{KL}}(p(x_{t_\Delta t}|x_t)\lVert p_k (x_{t+\Delta t}|h)))
    \\
    p_{k+1}(h) = \sum_{x_t} p(x_t)p_{k+1}(h|x_t)
    \\
    p_{k+1}(x_{t+\Delta t}|h) = \sum_{x_t} p(x_{t+\Delta t}|x_t)p_{k+1}(x_{t}|h).
\end{cases}
\end{align*}
The first line simply plugs the previous estimate of the encoder in to Eq.~\ref{optimal_encoding_DKL} (and normalizes the distribution), while the following two lines update marginal and conditional distributions using the new estimate of the encoder. In Refs.~\cite{tishby_IB, slonimphd} it is shown that this algorithm converges.

The BA algorithm requires access to the conditional distribution $p(x_{t+\Delta t}|x_t)$ for each $x_t$, as well as the steady state $p(x_t)$.
To solve the IB optimization problem we therefore need a numerical approximation of the transfer operator, which we obtain by an Ulam approximation \cite{ulam1960,li1976_ulam}. 
In brief, one divides space into bins and computes a finite-dimensional approximation to the conditional distribution as
\begin{align}
    \label{P_ulam_approx}
    P_{ij} = P(X_{t+\Delta t}=x_j | X_t=x_i) = N_{i\rightarrow j} / N_i,
\end{align}
where $N_i$ is the number of trajectories starting in bin $i$ and $N_{i\rightarrow j}$ the number of observed transitions from bin $i$ to bin $j$.
The transfer operator is then approximately given by
\begin{align}
     Up(x_j)\approx\sum_{x_i}P_{ij}p(x_i),
\end{align}
and eigenvalues and eigenvectors of $U$ can be computed by diagonalizing $P$.

%The Fokker-Planck equation may be written

\section{Dimensionality reduction through the lens of information theory}
\label{si_sec:dim_reduction_in_general}

\label{main_apx:model_reduction}
Consider a dynamical system which can be described by the evolution of the random variable $X_t$. 
In the case where $X_t$ is high-dimensional but undergoes stereotyped or regular behavior, it often suffices to describe the system with a reduced dynamical variable $H_t$, where $H_t$ is lower dimensional than $X_t$.
Schematically, a reduced model would be used in the following way to predict the system's evolution over a time $\Delta t$, for any choice of starting time $t$:
\begin{figure}[ph]
    \centering
    \includegraphics[width=0.8\linewidth]{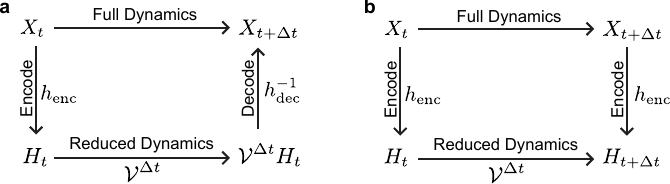}
    \caption{\textbf{Two views of model reduction.}
    (a) In one view, the reduced model should act as a useful surrogate of the full system.
    After reducing the system via an encoder $h_{\text{enc}}$, the reduced state is evolved forward in time via $\mathcal{V}^{\Delta t}$ and subsequently decoded by $h_{\text{dec}}^{-1}$ to obtain an estimate of the full system $X_{t+\Delta t}$.
    In general, if significant information has been lost during encoding, the decoding will be probabilistic to produce a distribution of possible future states.
    (b) Alternatively, one may insist only that the encoding is ``consistent'' with respect to the dynamics (and is not necessarily informative about the full system). In this picture, it should be equivalent to either encode the future state, or to first encode and then evolve the encoded state.
    }
    \label{fig:two_diagrams}
\end{figure}

A reduced description is useful only if it can be used to simulate the system forward in time and make predictions about the future. In other words, one aims to find a reduction $h(X)$ and an evolution rule $\mathcal{V}$ so that the above diagram commutes. Reducing the initial state and evolving it should be equivalent to first evolving the full system, and then reducing the final state.

The requirements of a reduced model may then be enumerated as follows:
\begin{enumerate}
    \item \textbf{Reduction}
    \\
    One hopes that the reduced state is more ``compact'' than the full state. One way to achieve this is to require, for example, that 
    \begin{align}
        \dim H_t\ll \dim X_t.
    \end{align}
    \item \textbf{Predictiveness}
    \\
    There are two ways to enforce predictiveness, which are shown in Fig.~\ref{fig:two_diagrams}. On the one hand is the view that the reduced variables should be useful as a surrogate for the full system, which are simple enough to evolve forward for long times. The hope is that the forecasted latent variable may still contain sufficient information to approximately predict the future full state of the system \cite{jiang2024training,Vlachas2022}.
    On the other hand, in physics often one instead aims to find variables which are simple to predict on their own. Instead of using them to decode into a prediction of the full state, one hopes that the evolved latent variable is equivalent to first evolving the full state, and then reducing it to its latent representation. 
    Mathematically, the first view leads to the requirement that the quantity
    \begin{align}
        d\big(h_{\text{dec}}^{-1}(\mathcal{V}^{\Delta t}h_{\text{enc}}(X_t)), X_{t+\Delta t}\big)
    \end{align}
    is small, where $d$ is some choice of distance that measures how the two predictions differ from each other.
    The second view leads to the requirement that 
    \begin{align}
        d\big(\mathcal{V}^{\Delta t}h_{\text{enc}}(X_t), h_{\text{enc}}(X_{t+\Delta t})\big)
    \end{align}
    is small. 
\end{enumerate}
In general, one may determine the functions $h_{\text{enc}}$, $h_{\text{dec}}$ and the operator $\mathcal{V}^{\Delta t}$ as the solution to an optimization problem
\begin{align}
    \min_{h_{\text{enc}}, \mathcal{V}, h^{-1}_{\text{dec}}} \quad \dim h_{\text{enc}}(X_t)  
    + 
    \beta_1 \,d\big(h_{\text{dec}}^{-1}(\mathcal{V}^{\Delta t}h_{\text{enc}}(X_t)), X_{t+\Delta t}\big)
    +
    \beta_2\,d\big(\mathcal{V}^{\Delta t}h_{\text{enc}}(X_t), h_{\text{enc}}(X_{t+\Delta t})\big)
    \label{eq:mr_obj_dists_full}
\end{align}
The parameters $\beta_1$ and $\beta_2$ must be \textit{chosen} to reflect the desired properties of the reduced model. 
The choice of $\beta_2=0$ or $\beta_1=0$ correspond to the two views of model reduction described above, but in principle both terms may be present. In both the main text and here, we focus on the cases where only one term is present, but comment briefly on the case $\beta_1,\beta_2\neq 0$ below.
It turns out that the case $\beta_1=0$ is uninteresting. The reason is that the objective function may then be trivially solved by $h(X_t)=\text{const}$ and $\mathcal{V}^{\Delta t} = \text{id}$.
For any choice of dimension, this will minimize the second term of the objective.
Consequently, we take our model reduction objective function to be
\begin{align}
    \min_{h_{\text{enc}}, \mathcal{V}, h^{-1}_{\text{dec}}} \mathcal{L}
    \equiv 
    \min_{h_{\text{enc}}, \mathcal{V}, h^{-1}_{\text{dec}}} \quad \dim h_{\text{enc}}(X_t) 
    + 
    \beta_1 \,d\big(h_{\text{dec}}^{-1}(\mathcal{V}^{\Delta t}h_{\text{enc}}(X_t)), X_{t+\Delta t}\big).
\end{align}

In this work, we replace measures of distance $d(X,Y)$ with the mutual information $I(X,Y)$. The distance $d(X,Y)$ measures how ``close'' $X$ is to $Y$, which depends inherently on the system's geometry. 
In contrast, the mutual information $I(X,Y)$ measures how precisely $Y$ can be determined given knowledge of $X$ and abstracts away the underlying geometry of the system.
With this choice, the objective is given by
\begin{align}
     \min_{h_{\text{enc}}, \mathcal{V}, h^{-1}_{\text{dec}}} \quad \dim h_{\text{enc}}(X_t) 
    -
    \beta_1 \,I\big(h_{\text{dec}}^{-1}(\mathcal{V}^{\Delta t}h_{\text{enc}}(X_t)), X_{t+\Delta t}\big)
\end{align}
where we have changed the sign of the second term; instead of minimizing distance, we want to maximize information.

As stated in the main text, this choice of ``distance'' has dramatic consequences for our optimization problem.
As we show in Appendix~\ref{main_apx:IB_in_general} in the main text, the objective then simplifies to
\begin{align}
     \min_{h_{\text{enc}}} \mathcal{L}
    =
     \min_{h_{\text{enc}}}\left[\dim h_{\text{enc}}(X_t) 
    -
    \beta_1 \,I\left(h_{\text{enc}}(X_t), X_{t+\Delta t}\right)
    \right]
    \label{eq:mr_obj_h_only}
\end{align}
Conceptually, this means that the purely information-theoretic framework we have taken does not tell us how to evolve the reduced variable forward in time, only how to choose the optimal reduced variables. However, the fact that any bijective $\mathcal{V}^{\Delta t}$ is a solution should not be taken to mean that $H_t$ has no dynamics, or that its dynamics are undefined. Instead, it simply means that the mutual information is insensitive to the choice of $\mathcal{V}^{\Delta t}$, because this transformation does not impact the information contained about $X_{t+\Delta t}$. The dynamics of $H_t$ may be obtained by encoding $X_t$ at each time $t$. 
To determine which $\mathcal{V}^{\Delta t}$ is optimal among all choices of bijective transformations, one must introduce and solve another objective function, for example $\min_{\mathcal{V}} \lVert H_{t+\Delta t} - \mathcal{V}^{\Delta t}H_t\rVert^2$, which again requires a choice of distance function.

In practice, one may directly solve Eq.~\eqref{eq:mr_obj_h_only} by fixing $\dim h_{\text{enc}}$ to various different values, and for each solving for the $h_{\text{enc}}$ which optimizes the mutual information with $X_{t+\Delta t}$. 
To gain an analytical understanding of what this objective is solving, we instead replace the dimensionality reduction term with a term that similarly restricts the capacity of the reduced variable; namely, the mutual information $I(X_t, h_{\text{enc}}(X_t))$. As a result of this choice, the objective becomes 
\begin{align}
     \min_{h_{\text{enc}}} \mathcal{L}_{\text{IB}} 
    =
     \min_{h_{\text{enc}}}\left[I(X_t, h_{\text{enc}}(X_t))
    -
    \beta_1 \,I\left(h_{\text{enc}}(X_t), X_{t+\Delta t}\right)
    \right].
    \label{eq:mr_obj_ib}
\end{align}
This is the information bottleneck objective \cite{tishby_IB,creutzig2008} (Appendix~\ref{main_apx:IB_in_general} in the main text). The benefit of working with this objective is that it is analytically tractable, allowing us gain theoretical insight into the form of solutions of Eq.~\eqref{eq:mr_obj_ib} which we expect to be similar to solutions of Eq.~\eqref{eq:mr_obj_h_only}. Indeed in practice we see that, for fixed $\dim h_{\text{enc}}$, solutions to \ref{eq:mr_obj_ib} (which may be found by setting $\beta_1\gg 1$) are qualitatively similar to the solution to Eq.~\eqref{eq:mr_obj_ib}.

\subsection*{Other variants of the objective function}
\subsubsection*{InfoMax} 
We argued above that the $\beta_1=0$ case for the objective function  Eq.~\eqref{eq:mr_obj_dists_full} is uninteresting because of the existence of a trivial solution. However, the situation changes when distances are replaced by mutual information. In that case, the objective becomes 
\begin{align}
     \min_{h_{\text{enc}}} \mathcal{L}_{\text{InfoMax}} 
    =
     \min_{h_{\text{enc}}}\left[\dim h_{\text{enc}}(X_t) 
    -
    \beta_2 \,I\left(h_{\text{enc}}(X_t), h_{\text{enc}}(X_{t+\Delta t})\right)
    \right]
    \label{eq:mr_infomax}
\end{align}
This objective is known as the InfoMax objective \cite{becker_infomax} and generally it does not have a trivial solution because mutual information is not a true metric function. For linear encodings $h_{\text{enc}}$ it reduces to the so-called canonical correlation analysis objective (CCA) \cite{hotellingcca}.
Interestingly, for discrete-state discrete-time Markov chains being partitioned into a binary reduced state $h_{\text{enc}}(X_t)\in\{0,1\}$ it has been shown that the optimal solution is one which assigns classes based on the first subleading left eigenvector of the transfer operator \cite{deng2011}.

\subsubsection*{The symmetric information bottleneck}
We consider here the situation where $\beta_1=0$, as in the case of InfoMax, but with the dimensionality reduction term replaced by a term encouraging compression, $I(X_t, h_{\text{enc}}(X_t))$ as in the main text. 
For simplicity let us assume now that the dynamics are stationary. Then $p(X_t=x)=p(X_{t+\Delta t}=x)$ and $I(X_t, h_{\text{enc}}(X_t))=I(X_{t+\Delta t}, h_{\text{enc}}(X_{t+\Delta t}))$.
The objective function can consequently be written 
\begin{align}
     \min_{h_{\text{enc}}} \mathcal{L}_{\text{SymIB}} 
    &=
    \min_{h_{\text{enc}}}\left[
     I(X_t, h_{\text{enc}}(X_t))
     +
    \beta_2 \,I\left(h_{\text{enc}}(X_t), h_{\text{enc}}(X_{t+\Delta t})\right)
    \right]
    \\
     &=
     \min_{h_{\text{enc}}}\frac{1}{2}\left[
     I(X_t, h_{\text{enc}}(X_t))
     +
     I(X_{t+\Delta t}, h_{\text{enc}}(X_{t+\Delta t}))
    -
    2\beta_2 \,I\left(h_{\text{enc}}(X_t), h_{\text{enc}}(X_{t+\Delta t})\right)
    \right].
\end{align}
This objective is known as the symmetric information bottleneck \cite{symIB_slonim}. In settings where $p(X_t, X_{t+\Delta t})$ may be estimated, a solution to this objective may be found using a Blahut-Arimoto-like algorithm just as the regular IB problem.

\subsubsection*{The deterministic information bottleneck}
The compression term we choose as a proxy for $\dim h_{\text{enc}}(X_t)$ is not unique. In particular, simply minimizing the Shannon entropy $S( h_{\text{enc}}(X_t))$ achieves a similar goal. 
This problem has been previously studied in Ref.~\cite{strouse2016deterministicIB}. Interestingly, this change leads the resulting encoder to be deterministic, $p(H_t|X_t)\sim\delta (H_t-f(X_t))$ where $f(X_t)$ is a deterministic function of $X_t$.

\subsubsection*{Other variants}
The last variant we consider is the case $\beta_1\neq0$, $\beta_2\neq 0$, but with the compression term $I(X_t, h_{\text{enc}}(X_t))$. The full objective function is
\begin{align}
     \min_{h_{\text{enc}}} \mathcal{L}_{\text{SymIB}} 
    =\min_{h_{\text{enc}}}
    \left[
     I(X_t, h_{\text{enc}}(X_t))
    -
    \beta_1 \,I\left(h_{\text{enc}}(X_t), X_{t+\Delta t}\right)
    -\beta_2 \,I(h_{\text{enc}}(X_t), h_{\text{enc}}(X_{t+\Delta t}))
    \right].
    \label{eq:mr_symib}
\end{align}
This objective may be thought of as an information bottleneck, but with an additional term $I(H_t, H_{t+\Delta t})$ that encourages the encoding to learn a variable which is predictive of its own future.
We do not consider this case further beyond noting that the addition of such a term does not affect our perturbative analysis in Section~\ref{si_sec:derivations}.

\section{Information-theoretic quantities for dynamical systems}
\label{si_sec:information_theory_for_dynamics}

\subsection*{Rate of information decay}
\label{apx:info_decay}

\begin{figure}[tp]
    \centering
    \includegraphics[width=0.5\textwidth]{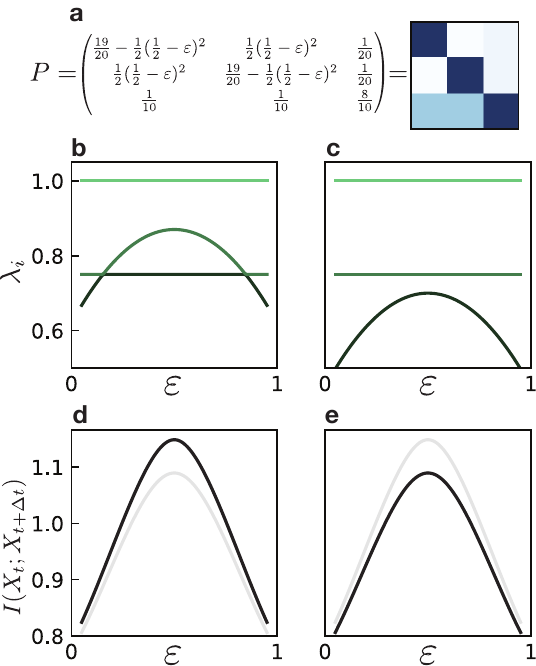}
    \caption{\textbf{Information gain due to an eigenvalue pile-up.}
    (a) To study the role of a gap closing compared to the pile-up of eigenvalues (beyond the dominant one) in the mutual information $I(X_{t}; X_{t+\Delta t})$ we build a matrix $P$ with which we can tune the eigenvalues via a parameter $\varepsilon$.
    (b) Eigenvalues for the matrix shown in (a) with varying parameter values $\varepsilon$; for $\varepsilon \rightarrow 0.5$, the spectral gap $\infev_0 - \infev_1$ closes. 
    (c) Eigenvalues for the matrix shown in (a), but with all $\frac{1}{2}(\frac{1}{2}-\varepsilon)^2$ terms subtracted with a constant (here $1/8$). Here, the spectral gap $\infev_0 - \infev_1$ does not close but there is an accumulation of eigenvalues at $\varepsilon\rightarrow 0.5$. 
    (d-e) Mutual information for $\Delta t=1$ for the matrices in (b-c), respectively, as a function of $\varepsilon$. The curve from the other scenario is shown in gray for comparison. Both exhibit a clear peak in information which differs only slightly in magnitude, showing that the information peak in Fig.~2 of the main text is not due to the closing gap alone, but rather due to contributions from subdominant eigenvalues. 
    }
    \label{si_fig:gapfig}
\end{figure}

Here we show how the decay of information in the system may be related to the spectrum of its transfer operator.
We consider the mutual information 
\begin{align}
    I(X_t, X_{t+\Delta t}) = \sum_{x,y}p(y|x)p(x)\log\frac{p(y|x)}{p(y)}
\end{align}
where $x$ denotes values of the random variable $X_t$ and $y$ the values of $X_{t+\Delta t}$. 
We further assume that the system's dynamics are given by a transfer operator $U$ with integral kernel $p(y|x)$ that can be spectrally decomposed as
\begin{align}
    p(y|x)=\sum_n e^{\infev_n t}\rho_n(y)\phi_n(x),
\end{align}
where $\infev_0=0$, $\phi_0 = \text{const}$ and $\rho_0(y)$ is the steady state distribution.
The mutual information can then be written
\begin{align}
    I(X_t, X_{t+\Delta t}) &= \sum_{x,y}\sum_n e^{\infev_n \Delta t}\rho_n(y)\phi_n(x)p(x)\log \left(1 + \sum_{m>0} e^{\infev_m \Delta t}\frac{\rho_m(y)}{p(y)}\phi_m(x)\right)
    \\
    &= \sum_{n,m>0} e^{(\infev_n +\infev_m)\Delta t}\sum_{x,y}\phi_n(x)\phi_m(x)p(x) \frac{\rho_n(y)\rho_m(y)}{p(y)}
    \\
    &\quad +
    \sum_{n,m,\ell>0} e^{(\infev_n +\infev_m+\infev_\ell)\Delta t}\sum_{x,y}\phi_n(x)\phi_m(x)\phi_\ell(x)p(x) \frac{\rho_n(y)\rho_m(y)\rho_\ell(y)}{p(y)}
    +
    ...
\end{align}

For long times, the contribution of $\infev_1$ dominates, with the remaining terms decaying as $\ee^{(\infev_n-\infev_1)\Delta t}$ for $n\geq 2$. Retaining only the first term, we see 
\begin{align}
    I(X_t, X_{t+\Delta t})
    &= e^{2\infev_1\Delta t}\sum_{x}\phi^2_1(x)p(x) \sum_y\frac{\rho_1(y)^2}{p(y)}.
\end{align}
For short times, other eigenvalues will also contribute to the mutual information (Fig.~\ref{si_fig:gapfig}).

\subsection*{Information-theoretic measures of Markovianity}
\response{
Here we discuss measures of Markovianity, both in the context of the original state $X_t$ as well as the encoded representation $H_t$.
For concreteness we assume that our system is observed at regularly spaced intervals of size $\Delta t$ and denote $x_{n}=x(t_0 + n\Delta t)$. 
\subsubsection*{(Non-)Markovianity of the measured state \texorpdfstring{$X_t$}{Xt}}
A system is non-Markovian if the transition probability depends on the previous state,
$p(x_{n+1}|x_{n}, x_{n-1}) \neq p(x_{n+1}|x_n)$. 
The deviation of a system from Markovianity can be quantified by measuring the difference between these distributions, given in terms of the Kullback-Leibler divergence
\begin{align*}
    D_{\text{KL}}\left(p(x_{n+1}|x_{n},x_{n-1})
    \lVert p(x_{n+1}|x_n)\right).
\end{align*}
If these distributions are (approximately) equal, this quantity will be (close to) zero.
In general this divergence will depend on the precise values of $x_n$ and $x_{n-1}$.
For that reason, we instead consider the average over all possible values of $x_n$ and $x_{n-1}$.
This quantity coincides with the conditional mutual information,
\begin{align}
    I\left(X_{n+1},X_{n-1} | X_{n} \right)
    &=\sum_{\substack{
    x_{n+1} \\
    x_{n}\\x_{n-1}}}
    p(x_{n+1},x_{n},x_{n-1})
    \log\frac{
    p(x_{n+1},x_{n-1}|x_n)}{p(x_{n+1}|x_n)p(x_{n-1}|x_n)}
    \\
    &=\sum_{\substack{
    x_{n}\\x_{n-1}}}p(x_n,x_{n-1})
    \sum_{x_{n+1}}
    p(x_{n+1}|x_n,x_{n-1})
    \log\frac{p(x_{n+1}|x_{n},x_{n-1})}{p(x_{n+1}|x_{n})}
    \\
    &=\mathbb{E}_{
    x_{n},x_{n-1}}D_{\text{KL}}\left(p(x_{n+1}|x_{n},x_{n-1})
    \lVert p(x_{n+1}|x_{n})\right).
\end{align}
Because the Kullback-Leibler divergence is non-negative, this conditional mutual information is zero only if $X_{n+1} - X_n - X_{n-1}$ forms a Markov chain \cite{Polyanskiy_Wu_2025}.
Using the ``chain rule'' for conditional information, we can express this as a difference in mutual informations \cite{cover2012elements}
\begin{align}
     I\left(X_{n+1},X_{n-1} | X_n \right) 
     =
     I\left(X_{n+1}, \{X_n, X_{n-1}\} \right) - I\left(X_{n+1}, X_n \right).
     \label{eq:condMI_as_diff}
\end{align}
}

\noindent
\response{
A system which is non-Markovian can often be made Markovian by introducing time-lagged variables.%, which follows from Taken's embedding theorem \cite{takens}.
One constructs a new dynamical variable which is a concatenation of the previous states of the system,
\begin{align*}
    X^{(k)}_n=\{X_{n-k},...,X_{n-1},X_n\}.
\end{align*}
However, a common problem in practice is that of finding the minimal sufficient $k$ such that the dynamics are Markovian.
This value of $k$ can be identified by finding the value of $k$ for which $\Delta I_{k}\approx0$, where we define
\begin{align*}
    \Delta I_{k}=I(X^{(k+1)}_{n+1},X^{(k+1)}_n) - I(X^{(k)}_{n+1},X^{(k)}_n).
\end{align*}
To see this, note that 
\begin{align*}
    I(X^{(k)}_{n+1},X^{(k)}_{n}|,X^{(k)}_{n-1}) 
    &= I(X^{(k)}_{n+1},X^{(k+1)}_{n}) - I(X^{(k)}_{n+1},X^{(k)}_{n})
    \\
    &\leq I(X^{(k+1)}_{n+1},X^{(k+1)}_{n}) - I(X^{(k)}_{n+1},X^{(k)}_{n})
    \\
    &= \Delta I_{k}
\end{align*}
Thus a small $\Delta I_{k}$ guarantees a small conditional entropy, and hence Markovianity. 
We demonstrate the utility of this approach by considering a Lorenz system in which only the $x$-variable is observed. By increasing the number of time lags, we see that the mutual information $I(X^{(k)}_{n+1},X^{(k)}_{n})$ plateaus at $k\approx 6$, implying Markovianity by the above argument (Fig.~\ref{si_fig:lorenz_memory}a). 
In Fig.~\ref{si_fig:lorenz_memory}b we compare our approach to that introduced in Ref.~\cite{costa2023}, which looks for a plateau in the transfer entropy of the system.
A brief comparison of the benefits of each approach is worth mentioning. 
The approach of Ref.~\cite{costa2023} relies on a binning procedure which can be very computationally costly for when the observed state variable is high dimensional. However, for low-dimensional systems this method can be very effective and may be more efficient than training a neural network to estimate mutual information.
Our approach is particularly attractive in situations where the observed system is high dimensional, but the total information quantity is low. 
This is because the neural network which estimates the mutual information is free to reduce the full state to a lower-dimensional space before estimating the mutual information, which potentially resolves some of the problems of high-dimensionality faced by approaches based on binning.
}

\subsubsection*{(Non-)Markovianity of the encoding variable \texorpdfstring{$H_t$}{Ht}}
\response{
If the full system is non-Markovian its dynamics are no longer given by a global transfer operator. If Markovianity is violated so that
$p(x_{t+\Delta t}|x_t, x_{t-\tau}) \neq p(x_{t+\Delta t}|x_t)$,
the transfer operator $U$ will no longer be independent of time and the eigenvectors corresponding to its dominant eigenvalues will not necessarily have any relation to the true long-time dynamics. 
What will IB learn in this case?
The only quantity arising in the optimal encoder is the one-time conditional transition probability $p(x_{t+\Delta t}|x_{t})$. If there is memory in the system, then this probability can be computed by integrating out the memory
\begin{align}
    p(x_{t+\Delta t}|x_{t})
    &=\int dx_{t-\tau} p(x_{t+\Delta t},x_{t-\tau}|x_t)
    \\
    &=\int dx_{t-\tau} p(x_{t+\Delta t}|x_t,x_{t-\tau})p(x_{t-\tau}|x_t)
    \\
    &=\int dx_{t-\tau} p(x_{t+\Delta t}|x_t,x_{t-\tau})\frac{p(x_t|x_{t-\tau})}{p(x_t)}p(x_{t-\tau})
\end{align}
Where we assume for simplicity that the memory manifests as a dependence on the state at a single fixed time $\tau$ in the past.
This operator may also be decomposed into outer products of right and left eigenvectors, as in the Markovian case. However, the eigenvalues will no longer correspond to a rate of decay of correlations, and the dominant eigenvectors will not necessarily correspond to slow dynamics.
While the study of the spectrum of this object may shed light on relevant processes in the system, we instead chose to deal with potential non-Markovianity by introducing time lags to make the dynamics approximately Markovian.
}

\response{
%Diagnosing and alleviating problems stemming from non-Markovianity in the full system is straightforward 
In general, the encoding $h(t)$ learned by IB will not exhibit Markovian dynamics. 
To see this, consider an encoder $p(h_t|x_t)$. The dynamics induced on the encoding will have transitions governed by the transition probabilities
\begin{align}
    p(h_{t+\Delta t}|h_t) &= \int \dd x_{t+\Delta t}\dd x_t\,
    p(h_{t+\Delta t},x_{t+\Delta t},x_t|h_t)
    \\
    &= \int \dd x_{t+\Delta t}\dd x_t\,
    p(h_{t+\Delta t}|x_{t+\Delta t})p(x_{t+\Delta t}|x_t)p(x_t|h_t)
\end{align}
where we used the fact that the variables form the Markov chain $H_t - X_t - X_{t+\Delta t} - H_{t+\Delta t}$
\footnote{
The notation $A-B-C$ means that $C$ is conditionally independent of $A$, given $B$, or equivalently $p(c|b,a)=p(c|a)$.
}
For more than one step, 
\begin{align}
    p(h_{t+2\Delta t}|h_{t+\Delta t},h_t) 
    %&= \int \dd x_{t+2\Delta t}\dd x_{t+\Delta t}\dd x_t\,
    %p(h_{t+2\Delta t},x_{t+2\Delta t},x_{t+\Delta t},x_t|h_{t+\Delta t},h_t)
    %\\
    &= \int \dd x_{t+2\Delta t}\dd x_{t+\Delta t}\dd x_t\,
    p(h_{t+2\Delta t}|x_{t+2\Delta t})
    p(x_{t+2\Delta t}|x_{t+\Delta t})
    p(x_{t+\Delta t}|x_t,h_{t+\Delta t})
    p(x_t|h_t)
    \\
    &\neq p(h_{t+2\Delta t}|h_{t+\Delta t}).
\end{align}
We show a concrete example of the difference between these two distributions for the encoding of a particle in a double-well in Fig. 4a,b of the main text.
Under what conditions are the encoding dynamics also Markovian? 
A sufficient condition is $p(x_{t+\Delta t}|x_t,h_{t+\Delta t})=p(x_{t+\Delta t}|h_{t+\Delta t})$. This is the case when the full state $x_{t+\Delta t}$ can be determined from the encoding $h_{t+\Delta t}$ as accurately as from the previous full state $x_{t}$. 
This is true, for example, when a (deterministic) high-dimensional system occupies a low-dimensional manifold.
However, for many systems of interest this will not be strictly true. Consider the dynamics given by a noisy Hopf normal form as in the main text. Although the dynamics essentially take place on a circular limit cycle, the current state $(r_t, \theta_t)$ does contain information about the future radius $r_{t+\Delta t}$ which is not accounted for by the angular coordinate alone. 
Instead of hoping that Markovianity holds exactly, one instead hopes that the encoded dynamics undergo \textit{approximately} Markovian dynamics.
}

\response{
A quantitative measure of Markovianity is given by $I\left(H_{t+2\Delta t},H_{t} | H_{t+\Delta t} \right)$.
Asymptotically for small compression, this quantity will go to zero if the dynamics of $X_t$ are Markovian because the encoding approaches a completely informative copy of $X_t$.
However, the above expression offers no guarantees about how $I\left(H_{t+2\Delta t},H_{t} | H_{t+\Delta t} \right)$ approaches zero. In fact, as shown in Fig. 4 of the main text, the approach may be far from monotonic.
}

\response{It is important to note that just because the encoded dynamics are Markovian does not mean that one has found a ``correct'' low dimensional description of the system.
We illustrate this on driven sine-Gordon dynamics. 
This system evolves according to a one-dimensional PDE. For certain driving amplitudes it exhibits quasiperiodic dynamics, and hence its state can be described by a point on a torus.
In Fig.~6 of the main text we show that the information shared between the encoding and the future state plateaus already after two dimensions, $h\in\mathbb{R}^2$.
Indeed, this description does suffice to essentially completely recover the full state from the encoding. 
To do this, the torus on which the system lives must be ``cut'' so that it can be flattened out in two dimensions (Fig.~6b).  
}

\subsubsection*{Estimating memory with mutual information}

\begin{figure}[tp]
    \centering
    \includegraphics[width=0.75\linewidth]{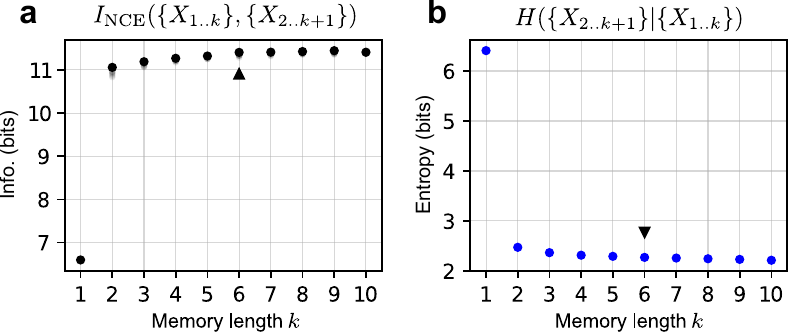}
    \caption{
    \response{\textbf{Testing for Markovianity in the data}
    Information can be used as a test of how far the input data deviates from Markovianity.
    Here we take data from the Lorenz system \cite{Lorenz1963} which was simulated with a time step $\delta t=0.02$, and we consider the state of our system as the $x$-coordinate, discarding the $y$ and $z$ measurements.
    (a) Mutual information between consecutive states as a function of memory length $k$. Mutual information is approximated with the noise contrastive estimate described in Methods.
    (b) Transfer entropy between states as a varying function of memory length. Entropy is approximated using the approach in Ref.~\cite{costa2023}.}}
    \label{si_fig:lorenz_memory}
\end{figure}
\response{
Markovianity is a crucial assumption underlying many approaches to data-driven inference and analysis of dynamical systems \cite{costa2023, ronceray2024, williams2015}. 
Deviations from Markovianity are commonly identified by studying the reduction in error of an inferred model when the system's state is augmented with time-lagged measurements \cite{williams2015}.
An alternative approach based on information theory is to instead consider how much information a measurement of the current state contains about the future state \cite{costa2023}.
The benefit of such an approach is that it doesn't require any assumptions about the underlying model governing a system's evolution. 
The downside is that estimating entropy and related quantities typically relies on binning, which for high-dimensional systems is rendered difficult by the curse of dimensionality \cite{roldanArrowOfTime}.
}

\response{
Using the approach developed in Ref.~\cite{costa2023} one can estimating the degree of non-Markovianity directly from data.
However, this requires an approximation of the full transfer operator.
This prompted us to ask whether direct measurements of the mutual information can identify the amount of Markovianity without explicitly constructing the transition probabilities.
To do this, we consider the Lorenz system \cite{Lorenz1963}.
%which was presented in Ref.~\cite{costa2023} as a test of the authors' approach. 
While the dynamics of the full state $X_t =(x_t,y_t,z_t)$ are Markovian, if one observes only one degree of freedom, for example $X_t=x_t$, the dynamics become non-Markovian. 
By constructing a new state using previous measurements, $X_t^{(k)}=(x_{t-k},...,x_{t-1},x_t)$, one can recover a Markovian system by taking sufficiently large $k$.
}

In Ref.~\cite{costa2023}, the memory length $k$ is chosen to be the value for which the transfer entropy $H(X_{t+1}^{(k)}|X_t^{(k)})$ reaches its minimum value. 
This indicates a variable which is maximally predictive of its future. 
For the Lorenz system under consideration (simulated with a time step of $\delta t=0.02$), we show the results of this approach in Fig.~\ref{si_fig:lorenz_memory}b, which indicates that a memory $k=5$ is sufficient to obtain a Markovian system.

\response{
Instead of considering the entropy, we identify the necessary amount of memory by seeing when the mutual information $I(X_{t+1}^{(k)},X_t^{(k)})$ plateaus as a function of $k$.
One notable benefit of this approach is that it does not require an approximation of the full transfer operator, as above, but can be directly estimated from data using the noise contrastive estimate of the mutual information \cite{oord_infonce} discussed in the main text.
We find that this metric returns the same critical value of $k=6$ after which the state is maximally informative of the future.}

\section{The optimal encoder at the first IB transition}
\label{si_sec:derivations}
Here we derive the form of the optimal encoder at the first IB transition,
\begin{align}
    p^*_\beta(h_t|x_t) \approx \frac{1}{N(x_t)} p^*_\beta(h_t) \, \exp\left(\beta \, \ee^{\infev_1 \Delta t} \phi_1(x_t) f_1(h_t)\right)
    \label{si:encoder_mostinformativefeature}
\end{align}
In Ref.~\cite{tishby_IB} (and reproduced above in Eq.~\eqref{ib_solution}), it is shown that the encoding minimizing the IB objective satisfies the implicit equation
\begin{equation}
    \label{optimal_encoding_DKL}
    \!\!\!\!p_{\beta}^*(h_t|x_t) = \frac{p_{\beta}^*(h_t)}{N(x_t)} \, \exp\left( - \beta D_{\text{KL}} \left[ p(x_{t+\Delta t}|x_{t}) \lVert p(x_{t+\Delta t}|h_t) \right] \right)
    %= 
    %\mathcal{N}' p(h) \exp\left[ - \beta \int \dd x' p_{X_{t+\Delta t}|X_{t}}(x'|x) \log p_{X_{t+\Delta t}|H_t}(x'|h) \right].
\end{equation}
where $N$ is a normalization factor ensuring {$\int \dd h\, p_{\beta}^*(h_t|x_t) = 1$}, and
\begin{equation}
    p_{\beta}^*(h_t) \equiv p^*_{\beta}(h_t) = \int \dd x_t \, p_{\beta}^*(h_t|x_t) p(x_t).
\end{equation}
The Kullback-Leibler divergence between two probability distributions $p(x)$ and $q(x)$ is defined by
\begin{equation}
    D_{\text{KL}}[p||q] = \int \dd x \, p(x) \, \log \frac{p(x)}{q(x)}.
\end{equation}
In the $D_{\text{KL}}$ in Eq.~\eqref{optimal_encoding_DKL}, the integration is done with respect to the common random variable $X_{t+\Delta t}$ of the two conditional distributions.

Equation~\ref{optimal_encoding_DKL} contains the conditional distribution $p(x_{t+\Delta t}|h_t)$, which is the composition of two operations: decoding (going from $H_t$ to $X_t$) and time evolution (going from $X_t$ to $X_{t+\Delta t}$).
The decoding is performed using Bayes' theorem $p(x_t|h_t) p(h_t) = p(h_t|x_t) p(x_t)$, so that  we have 
\begin{equation}
    p(x_{t+\Delta t}|h_t) 
    = \int \dd{x_t} p(x_{t+\Delta t}|x_t) \, p(h_t|x_t) \, \frac{p(x_t)}{p(h_t)}.
\end{equation}
Note the appearance of the encoder $p(h_t|x_t)$, making Eq.~\eqref{optimal_encoding_DKL} only an implicit solution of the optimization problem.
%The time evolution in Eq.~Eq.~\eqref{Uop} can be expressed in a probabilistic language as

The conditional distribution $p(x_{t+\Delta t}|x_t)$ can be written in terms of right and left eigenvectors of $U$.
Here we neglect $U_{\text{ess}}$, which is the operator corresponding to the essential spectrum.
The essential spectrum is the part of the spectrum that is not the discrete spectrum; by neglecting it, we are assuming that the essential radius $\rho_{\text{ess}}$ (the maximum absolute value of eigenvalues in the essential spectrum) is small enough compared to the first few eigenvalues $|\infev_n|$ in the point spectrum. While purely deterministic systems may exhibit a large essential radius, the introduction of noise causes the essential spectrum to shrink or disappear \cite{gaspard1995,froyland2013}.
We can then write the conditional distribution in terms of right and left eigenvectors of $U$,
\begin{align}
    p(x_{t+\Delta t} | x_{t}) &= \sum_n \ee^{\infev_n \Delta t} \rho_n(x_{t+\Delta t}) \phi_n(x_t). %+ U_{\text{ess}}.
    \label{int_kernel_decomp}
\end{align}
The Kullback-Leibler divergence in Eq.~\eqref{optimal_encoding_DKL} can be represented as a sum of two terms,
\begin{align}
    D_{\text{KL}}[...] &= \int \dd x_{t+\Delta t} \, p(x_{t+\Delta t}|x_t) \, \log p(x_{t+\Delta t}|x_t) %\nonumber
    -
    \int \dd x_{t+\Delta t} \, p(x_{t+\Delta t}|x_t) \, \log p(x_{t+\Delta t}|h_t).
\end{align}
The first term has no dependence on $h_t$ and can hence be absorbed into the normalization $N(x)$.
Plugging the decomposition Eq.~\eqref{int_kernel_decomp} into the second term leads directly to
\begin{equation}
    \label{si:optencoder_spectral}
    p_{\beta}^*(h_t|x_t) = \frac{1}{N(x_t)} \, p_{\beta}^*(h_t) \, \exp\left[ \beta \sum_{n} \ee^{\infev_n \Delta t} \phi_n(x_t) f_n(h_t) \right]
\end{equation}
where
\begin{align}
    f_n(h_t) = \int \dd x_{t+\Delta t}\,\rho_n(x_{t+\Delta t})\log p(x_{t+\Delta t}|h_t),
\end{align}
which can be understood as a quasi-cross entropy between the right eigenvector $\rho_n$ and the distribution on $x_{t+\Delta t}$ obtained by decoding $h$.
Equation~\ref{si:optencoder_spectral} depends only on the existence of the spectral decomposition; we have made no other assumptions on the dynamics (beyond Markovianity) up until this point.

\subsection{Perturbation theory at the uniform encoder}
\label{si_sec:stability_and_hessian}

In this section we derive our main result for the optimal encoder in the limit of high compression (small $\beta$). 
We begin by deriving the result for discrete latent states $h_t\in\{0,..,N_H-1\}$, before discussing continuous latent variables below.
In this section, for notational simplicity we will drop some subscripts and rename $x\equiv x_t$, $h\equiv h_t$, $y\equiv x_{t+\Delta t}$.

\textit{Theorem --
Consider a transfer operator $U$ with infinitesimal generator $\mathcal{L}_{U}$. Assume that $\mathcal{L}_{U}$ has a discrete spectrum with eigenvalues satisfying $0=\text{Re }\infev_0>\text{Re }\infev_1>\text{Re }\infev_2\gg\text{Re }\infev_3...$. 
Then, for $\beta$ just above the first IB transition $\beta_1$ so that $f_n(h)\rightarrow 0$, the optimal encoder is given approximately by}
\begin{equation}
    p^*_\beta(h|x) = \frac{1}{\mathcal{N}(x)} p^*_\beta(h) \, \exp\left(\beta \, \ee^{\infev_1 \Delta t} \phi_1(x) f_1(h)\right)
    \label{apx:encoder_mostinformativefeature}
\end{equation}
\textit{with corrections due to the second eigenfunction given by $f_2(h) \approx f_1(h) \ee^{-\Gamma\Delta t} + \mathcal{O}(\ee^{-2\Gamma \Delta t})$ where $\Gamma=\infev_1-\infev_2>0$ denotes the spectral gap.
For equilibrium systems exhibiting no probability fluxes in the steady state, this result is exact.}

\textit{Proof} - 
To show this we compute the $f_n(h)$ coefficients at the onset of instability. 
The instability at the first IB transition at $\beta=\beta_1$ corresponds to the emergence of negative eigenvalues in the Hessian of the IB loss, 
\begin{align}
    \frac{\partial^2 }{\partial f_n(h_i)\partial f_m(h_j)}\mathcal{L}_{\text{IB}}.
    \label{ib_hess}
\end{align}
Similar perturbative approaches have been studied in other contexts within IB in \cite{gedeon2002,PhysRevLett.126.240601, ngampruetikorn2021schwab}. 
For concreteness, we consider a discrete encoding $H_t\in\{0,...,N_H-1\}$. We express the encoder as in Eq. \ref{si:optencoder_spectral} where we now expand the normalization terms
\begin{equation}
    \label{eqapx:encoder_fn}
    p_{\beta}^*(\hm|x) = 
    \frac{f_0( \hm) \exp\left[ \beta \sum_{n} \ee^{\lambda_n \Delta t} \phi_n(x) f_n(\hm) \right]}
    {\sum_{\hl}f_0( \hl) \exp\left[ \beta \sum_{n} \ee^{\lambda_n \Delta t} \phi_n(x) f_n(\hl) \right]}.
\end{equation}
This conditional probability distribution is shorthand for $p_\beta^*(H_t=h_\mu|X_t=x)$.
In the following, we simplify notation as
\begin{equation}
    \label{eqapx:encoder_fn_condensed}
    p(\hm|x) = 
    \frac{f_0^{\mu} \exp\left[ \sum_{n}  \tphi_n(x) f_n^{\mu} \right]}
    {\sum_{\hl}f_0^{\lambda} \exp\left[ \sum_{n} \tphi_n(x) f_n^{\lambda} \right]}
\end{equation}
where $\tphi_n(x)= \beta \ee^{\lambda_n \Delta t} \phi_n(x)$. For the uniform encoder, $f_n^{\mu}=0$ for all $n\neq 0$.
%To understand how the $\phi_n$ are learned sequentially as $\beta$ increases, we consider $f_n^\mu$ as free parameters which can be varied and ask which are the first to become unstable at $f_n^\mu=0$.

Hereafter we assign $\tphi_0=\frac{1}{f_0^{\mu}}$, taking care to track the correct Greek index which otherwise is not present for $\tphi_n$ (in the following, $n$ is always paired with $\mu$ and $m$ always with $\nu$).
For example, a single derivative of the marginal is given by
\begin{align}
    \partial_n^\mu p(\hn)
    &=\int\dd x\,p(x)p(\hn|x)\tphi_n(x)\left(\delta_{\mu\nu}
    -
    p(\hm|x)\right).
\end{align}
To compute the Hessian, we use the fact that $\sum_{\lambda}\partial^{\mu}_n p(h_{\lambda})=0$ (by exchanging sum and derivative), and that $p(\hm|x)|_{\vec{f}=0}=p(\hm)$. 
We note in passing that we find the first derivative terms of $\mathcal{L}_{\text{IB}}$ vanish, see also \cite{wu2020, ngampruetikorn2021schwab}.
The compression term is given by
\begin{align*}
    \partial_n^\mu \partial_m^\nu I(X,H)
    &=
    \left(\langle \tphi_n \tphi_m \rangle - \langle \tphi_n\rangle\langle \tphi_m\rangle\right)
    \left(\delta_{\mu\nu}p(\hm) - p(\hn)p(\hm)\right) 
\end{align*}
and the information maximization term is
\begin{align*}
    \partial_n^\mu \partial_m^\nu I(Y,H)
    &= 
    \left(\left\langle \langle\tphi_n \rangle_{p(\cdot|y)}\langle\tphi_m \rangle_{p(\cdot|y)} \right\rangle_y - \langle \tphi_n\rangle\langle \tphi_m\rangle\right)
    \left(\delta_{\mu\nu}p(\hm) - p(\hn)p(\hm)\right) 
\end{align*}
Angled brackets with no subscript correspond to an average with respect to the steady state distribution, $\langle\cdot\rangle = \int \dd x\,\cdot \,p(x)$. 
Written out more explicitly, the second term is 
\begin{align}
    \left\langle \langle\tphi_n \rangle_{p(\cdot|y)}\langle\tphi_m \rangle_{p(\cdot|y)}\right\rangle_y
    &=\int \dd y\,p(y)\left(\int\dd x\tphi_n(x)p(y|x)\right)\left(\int\dd x'\tphi_m(x')p(y|x')\right)
\end{align}
where we retain the integration variable $y$ as a subscript on the angled brackets for clarity.
In sum, the Hessian of the Lagrangian is given by

\begin{align}
    H_{(\mu,n),(\nu,m)} &= 
    \left(\langle \tphi_n \tphi_m \rangle 
    - \beta \left\langle \langle\tphi_n \rangle_{p(\cdot|y)}\langle\tphi_m \rangle_{p(\cdot|y)} \right\rangle_y 
    - (1-\beta)\langle \tphi_n\rangle\langle \tphi_m\rangle\right)
    \left(\delta_{\mu\nu}p(\hm) - p(\hn)p(\hm)\right) 
    \label{eq:hessian}
\end{align}
where we index $H$ with a multi-index $(\mu,n)$.
From the form of Eq. \ref{eq:hessian}, we see that $H$ is given by a Kronecker (tensor) product
\begin{align*}
    H_{(\mu,n),(\nu,m)} &= A^{\beta}_{nm}\otimes G_{\mu\nu}.
\end{align*}
Eigenvalues of $H$ are given by products of eigenvalues of $A^{\beta}$ and $G$, while eigenvectors are given by the tensor product of eigenvectors of $A^{\beta}$ and $G$. 
Note that $G$ is nothing other than the covariance matrix for a multinomial distribution.
The appearance of unstable directions of the Hessian corresponds to the appearance of negative eigenvalues in its spectrum. 
As $G$ does not depend on $\beta$, zero-crossings of eigenvalues of $H$ therefore correspond to zero-crossings of eigenvalues in the spectrum of $A^{\beta}$.

\subsubsection*{Equilibrium systems}
We first study the stability of the matrix $A^{\beta}_{nm}$ in the case of dynamics which satisfy detailed balance. In particular, we consider a Fokker-Planck operator $\mathcal{L}_{\text{FP}}$ of the form
\begin{align}
    \partial_t p(x) = \mathcal{L}_{\text{FP}}p(x)&= -\partial_i(f_i(x) p(x)) + D\partial_i^2 p(x).
\end{align}
The steady state distribution $\rho_0(x)$ satisfies
\begin{align}
    \mathcal{L}_{\text{FP}}\rho_0(x) = -\partial_i J_i(x) = 0
\end{align}
where $J_i = f_i \rho_0 - D\partial_i \rho_0$ is a flux. Our main assumption in this subsection is that in the steady state, fluxes vanish: $J_i = 0$.
In this case, left eigenfunctions $\phi_n$ of $\mathcal{L}_{\text{FP}}$ become \textit{right} eigenfunctions when multiplied by $\rho_0$. 
To see this, note that 
\begin{align}
    \nonumber
    \mathcal{L}_{\text{FP}}(\rho_0\phi_n)&=\phi_n \mathcal{L}_{\text{FP}}\rho_0 + \rho_0 \mathcal{L}_{\text{FP}}^{\dagger}\phi_n - 2\partial_i\phi_n\underbrace{\left(f_i\rho_0 - D\partial_i \rho_0\right)}_{J_i} 
    \\
    %&= \lambda_n\rho_0 \phi_n - 2\vec{J}\cdot\nabla\phi_n
    &= \lambda_n\rho_0 \phi_n
\end{align}
where the first term disappears because $\rho_0$ is the steady-state distribution, and the third term disappears because of our above assumption on disappearing fluxes, $J_i=0$. 
The following inner product then satisfies
\begin{align*}
    \langle \phi_n \rho_0, \mathcal{L}_{\text{FP}}^{\dagger}\phi_m \rangle &=
    \langle \mathcal{L}_{\text{FP}}\phi_n \rho_0, \phi_m \rangle 
    \\
    \lambda_m \langle \phi_n \rho_0, \phi_m \rangle &=
    \lambda_n \langle \phi_n \rho_0, \phi_m \rangle
    \\
    \rightarrow (\lambda_m-\lambda_n) \langle \phi_n \rho_0, \phi_m \rangle &= 0.
\end{align*}
This shows that if $\lambda_n\neq\lambda_m$, the inner product vanishes. 
Similarly, one can show that the same must be true for $\langle \frac{\rho_n}{\rho_0}, \frac{\rho_m}{\rho_0}\rangle$.
These are precisely the types of terms appearing in the Hessian.
Consequently, for equilibrium (no flux) systems the Hessian is diagonal. The first term follows directly from the above, while the second is given by
\begin{align*}
    \sum_{y}&\frac{1}{p(y)}
        \left(\sum_x p(y,x) \tphi_m(x) \right)
        \left(\sum_{x'} p(y,x')\pl \tphi_n(x') \right)
        \\
        &=
        \sum_{ij}\sum_{y}\frac{1}{p(y)}\rho_i(y)\rho_j(y)
        \ee^{(\lambda_i+\lambda_n)\Delta t}\langle  \phi_i\rho_0, \phi_n \rangle
        \\
        &\quad\quad\quad\quad\times
        \ee^{(\lambda_j+\lambda_m)\Delta t}\langle  \phi_j\rho_0, \phi_m \rangle
        \\
        &=
        \sum_{ij}\ee^{(\lambda_i+\lambda_n+\lambda_j+\lambda_m)\Delta t}\delta_{ij}\delta_{ni}\delta_{mj}
        \\
        &=
        \delta_{nm}\ee^{(2\lambda_n+2\lambda_m)\Delta t}
\end{align*}
The full matrix $A$ appearing in the Hessian then takes the form
\begin{align}
    A_{nm}=\delta_{nm}\ee^{2\lambda_n\Delta t}\left(1 - \beta \ee^{2\lambda_n\Delta t}\right)
    \label{si_eq:stability_mat_equilib}
\end{align}
except for the $n=m=0$ term, which is $A_{00}=0$.
The eigenvalues are given directly by these diagonal elements. For small $\beta$ these are all positive, and become unstable one after the other at 
\begin{align*}
    \beta=\ee^{-2\lambda_1\Delta t}, \ee^{-2\lambda_2\Delta t},...
\end{align*}
which are increasing in order (remember $\lambda_i\leq 0$). It follows that at the first transition, only $f_1$ becomes non-zero, and hence the encoder takes the form given by Eq.~\eqref{si:encoder_mostinformativefeature}. 
In the equilibrium case, the encoder learns \textit{exclusively} the first eigenfunction, with no correction due to the second eigenfunction.

\subsubsection*{General Case}
We now show that the first component $f_1$ is selected at the first IB transition even when the flux $\vec{J}$ is non-zero.
From our assumption on the spectrum of $\mathcal{L}_U$, namely $0>\text{Re}\infev_1>\text{Re}\infev_2\gg \text{Re}\infev_3...$
it follows that near the first IB transition the IB loss is given by 
\begin{align*}
    \mathcal{L}_{\text{IB}}&=
    %(\beta-\beta_{\text{crit}})^2 
    \begin{pmatrix}
        f_1 \\ f_2 
    \end{pmatrix}^T
    \begin{pmatrix}
        a_{11} & a_{12} \\
        a_{12} & a_{22} 
    \end{pmatrix}
    \begin{pmatrix}
        f_1 \\ f_2 
    \end{pmatrix}
     +\mathcal{O}(f_n^3, \,\ee^{\infev_3\Delta t})
\end{align*}
with
\begin{align*}
    a_{11} &= \ee^{2\infev_1\Delta t}\left(\langle\phi_1^2\rangle - \beta B_{11}\right) = \ee^{2\infev_1\Delta t}\hat{a}_{11}
    \\
    a_{12} &= \ee^{(\infev_1 + \infev_2)\Delta t}\left(\langle\phi_1\phi_2\rangle - \beta B_{12}\right) = \ee^{(\infev_1 + \infev_2)\Delta t}\hat{a}_{12}
    \\
    a_{22} &= \ee^{2\infev_2 \Delta t}\left(\langle\phi_2^2\rangle - \beta B_{22}\right) = \ee^{2\infev_2 \Delta t}\hat{a}_{22},
\end{align*}
where we have introduced the shorthand $B_{ij} = \left\langle \langle\tphi_i \rangle_{p(\cdot|y)}\langle\tphi_j \rangle_{p(\cdot|y)} \right\rangle_y $ and all angled brackets denote averaging with respect to the steady state distribution. 
The stability of the uniform encoder at $f_n=0$ is given by the stability of the 2$\times$2 matrix above.
The eigenvalues $\omega_i$ and eigenvectors $\vec{v}_i$ of this matrix can be computed explicitly,
\begin{align*}
    \eta_\pm &= \frac{1}{2}
    \left(
    \ee^{2\infev_1\Delta t}\hat{a}_{11} + \ee^{2\infev_2 \Delta t}\hat{a}_{22}
    \pm D
    \right)
    \\
    \vec{v}_{\pm} &= 
    \begin{pmatrix}
        \frac{-1}{2\ee^{(\lambda_1 + \lambda_2)\Delta t}\hat{a}_{12}}
    \left(-\ee^{2\lambda_1\Delta t}\hat{a}_{11} + \ee^{2\lambda_2 \Delta t}\hat{a}_{22} \mp D\right)
    \\
    1
    \end{pmatrix}
\end{align*}
where 
\begin{align*}
    D &=(\ee^{4\lambda_1\Delta t}\hat{a}_{11}^2 + 2\ee^{2(\lambda_1 + \lambda_2)\Delta t}(2\hat{a}_{12}^2 - \hat{a}_{11}\hat{a}_{22}) 
    +
    \ee^{4\lambda_2 \Delta t}\hat{a}_{22}^2)^{1/2}.
\end{align*}
In what follows, we will express quantities in terms of the spectral gap $\Gamma=\lambda_1-\lambda_2 > 0$. For example, the expression above can be written
\begin{align*}
    D &= \ee^{2\lambda_1\Delta t}\left(\hat{a}_{11}^2 + 2\ee^{-2\Gamma\Delta t}(2\hat{a}_{12}^2 - \hat{a}_{11}\hat{a}_{22})  + \ee^{-4\Gamma\Delta t}\hat{a}_{22}^2\right)^{1/2}
    \\
    &= \ee^{2\lambda_1\Delta t}\left(\hat{a}_{11} + \mathcal{O}(\ee^{-2\Gamma\Delta t})\right).
\end{align*}
The first eigenvalue to become negative is $\eta_+$. We are interested in the ratio of the corresponding eigenvector components, $f_1/f_2 = \vec{v}_{+,1}/\vec{v}_{+,2}$, which tells us how much the encoder will depend on the first eigenfunction $\phi_1(x)$ compared to the second. 
This ratio is given by
\begin{align}
    \frac{f_1}{f_2}&=\frac{-\ee^{(\lambda_1-\lambda_2)\Delta t}}{2\hat{a}_{12}}
    \left(-\hat{a}_{11} + \ee^{2(\lambda_2-\lambda_1) \Delta t}\hat{a}_{22} - D\right)
    \nonumber
    \\
    &=\frac{\ee^{\Gamma\Delta t}}{2\hat{a}_{12}}
    \left(2\hat{a}_{11} + \mathcal{O}(\ee^{-2\Gamma\Delta t})\right)
    \nonumber
    \\
    &=\ee^{\Gamma\Delta t}\frac{\hat{a}_{11}}{\hat{a}_{12}}
     + \mathcal{O}(\ee^{-2\Gamma\Delta t}).
     \label{si_eq:stability_mat_noneq}
\end{align}
This suggests that we must make one additional assumption, which is that the factor $\hat{a}_{11}/\hat{a}_{12}$ is not small.
%For large times $\Delta t\gg 0$, we see that the ratio grows exponentially with a rate given by the spectral gap.
This is true whenever the flux is small, as 
\begin{align}
    \langle\phi_n\phi_m\rangle_{\rho_0} = -\frac{2}{\lambda_n-\lambda_m}\langle (\vec{J}\cdot\nabla\phi_n )\phi_m\rangle_{\rho_0}.
\end{align}

Similar to the equilibrium case, we see that at the first transition the encoder depends only on $f_1$, giving Eq.~\eqref{si:encoder_mostinformativefeature}. 
In contrast to the equilibrium case, there may be a small correction due to the second eigenfunction, however this  becomes exponentially small for long times $\Delta t$.

Note that in this calculation, $\phi_0(x)$ is constant (following from the assumption of a non-degenerate eigenvalue at 0) so that the $f_0(h)$ factors can be absorbed into $p(h)$. 
This changes in the case of a degenerate ground state, which corresponds a situation where there are decoupled sectors in which the dynamics evolve independently. 
Then, each eigenfunction corresponding to the zero eigenvalue is piecewise constant on one of the independent sectors.
The optimal encoder Eq.~\ref{apx:encoder_mostinformativefeature} will depend instead on $\phi_0(x)$, which identifies the independent sectors. 

\subsection{Optimal IB encoder after the transition}
\label{si_sec:discrete_fourth_order}

We characterize the transition by expanding to fourth order and showing that it is continuous. 
We take 
\begin{equation}
    p(\hm|x) = 
    \frac{f_0^{\mu} \exp\left[ \tphi_1(x) f_1^{\mu} \right]}
    {\sum_{\hl}f_0^{\lambda} \exp\left[ \tphi_1(x) f_1^{\lambda} \right]}
\end{equation}
and expand to fourth order in $f_1^{\mu}$. We have 
\begin{align}
p(h_\mu|x) &\approx f_0^{\mu} \Big{[} 1 + \tphi_1(x) f_1^{\mu} 
            \\
            &\qquad+\frac{1}{2}\tphi_1(x)^2 \left((f_1^{\mu})^2 - \langle f_1^2 \rangle_h\right) \\
            &\qquad+ \frac{1}{6}\tphi_1(x)^3 \left((f_1^{\mu})^3 - 3f_1^{\mu} \langle f_1^2 \rangle_h - \langle f_1^3 \rangle_h\right) 
            \\
            &\qquad+\frac{1}{24}\tphi_1(x)^4 \left((f_1^{\mu})^4 - 4f_1^{\mu} \langle f_1^3 \rangle_h - 6(f_1^{\mu})^2 \langle f_1^2 \rangle_h + 6\langle f_1^2 \rangle_h^2 - \langle f_1^4 \rangle_h\right) \Big{]}
            \\
        &\equiv f_0^{\mu} \Big{[} 1 + \tphi_1(x) f_1^{\mu} +\frac{1}{2}\tphi_1(x)^2 A_{\mu} + \frac{1}{6}\tphi_1(x)^3 B_{\mu} +\frac{1}{24}\tphi_1(x)^4 C_{\mu}\Big{]}
\end{align}
where $\langle G\rangle_h=\sum_\mu f_0^{\mu}G(h_{\mu})$ denotes the average over all $h$ (recall that $p(h_{\mu})=f_0^{\mu}$).
The expression for $p(h_{\mu})$ is similar, except all $\phi(x)$ are replaced with averages over $p(x)$: $\langle\phi\rangle_x$.
Similarly, we have
\begin{align}
    \log p(h_\mu|x)
    &= \log f_0^{\mu} + \tphi_1(x) f_1^{\mu} - \frac{1}{2}\tphi_1(x)^2 \langle f_1^2 \rangle_h - \frac{1}{6}\tphi_1(x)^3 \langle f_1^3 \rangle_h - \frac{1}{24}\tphi_1(x)^4 \left( \langle f_1^4 \rangle_h - 3\langle f_1^2 \rangle_h^2 \right)
\end{align}
We can now readily compute the mutual information 
\begin{align}
    I(X_t,H_t)
    &=\sum_\mu\int\dd x \,p(h_\mu|x)p(x)\left[\log p(h_\mu|x) - \log p(h_\mu)\right] \nonumber \\
    &= \left[ \langle \tphi_1^2 \rangle_x \langle f_1^2 \rangle_h + \frac{1}{2}\langle \tphi_1^3 \rangle_x \langle f_1^3 \rangle_h + \frac{1}{6}\langle \tphi_1^4 \rangle_x \left( \langle f_1^4 \rangle_h - 3\langle f_1^2 \rangle_h^2 \right) \right] \nonumber \\
    &\qquad - \left[ \frac{1}{2}\langle \tphi_1^2 \rangle_x \langle f_1^2 \rangle_h + \frac{1}{6}\langle \tphi_1^3 \rangle_x \langle f_1^3 \rangle_h + \frac{1}{24}\langle \tphi_1^4 \rangle_x \left( \langle f_1^4 \rangle_h - 3\langle f_1^2 \rangle_h^2 \right) \right] \nonumber \\
    &\qquad - \frac{1}{8} \langle \tphi_1^2 \rangle_x^2 \left( \langle f_1^4 \rangle_h - \langle f_1^2 \rangle_h^2 \right) \nonumber \\
    &= \frac{1}{2}\langle \tphi_1^2 \rangle_x \langle f_1^2 \rangle_h + \frac{1}{3}\langle \tphi_1^3 \rangle_x \langle f_1^3 \rangle_h + \frac{1}{8}\langle \tphi_1^4 \rangle_x \left( \langle f_1^4 \rangle_h - 3\langle f_1^2 \rangle_h^2 \right) - \frac{1}{8} \langle \tphi_1^2 \rangle_x^2 \left( \langle f_1^4 \rangle_h - \langle f_1^2 \rangle_h^2 \right).
\end{align}
The computation for the prediction term $I(H_t,X_{t+\Delta t})$ proceeds similarly, if slightly messier:
\begin{align}
    I(H_t,Y) &= \frac{1}{2}\langle f_1^2 \rangle_h \left\langle \langle\tphi_1\rangle_{x|y}^2 \right\rangle_y \nonumber \\
    &\quad + \langle f_1^3 \rangle_h \left[ \frac{1}{2} \left\langle \langle\tphi_1\rangle_{x|y} \langle\tphi_1^2\rangle_{x|y} \right\rangle_y - \frac{1}{6} \left\langle \langle\tphi_1\rangle_{x|y}^3 \right\rangle_y \right] \nonumber \\
    &\quad + \langle f_1^4 \rangle_h \left[ \frac{1}{6}\left\langle \langle\tphi_1\rangle_{x|y} \langle\tphi_1^3\rangle_{x|y} \right\rangle_y + \frac{1}{8}\left\langle \langle\tphi_1^2\rangle_{x|y}^2 \right\rangle_y - \frac{1}{4}\left\langle \langle\tphi_1\rangle_{x|y}^2 \langle\tphi_1^2\rangle_{x|y} \right\rangle_y - \frac{1}{8}\langle\tphi_1^2\rangle_x^2 + \frac{1}{12}\left\langle \langle\tphi_1\rangle_{x|y}^4 \right\rangle_y \right] \nonumber \\
    &\quad + \langle f_1^2 \rangle_h^2 \left[ - \frac{1}{2}\left\langle \langle\tphi_1\rangle_{x|y}\langle\tphi_1^3\rangle_{x|y} \right\rangle_y - \frac{1}{8}\left\langle \langle\tphi_1^2\rangle_{x|y}^2 \right\rangle_y + \frac{1}{4}\left\langle \langle\tphi_1\rangle_{x|y}^2\langle\tphi_1^2\rangle_{x|y} \right\rangle_y + \frac{1}{8}\langle\tphi_1^2\rangle_x^2 \right]
\end{align}
Together, these yield the full Lagrangian:
\begin{align}
    \mathcal{L}_{\text{IB}} &= \frac{1}{2}\langle f_1^2 \rangle_h \Big[ \langle\tphi_1^2\rangle_x - \beta \left\langle \langle\tphi_1\rangle_{x|y}^2 \right\rangle_y \Big] \nonumber \\
    &\quad + \langle f_1^3 \rangle_h \left[ \frac{1}{3}\langle\tphi_1^3\rangle_x - \beta \left( \frac{1}{2} \left\langle \langle\tphi_1\rangle_{x|y} \langle\tphi_1^2\rangle_{x|y} \right\rangle_y - \frac{1}{6} \left\langle \langle\tphi_1\rangle_{x|y}^3 \right\rangle_y \right) \right] \nonumber \\
    &\quad + \langle f_1^4 \rangle_h \Bigg[ \frac{1}{8}\left(\langle\tphi_1^4\rangle_x - \langle\tphi_1^2\rangle_x^2\right) \nonumber \\
    &\qquad\qquad\quad - \beta \left( \frac{1}{6}\left\langle \langle\tphi_1\rangle_{x|y} \langle\tphi_1^3\rangle_{x|y} \right\rangle_y + \frac{1}{8}\left\langle \langle\tphi_1^2\rangle_{x|y}^2 \right\rangle_y - \frac{1}{4}\left\langle \langle\tphi_1\rangle_{x|y}^2 \langle\tphi_1^2\rangle_{x|y} \right\rangle_y - \frac{1}{8}\langle\tphi_1^2\rangle_x^2 + \frac{1}{12}\left\langle \langle\tphi_1\rangle_{x|y}^4 \right\rangle_y \right) \Bigg] \nonumber \\
    &\quad + \langle f_1^2 \rangle_h^2 \Bigg[ \frac{1}{8}\left(\langle\tphi_1^2\rangle_x^2 - 3\langle\tphi_1^4\rangle_x\right) \nonumber \\
    &\qquad\qquad\quad - \beta \left( - \frac{1}{2}\left\langle \langle\tphi_1\rangle_{x|y}\langle\tphi_1^3\rangle_{x|y} \right\rangle_y - \frac{1}{8}\left\langle \langle\tphi_1^2\rangle_{x|y}^2 \right\rangle_y + \frac{1}{4}\left\langle \langle\tphi_1\rangle_{x|y}^2\langle\tphi_1^2\rangle_{x|y} \right\rangle_y + \frac{1}{8}\langle\tphi_1^2\rangle_x^2 \right) \Bigg]
\end{align}
For further analysis, we consider the case where $N_H=2$. This is sufficient to study the behavior at the transition because more than one encoding degree of freedom would be redundant. 
We further consider the case of a symmetric perturbation, in which $f_1^0=-f_1^1\equiv f$. This simplifies the moments of $f$ significantly, yielding
\begin{align}
    \langle f_1^2\rangle_h &= f^2
    \\
    \langle f_1^3\rangle_h &= 0
    \\
    \langle f_1^4\rangle_h &= f^4
\end{align}
Then the IB loss simplifies dramatically to
\begin{align}
    \mathcal{L}_{\text{IB}} &= \frac{1}{2}f^2 \left[ \langle\phi_1^2\rangle_x - \beta \left\langle \langle\phi_1\rangle_{x|y}^2 \right\rangle_y \right] - \frac{1}{12}f^4 \left[ 3\langle\phi_1^4\rangle_x - \beta \left( 4\left\langle \langle\phi_1\rangle_{x|y} \langle\phi_1^3\rangle_{x|y} \right\rangle_y - \left\langle \langle\phi_1\rangle_{x|y}^4 \right\rangle_y \right) \right]
\end{align}
which has minima at
\begin{equation}
    f^* = \pm \sqrt{ \frac{3 \left( \langle\phi_1^2\rangle_x - \beta \left\langle \langle\phi_1\rangle_{x|y}^2 \right\rangle_y \right)}{ 3\langle\phi_1^4\rangle_x - \beta \left( 4\left\langle \langle\phi_1\rangle_{x|y} \langle\phi_1^3\rangle_{x|y} \right\rangle_y - \left\langle \langle\phi_1\rangle_{x|y}^4 \right\rangle_y \right)} }.
\end{equation}
A stable solution emerges when 
\begin{align}
    \beta > \frac{\langle\phi_1^2\rangle_x}{\left\langle \langle\phi_1\rangle_{x|y}^2 \right\rangle_y},
\end{align}
which matches the transition criterium for a Gaussian encoder (see below). Together, this shows that the value of $f$ changes continuously at the transition.

\subsection{Stability analysis for continuous encodings}
\label{si_sec:gaussian_stability}

Here we compute the stability of the uniform encoder for continuous latent variables $h$, where we constrain $p(h|x)$ to be Gaussian. This is motivated by the fact that variational IB is performed with precisely such an Ansatz. For simplicity, we assume a fixed and isotropic standard deviation $\sigma$:
\begin{align}
    p(h|x)= N\exp\left[-\frac{1}{2\sigma^2}(h-\mu(x))^2\right]
\end{align}
We further consider the following expansion for $\mu$
\begin{align}
    \mu(x)=\sum_i m_i\phi_i(x).
\end{align}
Our goal is to study the stability of the IB objective as a function of the coefficients $m_i$, analogous to the section above where we consider variations of the coefficients $f_i$. The trivial encoder again corresponds to $m_i=0$ for all $i$.

We start by computing 
\begin{align}
    \frac{\partial}{\partial m_i}p(h|x) &= \frac{-1}{\sigma^2}\phi_i(x)\left(h-\mu(x)\right)p(h|x)
\end{align}
which, at the uniform encoder, evaluates to
\begin{align}
    \frac{\partial}{\partial m_i}p(h|x) &= \frac{-1}{\sigma^2}h\phi_i(x)p(h)
\end{align}
The compression term is given by
\begin{align*}
    \partial_n \partial_m I(X,H)
&=  \int \dd x' \dd h
    \frac{p(x')}{p(h)}\partial_m p(h|x')\partial_n p(h|x')
    - 
    \underbrace{\int \dd h \frac{1}{p(h)}\partial_m p(h) \partial_n p(h)}_{\partial_m\partial_n 1=0}
    \\
    &=  \frac{1}{\sigma^4}\int \dd x' \dd h
    p(x')\phi_n(x')\phi_m(x')h^2p(h)
    \\
    &=  \frac{1}{\sigma^2}\langle
    \phi_n\phi_m\rangle.
\end{align*}
The Hessian of the prediction term is, similarly to the above calculation for discrete $h$, given by
\begin{align*}
    \partial_n \partial_m I(Y,H)
    &= 
    \int \dd y \,\dd h\,\frac{1}{p(y)p(h)}
        \left(\int \dd x\, p(y,x)\partial_m p(h|x)\right)
        \left(\int \dd x' p(y,x')\partial_n p(h|x')\right)
    \\
    &= 
    \frac{1}{\sigma^4}\int \dd y \,p(y)
        \left(\int \dd x\, \phi_m(x)p(x|y) \right)
        \left(\int \dd x'\phi_n(x')p(x'|y)\right)
        \int \dd h \, p(h)h^2
    \\
    &= \frac{1}{\sigma^2}\left\langle\langle \phi_m\rangle_{p(\cdot|y)}\langle \phi_n\rangle_{p(\cdot|y)}\right\rangle_y    
\end{align*}
The Hessian of the full Lagrangian, given this Ansatz, is then
\begin{align}
    \partial_i\partial_j\mathcal{L}_{\text{IB}}
    =
    \frac{1}{\sigma^2}\left(\langle
    \phi_n\phi_m\rangle - \beta\left\langle\langle \phi_m\rangle_{p(\cdot|y)}\langle \phi_n\rangle_{p(\cdot|y)}\right\rangle_y  \right)
\end{align}
This is exactly the form that $A^{\beta}_{ij}$ takes above, without the $n,m=0$ terms that come from the normalization term (with our Gaussian Ansatz the normalization term is unaffected and hence these terms do not appear). Note that for discrete encodings, all information about $h$ was contained in the matrix $G_{\mu\nu}$, and this matrix did not affect the stability of the uniform encoder. Similarly, the prefactor here is nothing other than the inverse of the variance of the trivial encoder $p(h)$.

\subsection{Gaussian encoder after the first transition}
\label{si_sec:gaussian_fourth_order}

We now build on our result from above showing that at the first transition the $m_1$ direction becomes unstable. Assuming an Ansatz
\begin{align}
    p(h|x)=N\exp\left(\frac{-1}{2\sigma^2}(h-m_1\phi_1(x))^2\right),
\end{align}
we ask at what value of $m_1$ the IB objective obtains its minimum. To do so, we perform an asymptotic expansion of the objective in $m_1$ up to fourth order.
The conditional distribution can be expanded as
\begin{align*}
    p(h|x)&=p_0(h)\Big[1+\frac{h}{\sigma^2}\phi_1(x)m_1
    + \frac{(h^2-\sigma^2)}{2\sigma^2}\phi_1^2(x)m_1^2
    +\frac{h(h^2-3\sigma^2)}{6\sigma^2}\phi_1^3(x)m_1^3+
    \\
    &\quad\quad+\frac{(h^4-6h^2\sigma^2+3\sigma^4)}{24\sigma^8}\phi_1^4(x)m_1^4 + \mathcal{O}(m_1^5)
    \Big]
\end{align*}
We compute the integral to check normalization:
\begin{align*}
    \int p(h|x)dh = 1
    + \frac{(\sigma^2-\sigma^2)}{2\sigma^2}\phi_1^2(x)m_1^2
    +\frac{(3\sigma^4-6\sigma^4+3\sigma^4)}{24\sigma^8}\phi_1^4(x)m_1^4=1
\end{align*}
The compression term is
\begin{align*}
    I(X,H)&= 
    \int \dd x\, p(x)
    \frac{1}{2\sigma^2}\left(\phi_1(x)-\langle\phi_1\rangle\right)^2m_1^2
    +
    \\
    &\quad +\int \dd x\, p(x)\frac{1}{4\sigma^4}\left( 2(\phi_1(x)-\langle\phi_1\rangle)^2 + \langle\phi_1\rangle^2 -\langle\phi_1^2\rangle
   \right)(\langle\phi_1\rangle^2-\langle\phi_1^2\rangle)m_1^4
   \\
   &= 
    \frac{1}{2\sigma^2}(\langle\phi_1^2\rangle - \langle\phi_1\rangle^2)
    m_1^2
   + \frac{1}{4\sigma^4}
   \left( \langle\phi_1^2\rangle - \langle\phi_1\rangle^2 
   \right)\left(\langle\phi_1\rangle^2-\langle\phi_1^2\rangle\right)m_1^4 
   \\
   &= 
    \left(\frac{\langle\phi_1^2\rangle - \langle\phi_1\rangle^2}{2\sigma^2}\right)
    m_1^2
   - \left(\frac{\langle\phi_1^2\rangle - \langle\phi_1\rangle^2 }{2\sigma^2}\right)^2
   m_1^4 
\end{align*}
Note that the prefactors may be written
\begin{align*}
    \left(\frac{\langle\phi_1^2\rangle - \langle\phi_1\rangle^2}{2\sigma^2}\right)=\frac{\text{Var}(\phi_1)}{2\sigma^2}
\end{align*}
The fourth order term in the prediction term is
\begin{align*}
    %&= (f^4 - 2 f^2 f2 + f2^2 + 2 f2 g^2 - 1 g^4  - 2 f2 g2 + g2^2) 
    %\\&=  (f^2 - f2)^2  + f2^2 - f2^2 + 2 f2 g^2 - 1 g^4  - 2 f2 g2 + g2^2) 
    %\\&=  (f^2 - f2)^2  - (f2 - g^2)^2  + f2^2 - 2 f2 g2 + g2^2) 
    %\\&=  (f^2 - f2)^2  - (f2 - g^2)^2  + (f2 - g2)^2
    %\\&= ((f^2 - f2)^2  - (f2 - g^2)^2  + (f2 - g2)^2 )
    %\\
    %&= (...)-f2^2 + 2g2\,g^2 - g^4 + f2^2 - 2g2\,g2 + g2^2
    %\\
    %&= 2g2\,g^2 - g^4 - g2^2 
   &=-\frac{1}{4 \sigma^4} \int\dd y\,
    \left(
    \left( \langle\phi\rangle_{p(\cdot|y)}^2- \langle\phi^2\rangle_{p(\cdot|y)}\right)^2
    -
    \left( \langle\phi\rangle^2- \langle\phi^2\rangle_{p(\cdot|y)}\right)^2
    +
    \left( \langle\phi^2\rangle_{p(\cdot|y)}- \langle\phi^2\rangle\right)^2    
    \right)m_1^4
    \\
    &=\frac{1}{4 \sigma^4} 
    \left(
    \left\langle\left( \langle\phi\rangle_{p(\cdot|y)}^2- \langle\phi^2\rangle_{p(\cdot|y)}\right)^2\right\rangle_{y}
    -
    \left(\langle\phi\rangle^2- \langle\phi^2\rangle\right)^2  
    \right)m_1^4
    \\
    &=\left(
    \left\langle\left( \frac{\langle\phi\rangle_{p(\cdot|y)}^2- \langle\phi^2\rangle_{p(\cdot|y)}}{2\sigma^2}\right)^2\right\rangle_{y}
    -
    \left(\frac{\langle\phi\rangle^2- \langle\phi^2\rangle }{2\sigma^2}\right)^2\right)m_1^4
\end{align*}
While the second order term is of the same form as in $I(X,H)$
\begin{align}
    \left\langle\left(\frac{\langle\phi_1\rangle_{p(\cdot|y)} - \langle\phi_1\rangle}{2\sigma^2}\right)^2\right\rangle_y
    m_1^2
\end{align}
Put together, we find 
\begin{align*}
    \mathcal{L}_{\text{IB}}
    &=
    \frac{1}{2\sigma^2}\left[ \left(\langle\phi_1^2\rangle - \langle\phi_1\rangle^2\right)
    -\beta\left\langle\langle\phi_1\rangle_{p(\cdot|y)}^2 - \langle\phi_1\rangle^2\right\rangle_y\right]
    m_1^2
    \\
    &\quad
    -
    \frac{1}{(2\sigma^2)^2}\left[(1-\beta)\left(\langle\phi_1^2\rangle - \langle\phi_1\rangle^2\right)^2-\beta\left\langle\left( \langle\phi_1\rangle_{p(\cdot|y)}^2- \langle\phi_1^2\rangle_{p(\cdot|y)}\right)^2\right\rangle_{y}\right]m_1^4
\end{align*}
This can be neatly summarized as
\begin{align}
    \mathcal{L}_{\text{IB}}
    &\propto 2\sigma^2(\text{Var}(\phi_1)-\beta\langle\text{Var}(\phi_1|y)\rangle_y)m_1^2
        - \left((1-\beta)\text{Var}^2(\phi_1) -  \beta\langle\text{Var}^2(\phi_1|y)\rangle_y\right)m_1^4
\end{align}
which has minima at 
\begin{align}
    m_1 = 
    \begin{cases}
        0&\text{if $\text{Var}(\phi_1)\geq\beta\text{Var}(\phi_1|y)$}
    \\
        \sqrt{\frac{\sigma^2(\text{Var}(\phi_1)-\beta\text{Var}(\phi_1|y))}{(1-\beta)\text{Var}^2(\phi_1) -  \beta\text{Var}^2(\phi_1|y))}}
        &\text{otherwise.}
    \end{cases}
\end{align}
After the first transition, both the numerator and denominator in the radical will become negative so that the entire quantity is real.

\subsection{Stability analysis of \texorpdfstring{$I(H_t, H_{t+\Delta t})$}{I(Ht,Ht+dt)}}
\label{si_sec:IHH_stability}
Here we compute the stability of the InfoMax objective. For notational simplicity we write $H_t\equiv H_0$, $H_{t+\Delta t}\equiv H_t$. The mutual information we will calculate the Hessian of is given by
\begin{align*}
    I(H_0, H_t)=\int\dd h_0\dd h_t \,p(h_0,h_t)\log\frac{p(h_0,h_t)}{p(h_0)p(h_t)}.
\end{align*}
We want to compute the derivatives of this object with respect to $f^\mu_n$, where
\begin{align*}
    p(h_\mu|x)=\frac{p(h_\mu)}{N(x)}\exp\left[\sum_i\phi_i(x)f_i^\mu\right]
\end{align*}
As a reminder, Greek indices correspond to indices of the discrete latent variable $h_{\mu}\in\{0,...,N_H-1\}$ while latin indices correspond to eigenfunctions of the transfer operator, ordered by eigenvalue. From our computations above, we have
\begin{align}
    \partial_n^\mu p(\hn|x)&=p(\hn|x)\tphi_n(x)\left(\delta_{\mu\nu}
    -
    p(\hm|x)\right)
    \nonumber
    \\
    &=p(\hn)\tphi_n(x)\left(\delta_{\mu\nu}
    -
    p(\hm)\right)
    \label{eq:deriv_encoder_h_at_trivial}
\end{align}
where for $n=0$ we assign $\tphi_0=\frac{1}{f_0^{\mu}}$, taking care to track the correct Greek index which otherwise is not present for $\tphi_n$. The second line is the derivative evaluated at the trivial encoder ($f_i=0$ for all $i$).
Similarly,
\begin{align}
    \partial_n^\mu p(\hn)    &=\int\dd x\,p(x)p(\hn|x)\tphi_n(x)\left(\delta_{\mu\nu}
    -
    p(\hm|x)\right)
    \nonumber
    \\
    &= p(\hn)(\delta_{\mu\nu}-p(\hm))\delta_{n0}
    \label{eq:deriv_marginal_h}
\end{align}
also evaluated at the trivial encoder.
The joint distribution may be written 
\begin{align*}
    p(h_0,h_t)&=\int\dd x_0\dd x_t \, p(h_0,h_t|x_0,x_t)p(x_0,x_t)
    \\
    &=\int\dd x_0\dd x_t \, p(h_t|x_t)p(h_0|x_0)p(x_0,x_t)
\end{align*}
where we used the fact that the variables form the Markov chain $H_0-X_0-X_t-H_t$.
The derivative is thus
\begin{align*}
    \partial_n^\mu p(h_0,h_t)&=\int\dd x_0\dd x_t \, p(h_0,h_t|x_0,x_t)p(x_0,x_t)
    \\
    &=\int\dd x_0\dd x_t \, \left(\partial_n^\mu p(h_t|x_t) \right)p(h_0|x_0) p(x_0,x_t)
    +p(h_t|x_t)\left(\partial_n^\mu p(h_0|x_0) \right)p(x_0,x_t)
    \\
    &=\int\dd x_0\dd x_t \, p(\hn^t)\left(\delta_{\mu\nu}
    -
    p(\hn^t)\right)\tphi_n(x_0) p(\hl^0) p(x_0,x_t)
    +p(\hn^t)
    p(\hl^0)\left(\delta_{\mu\lambda}
    -
    p(\hl^0)\right)\tphi_n(x_t)p(x_0,x_t)
    \\
    &=
    \delta_{n0}
    p(\hn^t)p(\hl^0)
    \left(\delta_{\mu\nu}
    -
    p(\hn^t)
    +\delta_{\mu\lambda}
    -
    p(\hl^0)\right)
\end{align*}
To get to the last line, we evaluated everything at the trivial encoder and used the assumption that $p(x_0)$ and $p(x_t)$ are the same distribution (steady state assumption).
Note that the derivative depends only trivially on $n$, so that it does not ``see'' subleading eigenfunctions. 
The Hessian of the mutual information is, denoting $h^0_i=h_{i}$
\begin{align}
    \partial_n^\nu\partial_m^\mu I(H_0,H_t)
    &=
    \partial_n^\nu\sum_{h_i,h^t_j}
    \partial_m^\mu p(h_i,h^t_j)\left(\log\frac{p(h_i,h_j^t)}{p(h_i)p(h_j^t)} + 1\right)
    - \underbrace{\frac{p(h_i,h_j^t)}{p(h_i)}\partial_m^\mu p(h_i)}_{\sim\sum_i\partial_n p(h_i)=0}
    - \frac{p(h_i,h_j^t)}{p(h_j^t)}\partial_n^\nu p(h_i)
    \nonumber
    \\
    &=
    \sum_{h_i,h^t_j}
    \partial_n^\nu\partial_m^\mu p(h_i,h^t_j)\left(\log\frac{p(h_i,h_j^t)}{p(h_i)p(h_j^t)} + 1\right)
    +
    \partial_n^\nu p(h_i,h^t_j)\partial_m^\mu p(h_i,h^t_j)\frac{1}{p(h_i,h^t_j)}
    \nonumber
    \\
    &\quad\quad
    -\sum_{h_i,h^t_j}
    \partial_n^\nu p(h_i)\partial_m^\mu p(h_i,h^t_j)\frac{1}{p(h_i)}
    -
    \partial_n^\nu p(h_j^t)\partial_m^\mu p(h_i,h^t_j)\frac{1}{p(h_j^t)}
\end{align}
Evaluated at the uniform encoder, the first term disappears because we have the derivative of a constant. 
From the equations for the derivatives Eq.~\eqref{eq:deriv_encoder_h_at_trivial} and Eq.~\eqref{eq:deriv_marginal_h}, we can conclude $\partial_n^\nu\partial_m^\mu I(H_0,H_t)\propto \delta_{n0}\delta_{m0}$, which shows that we lose all information about the eigenfunctions.
The Hessian of the Lagrangian for IB modified by the term $I(H_0,H_t)$ will therefore have the same stability properties as the original IB problem.

\subsection{Numerical experiments}
\label{si_sec:perturbation_numerics}
%\label{sec:mostinformative}

Here we numerically study the behavior of the encoder in the high-compression regime.
By isolating its dependence on $x_t$, we may identify the most relevant features of the state variable and show that they coincide with left eigenfunctions of the transfer operator.

We consider to the simple example of a particle in a double well considered in the main text, which we map to a discrete variable $H_t\in \{0,...,N_H-1\}$. 
In this system the IB loss function can be optimized directly, as shown in Ref.~\cite{tishby_IB}, using the Blahut-Arimoto algorithm \cite{cover2012elements} described above.

\begin{figure}[tp]
    \centering
    \includegraphics[width=0.9\textwidth]{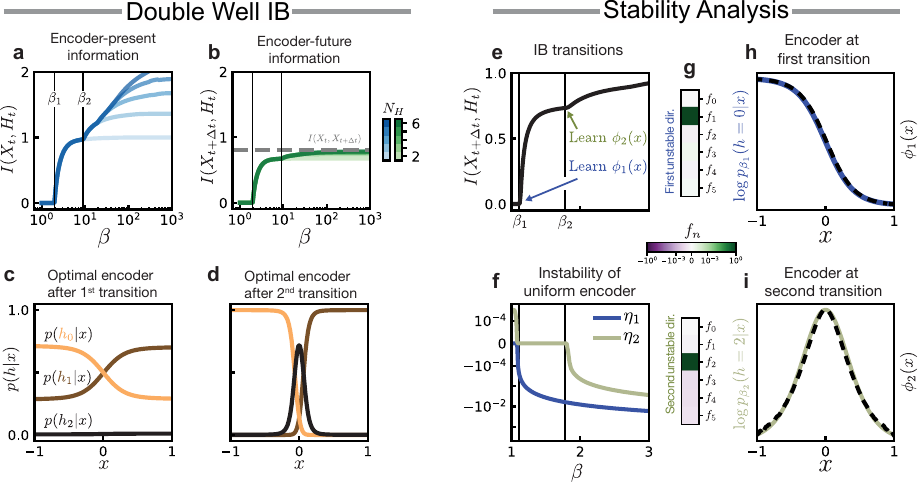}
    \caption{\textbf{Information bottleneck for a Brownian particle in a double well.}
    (a) Mutual information between the current state $X_t$ and its encoding $H_t$ for various alphabet sizes $N_H$ (color). 
    The first two IB transitions are denoted with black lines and occur at $\beta_1$ and $\beta_2$.
    (b) Mutual information between the current state's encoding $H_t$ and the future $X_{t+\Delta t}$. The gray dashed line denotes the maximum attainable information, which is the mutual information $I(X_t, X_{t+\Delta t})$. Color denotes encoding alphabet size.
    Black lines showing IB transitions are shown for reference.
    (c) Optimal encoder after the first transition ($\beta\gtrsim \beta_1$) with alphabet size $N_H=3$.
    (d) Optimal encoder after the second transition ($\beta\gtrsim \beta_2$) with alphabet size $N_H=3$.
    (e) When $\beta$ is increased, the IB objective favors less compression so that more and more information can go through the encoder.
    This occurs in steps, where the spectral content of the transfer operator is included starting from eigenvalues with largest magnitude (i.e., the slowest ones). 
    (f) Transitions are characterized by the appearance of negative eigenvalues in the spectrum of the Hessian of the IB loss function.
    Here we consider the Hessian evaluated at the uniform encoder $p(h|x)=N_H^{-1}$.
    The IB transitions $\beta_1 \approx 1.1$ and $\beta_2 \approx 1.8$  correspond to the appearence of negative eigenvalues of the Hessian.
    (g) The unstable directions are dominated by single components (note the color scale is logarithmic). 
    (h) At the first transition, the logarithm of the encoder is given by the eigenfunction $\phi_1(x)$, up to rescaling ($y$-axis is shown in arbitrary units).
    (i) Likewise, at the second transition the encoder is given by $\phi_2(x)$.
    }
    \label{si_fig:doublewell_stability}
\end{figure}

To focus on the properties of encodings for varying degrees of compression $\beta$, we consider a fixed set of dynamical parameters $\mu$ and $\sigma$. 
Increasing $\beta$ reduces the amount of compression, i.e. ``widens'' the bottleneck, allowing more information to pass into the encoder. 
This leads to a series of IB transitions (Fig.~\ref{si_fig:doublewell_stability}a,b).
%The information about the present state $I(X_{t}, H_t)$ is bounded by the entropy of the random variable $X_t$, which occurs when $H_t$ exactly encodes $X_t$.
The form of the optimal encoder changes qualitatively at these transitions. Before $\beta_1$, the optimal encoder has no dependence on $x$ so that $p(H_t=h_i|x_t)=\text{const}$ for all $h_i$.
After the first transition, the encoder begins to associate regions of $x$ to particular values of $h$ (Fig.~\ref{si_fig:doublewell_stability}c,d).
%which can be understood as a soft partition of the state space (SI Fig.~S2c). 
%In this case, $h$ essentially encodes the identity of the well which the particle is currently located in.
%When $\beta$ is increased further, the encoder %undergoes another transition and begins to distinguish states $x$ which are near the barrier of the double well (SI Fig.~S2d).
We are interested in the form of the encoder at $\beta\gtrsim\beta_1$, just above the first IB transition, as this reflects the \textit{most informative} features of the full state variable $x$ (Fig.~\ref{si_fig:doublewell_stability}e).
The dependence of $p(h_t|x_t)$ on $x$ can be explained by our stability analysis above.
Stability is governed by the eigenvalues $\eta_i$ of the Hessian of the IB Lagrangian with respect to the parameters $f_n(h_t)$. These parameters tell us how much the encoder ``weights'' each transfer operator eigenfunction; $f_n(h_t)=0$ (for $n>0$) corresponds to the uniform, or trivial encoder $p(h_t|x_t)=p(h_t)$.

For small $\beta$ all eigenvalues $\eta_i$ are positive, indicating that the uniform encoder is a stable minimum of the IB Lagrangian. 
In Fig.~\ref{si_fig:doublewell_stability}f we show the smallest two eigenvalues of the IB Hessian when evaluated at the uniform encoder. 
At the first transition one eigenvalue becomes negative, so that the uniform encoder is unstable.
The eigenvector corresponding to the unstable eigenvalue $\eta_1$ indicates how the weights $f_n(h_t)$ should be adjusted to lower the value of the IB Lagrangian.
As expected from our analysis, these are given by a vector which isolates $f_1$ (Fig.~\ref{si_fig:doublewell_stability}g, top).
By taking the logarithm of the encoder after the transition, we can independently confirm that the encoder depends only on $\phi_1(x)$ (Fig.~\ref{si_fig:doublewell_stability}h).

\begin{figure}[tp]
    \centering
    \includegraphics[width=0.6\textwidth]{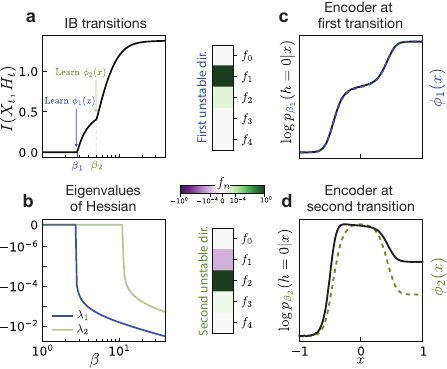}
    \caption{\textbf{IB transitions for a Brownian particle in a triple-well potential.}
    (a) Information transitions with varying $\beta$ for a particle in a triple-well potential.
    (b) Eigenvalues of the IB Hessian evaluated at the uniform encoder $p(h|x)=N_H^{-1}$.
    The appearance of an unstable direction at $\beta_1 \approx 3$  coincides with the first IB transition.
    The emergence of a second unstable direction doesn't correspond precisely to the IB transition at $\beta_2$ because stability is evaluated at the uniform encoder, but the true optimal encoder at $\beta_2$ is not uniform.
    (c) The unstable directions are dominated by single components (note the color scale is logarithmic). At the first transition, the logarithm of the encoder is given by the eigenfunction $\phi_1(x)$, up to rescaling ($y$-axis is shown in arbitrary units).
    (d) Likewise, at the second transition the encoder is given primarily by $\phi_2(x)$.
    }
    \label{si_fig:triplewell_stability}
\end{figure}

Our stability analysis predicts that a second mode becomes unstable at the second IB transition $\beta\approx \beta_2$. Here we see that this unstable mode selects $f_2$, and that the encoder correspondingly gains dependence on $\phi_2(x)$ (Fig.~\ref{si_fig:doublewell_stability}i).
Note that in general, $\eta_2$ must not necessarily become negative precisely at $\beta_2$ because the stability analysis is performed at the uniform encoder while the true optimal encoder has already deviated from uniformity. 
In Figure~\ref{si_fig:triplewell_stability}, we perform the same analysis for a triple-well potential where this difference is more apparent.

%Because the first IB transition is a second order transition \cite{gedeon2002}, we may write the optimal encoder near the transition as \cite{wu2020}
%\begin{align*}
%    p(h|x) = p(h) + (\beta - \beta^1)\phi_1(x)f_1(h). 
%\end{align*}

\subsection{Limitations of the above results}
\label{si_sec:expansion_limitations}

As mentioned above, our result is exact only for systems (such as equilibrium systems) that satisfy
\begin{align*}
    \langle \phi_n,\rho_0\phi_m\rangle=\delta_{nm}.
\end{align*}
Although our computation of the Hessian is still correct for non-equilibrium systems, it may become very far from a diagonal matrix in the presence of transient dynamics.
Unless otherwise stated, we assume that the systems in this work are close to a steady state, so that the initial state $x_t$ is sampled from the stationary distribution.

What happens when this assumption is violated, and we only seek to describe our system during some transient behavior?
One such setting is when the system is prepared in a state far from the equilibrium so that it undergoes strong transient dynamics.
In such regimes, pseudospectra may be more relevant than the operator's spectrum, and transfer operator eigenfunctions may cease to be a useful descriptor of the system.

\begin{figure}[tp]
    \centering
    \includegraphics[width=0.8\linewidth]{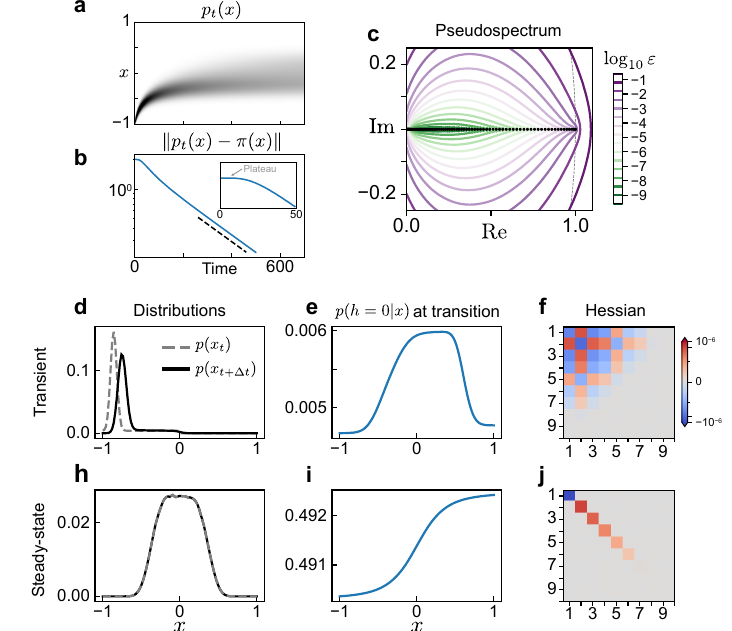}
    \caption{
    \response{\textbf{IB for transient dynamics}
    (a) Evolution of an initial distribution of particles prepared in a state highly concentrated state near $x\approx -1$ in a potential well centered around $x=0$. 
    (b) Relaxation of the distribution towards the steady state distribution $\pi(x)$. 
    For long times, this relaxation occurs on a time scale given by the subleading eigenvalue $\lambda_1$ of the transfer operator (dashed line has slope $\lambda_1$). For short times however (inset), the dynamics has nothing to do with this eigenvalue and instead exhibits a ``cut-off'' phenomenon \cite{trefethen2005spectra} where the distance from the steady state is constant until some time $t\approx 20$ where the distance begins to decay. 
    (c) Pseudospectrum of the transfer operator in (a,b), shown as level sets of the resolvent.
    (d) Initial and final distribution of transient dynamics.
    (e) IB-optimal encoder at the first IB transition for the transient dynamics in (d).
    (f) Hessian of the IB objective function parameterized in terms of coefficients for transfer operator eigenfunctions (as in SI).
    (h-i) Corresponding results for IB performed on steady-state dynamics.}
    }
    \label{si_fig:transients_exact_IB_pseudospectr}
\end{figure}

In our theory, this deviation from a description in terms of the transfer operator spectrum is reflected in a change in structure in the Hessian, which is clearly illustrated in the case of a Brownian particle in a potential well in Fig.~\ref{si_fig:transients_exact_IB_pseudospectr}. 
This means that the features learned by IB will not coincide exactly with the operator's eigenfunctions.
The Hessian is unstable with respect to deviations exactly in the direction of $\phi_1(x)$ only in the case that the overlap between eigenfunctions $\langle\phi_1(x),p(x)\phi_n(x)\rangle$ is small for $n\neq 1$. 
For equilibrium systems, this quantity is zero. However, if $p(x)$ is not the steady state but some other arbitrary distribution, this quantity may be far from zero. 
That is particularly clear in the potential well, where the eigenfunctions are nearly constant away from the center of the well. By choosing a distribution that is only supported in these regions where $\phi_1\approx C_1$ and $\phi_n\approx C_n$, one can get large values of the overlap. 
The encoder learned by IB for transient dynamics is shown in Fig.~\ref{si_fig:transients_exact_IB_pseudospectr}, where we see that it strongly depends on $x$ in the region of the transient.
This shows that IB learns a relevant feature even in the case of transient dynamics. 

\subsection{``Distillation'' of the optimal encoder}
\label{si_sec:distillation}
\response{
Here we elaborate on how to interpret the encoding variable $h$ and its relation to the transfer operator eigenfunctions.
As mentioned in the main text, in the limit of high compression (small $\beta$), the encoder is given by a small deviation $\delta p$ from the trivial encoder $p(h)$, 
\begin{align}
    p(h_t|x_t)=p(h_t)(1+\delta p(h_t|x_t)).
\end{align}
In this regime the dynamics of $h_t$ are dominated by noise. One can see this by considering
\begin{align}
    p(h_{t+\Delta t},h_t) &= \int \dd x_{t+\Delta t}\dd x_t\,
    p(h_{t+\Delta t},x_{t+\Delta t},x_t|h_t) p(h_t)
    \\
    &= \int \dd x_{t+\Delta t}\dd x_t\,
    p(h_{t+\Delta t}|x_{t+\Delta t})p(x_{t+\Delta t}|x_t)p(x_t|h_t)p(h_t)
    \\
    &= p(h_{t+\Delta t})p(h_t)(1+\mathcal{O}(\delta p^2)),
\end{align}
which shows that the variables $h_{t+\Delta t}$ and $h_t$ are nearly statistically independent of one another.
This means that $h$ may still be a nearly completely random variable; it is the small deviations from randomness that interest us.}

\response{
How can we extract the ``interesting'', predictive part of the encoder?
We are not interested in the variable $h$ itself, but rather how it couples to the variable $x$.
To isolate this, we disregard the noise by taking the weak coupling we discover with IB, and effectively letting the coupling become strong.
In practice, this is achieved by simply taking the most likely state for a particular value of $x_t$:
\begin{align*}
    h_t(x_t)=\arg\max_{h}p(h|x_t).
\end{align*}
For variational (continuous) IB this is straightforward: rather than showing the latent variable after ``reparameterization'' (which amounts to adding Gaussian noise; see Section 4C and Section 5) we show the variable before noise has been added.
The resulting $h_t$ is given by the mean (and, because we assume Gaussianity, the mode) of the distribution. 
For exact IB this is also straightforward because we have access to the full encoder $p(h|x)$, so we simply take the most-likely discrete state for each $x$. We note that this is equivalent to fixing a (small) dimensionality and taking large $\beta$, which is what was done for the discrete problems in the main text.}

\response{
For continuous variables learned in VIB this procedure directly delivers the transfer function eigenfunctions. For discrete IB this is not the case, however, because the learned variable and the eigenfunction live in different spaces: for example, $h_t\in \{1,...,N_H\}$ while $\phi(x_t)\in L^\infty(\mathbb{R}^n)$. In order to obtain the eigenfunctions from these encodings (as presented in e.g. Figs.~\ref{si_fig:doublewell_stability} and \ref{si_fig:triplewell_stability}), one must effectively extract the conditional probability. In the Blahut Arimto algorithm we use, we have direct access to this quantity. If one instead only has samples of $h_t$ it may be approximated as the mean $p(h_t|x_t)=\mathbb{E}[h_t|x_t]$.
}

The above distillation procedure is closely related to the deterministic information bottleneck \cite{strouse2016deterministicIB}, however the two procedures are not identical, as noted in Ref.~\cite{strouse2016deterministicIB}.
In short, deterministic IB yields a delta-distributed encoder
\begin{align*}
    p(h|x_t)=\delta (h - h^*(x_t))
\end{align*}
with 
\begin{align}
    h_{\text{DIB}}^*(x_t)=\underset{h}{\operatorname{argmax}} \left[\log p_{\text{DIB}}(h)-\beta D_{\text{KL}}(p(x_{t+\Delta t}|x)\lVert p_{\text{DIB}}(x_{t+\Delta t}|h)\right].
\end{align}
where the distributions $p_{\text{DIB}}(h)$ and $p_{\text{DIB}}(x_{t+\Delta t}|h)$ are determined self-consistently with the definition of $h_{\text{DIB}}^*(x_t)$.
Our distillation procedure is similar, but uses the IB-optimal encoder:
\begin{align}
    h_{\text{IB,distilled}}^*(x_t)
    &=\underset{h}{\operatorname{argmax}}\log\left(\frac{p_{\text{IB}}(h)}{\mathcal{N}(x)}\exp\left[-\beta D_{\text{KL}}(p(x_{t+\Delta t}|x)\lVert p_{\text{IB}}(x_{t+\Delta t}|h)\right]\right)
    \\&=\underset{h}{\operatorname{argmax}} \left[\log p_{\text{IB}}(h)-\beta D_{\text{KL}}(p(x_{t+\Delta t}|x)\lVert p_{\text{IB}}(x_{t+\Delta t}|h)\right]
\end{align}
Here, the distributions $p_{\text{IB}}(h)$ and $p_{\text{IB}}(x_{t+\Delta t}|h)$ are not determined self-consistently in relation to $h_{\text{IB,distilled}}^*(x_t)$, but instead from the IB objective. Hence, the solutions may differ in practice.

\section{Variational IB}
\label{si_sec:dvib}
Exactly solving the IB problem in principle requires access to the full distribution $p(x_{t},x_{t+\Delta t})$. 
One way around this is via so-called Deep Variational IB as introduced in Ref.~\cite{alemi2016_dvib} which optimizes an an upper bound on the IB objective 
\begin{align}
    \mathcal{L}_{\text{IB}} = I(X_t; H_t) - \beta I(X_{t+\Delta t}; H_t).
    \label{vibloss_methods}
\end{align}
To bound the first term, \cite{alemi2016_dvib} introduces a variational ansatz for the marginal $\hat{p}(h_t)$.
It follows from positivity of the Kullback-Leibler divergence $D_{\text{KL}}(p(h_t)\lVert \hat{p}(h_t))$ that
\begin{align}
    \int dh_t p(h_t)\log p(h_t) \geq \int dh_t p(h_t)\log \hat{p}(h_t),
\end{align}
and hence 
\begin{align}
    I(X_t,H_t) &\leq \int dx_t dh_t p(h_t|x_t)p(x_t) \log \frac{p(h_t|x_t)}{\hat{p}(h_t)}
    \\
    &= \mathbb{E}_{X_t} D_{\text{KL}}(p(h_t|x_t)\lVert \hat{p}(h_t)).
\end{align}
 If $p(h_t|x_t)$ and $\hat{p}(h_t)$ are chosen properly, the Kullback-Leibler can be expressed analytically which enables gradients to be effectively computed.
As in Ref.~\cite{alemi2016_dvib}, we take a Gaussian ansatz for $p(h_t|x_t)$ and let the marginal $\hat p (h_t)$ be a spherical unit-variance Gaussian.
%. as in Ref.~\cite{alemi2016_dvib}. 
More concretely, encoded variables $H_t$ are sampled from $p(h_t|x_t)$ by computing 
\begin{align}
    h_t = f_{W}(x_t) + \sigma_{W}(x_t)\eta,    
\end{align} 
where $f_{W}$ and $\sigma_{W}$ are deterministic functions modeled by neural networks with parameters (weights) $W$, and $\eta$ is a Gaussian random variable with unit variance.

\begin{figure*}[t]
    \centering
    \includegraphics[width=0.95\textwidth]{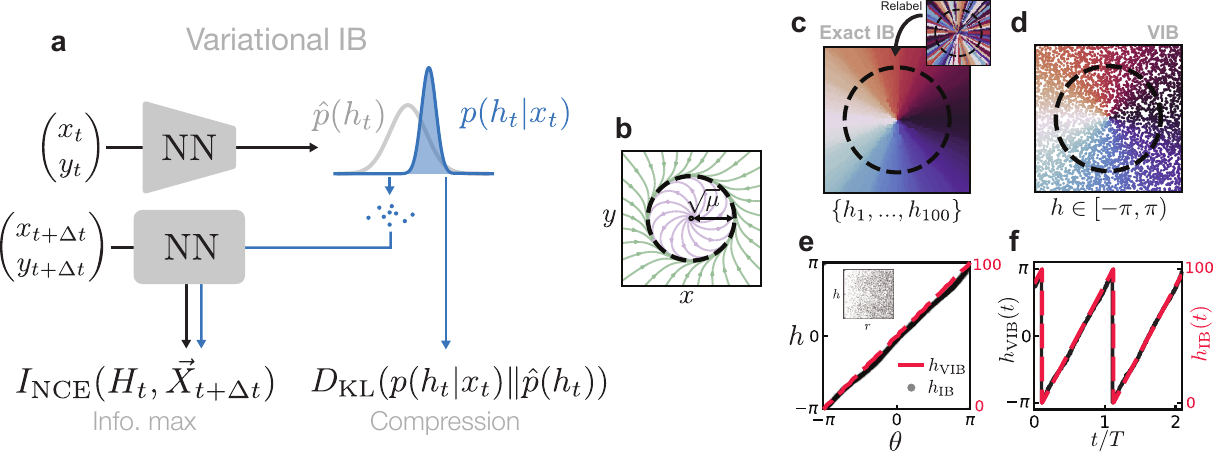}
    \caption{\textbf{Variational IB extends exact IB.}
    %(a) Exact IB requires access to the full conditional distribution $p(x_{t+\Delta t}|x_t)$ as well as $p(x_t)$.
    %The optimal stochastic encoder $p(h|x)$ is found via the Blahut-Arimoto algorithm, which iteratively computes $p(x_{t+\Delta t},h_t)$ and $p(h_t)$, and uses them to update the value of $p(h_t|x_t)$. 
    (a) Variational IB optimizes the variational objective directly from samples, in contrast to exact IB which requires an estimation of the full conditional distribution $p(x_{x+\Delta t}|x_t)$.
    Values of $h_t$ are drawn from a ``latent'' distribution $p(h|x)$, from which one can estimate the mutual information $I(X_{t+\Delta t},H_t)$. The compression term is approximated by the Kullback-Leibler divergence between the learned $p(h|x)$ and a variational ansatz for the marginal $\hat p(h)$.
    (b) Phase portrait for a dynamical system above a Hopf bifurcation.
    (c) Exact IB learns (up to permutations) an encoding $h$ which corresponds to the polar angle coordinate $\theta$.
    (d) Variational IB similarly learns the correct encoding, directly from samples.
    (e) Correspondence between IB encodings and the angle coordinate $\theta$. The encoding learned by VIB is independent of $r$ (inset). 
    (f) Time dependence of the encoded variable $h_t$.
    }
    \label{fig:Hopf}
\end{figure*}

To bound the entire loss from above, we must bound $I(X_{t+\Delta t},H_t)$ from below. We do this using the noise-contrastive estimate of the mutual information introduced in Ref.~\cite{oord2019infoNCE}.
This recasts the problem as one of distinguishing samples from the distributions $p(x_{t+\Delta t}|h_t)$ and $p(x_{t+\Delta t})$. 
Given a batch of $B$ pairs $(x_{t+\Delta t},h_t)$ and one particular sampled value $h_t^{(i)}$, one asks what the probability is that a  sampled value $x_{t+\Delta t}^{(j)}$ is from $p(x_{t+\Delta t}|h_t^{(i)})$ (is a \textit{positive} sample) and not $p(x_{t+\Delta t})$ (is a \textit{negative} sample). This probability is 
\begin{align}
    p(x_{t+\Delta t}^{(j)} =\text{pos}|h_t^{(i)})=\frac{\frac{p(x_{t+\Delta t}^{(j)}|h_t^{(i)})}{p(x_{t+\Delta t}^{(i)})}}{\sum_k^B \frac{p(x_{t+\Delta t}^{(k)}|h_t^{(i)})}{p(x_{t+\Delta t}^{(k)})}},
\end{align}
where $i$ is the index of the positive sample.
The log likelihood of these probabilities,
\begin{align}
    \mathbb{E}[-\log p(x_{t+\Delta t}^{(j)} =\text{pos}|h_t^{(i)})],
    \label{eq:true_loglik}
\end{align}
where the expectation is taken over indices $i$ and $j$, is closely related to the mutual information $I(X_{t+\Delta t},H_t)$:
in the limit of infinite samples $B\rightarrow \infty$, this quantity is given by
\begin{align}
    -\int dx_{t+\Delta t} dh_t \, p(h_t) p(x_{t+\Delta t}|h_t) \log p(x_{t+\Delta t} =\text{pos}|h_t) = \log B - I(X_{t+\Delta t},H_t).
\end{align}
If one had access to the probabilities $p(x_{t+\Delta t}^{(j)} =\text{pos}|h_t^{(i)})$ appearing in Eq.~\ref{eq:true_loglik}, the mutual information could thus be easily determined.
One attempts to estimate these probabilities by introducing a variational ansatz $f(x_{t+\Delta t},h_t)$ to approximate the density ratio $\frac{p(x_{t+\Delta t}|h_t)}{p(x_{t+\Delta t})}$.
Typically this $f$ is represented by a neural network. 
One can then obtain a bound on the mutual information by minimizing 
\begin{align}
    \mathcal{L}_B = \mathbb{E}\left[-\log \frac{f(x_{t+\Delta t}^{(i)}|h_t^{(i)})}{\sum_k^B f(x_{t+\Delta t}^{(k)}|h_t^{(i)})}\right],
\end{align}
from which the InfoNCE estimate of the mutual information can be calculated as
\begin{align}
    I_{\text{NCE}}(X_{t+\Delta t},H_t)=\log B - \mathcal{L}_B\leq I_{\text{true}}(X_{t+\Delta t},H_t).
\end{align}
The full objective to be minimized is given by 
\begin{align}
    \mathcal{L}_{\text{VIB}} = D_{KL}(p(H_t|X_t)\lVert \hat{p}(H_t)) - \beta I_{\text{NCE}}(X_{t+\Delta t}; H_t),
    \label{apxeq:vibloss}
\end{align}
an illustration can be seen in Fig.~\ref{fig:Hopf}.
This loss is evaluated on batches of sample pairs $\{(x^{(1)}_t, x^{(1)}_{t+\Delta t}), (x^{(2)}_t, x^{(2)}_{t+\Delta t}),...\}$. A minimum is found via stochastic gradient descent. Note that other variational loss functions inspired by the information bottleneck have been derived, for example \cite{abdelaleem2023deep}. 

The VIB loss Eq.~\ref{apxeq:vibloss} is very similar to the loss function for a time-lagged $\beta$-variational autoencoder ($\beta$-VAE) \cite{higgins2017betavae}, which have a form $\mathcal{L}=D_{KL}-\mathcal{L}_{\text{rec}}$, where the reconstruction term $\mathcal{L}_{\text{rec}}$ measures the deviation from the true future state $X_{t+\Delta t}$ and the reconstruction from the latent variable $H_t$.
In the VIB, this term is replaced by the mutual information $I_{\text{NCE}}$: rather than searching for a latent variable which can reconstruct the 
While a standard $\beta$-VAE with mean squared error (MSE) loss finds a latent variable that can reconstruct the full state at a time $\Delta t$ in the future, the VIB merely tries to reconstruct the statistics of $X_{t+\Delta t}$ conditioned on $X_t$.

Replacing reconstruction losses with information-theoretical loss functions makes sense in some scenarios, such as chaotic systems. Recent work in this direction has used the Kullback-Leibler divergence as a loss function in place of a $L_2$ loss which generated well-behaved long-term dynamics \cite{mikhaeil2022, koppe2019}. Other metrics which penalize the number of ``false neighbors'' in latent space have also shown to improve performance for chaotic systems \cite{gilpin2020}.

The extent to which our findings might apply to time-lagged VAEs with $L_2$ reconstruction losses is an interesting question for future work. In Fig.~\ref{apxfig:vae}, we observe that time-lagged VAEs learn a similar latent variable as VIB for the simulated fluids dataset.
Indeed, several works have noted the apparent similarity between modes learned by variational and regular autoencoders with linear methods such as principal component analysis (PCA) \cite{eivazi2022,brunton2020_review,Rolinek_2019_CVPR}.

\subsection*{VIB verification for noisy Hopf oscillator}
We now show that VIB learns an encoding consistent with exact IB, i.e. one which depends only on the dominant transfer operator eigenfunctions.
We consider the example of a Hopf oscillator given by the dynamical system
\begin{align}
    \dot{r} &= r(\mu - r^2), \nonumber
    \\
    \dot{\theta} &= \omega.
    \label{eq:hopf_eom}
\end{align}
Equation~\eqref{eq:hopf_eom} is the normal form (in polar coordinates) for dynamics near a Hopf bifurcation~\cite{Strogatz2018}.
For $\mu>0$ this system exhibits a circular limit cycle of radius $\sqrt{\mu}$ (Fig.~\ref{fig:Hopf}b).
In the following we show that this system has infinite purely imaginary eigenvalues $\lambda_n = in\omega$
and that as a result, encoders which exactly encode the angle coordinate, $p(h|r,\theta)\propto \delta (h - \theta)$,
are solutions of the IB optimization problem.

%First, we verify this by performing exact IB for simulated Hopf oscillator dynamics.
The exact IB calculation breaks down for perfectly deterministic dynamics, hence we add a small amount of white noise to the dynamics. While this slightly perturbs the spectral content of the transfer operator, we still find our results to be consistent with the those expected for the deterministic case (see the next section). 
As we increase the size of the encoding alphabet $N_H$, the encoder partitions space by finer and finer angular wedges.
The learned encoding is invariant under permutations of the encoded symbols. 
Upon reordering, we find that, for large alphabets $N_H\gg 1$, the encoder indeed approximates $p(h|r,\theta)\propto \delta(h - \theta)$ as expected (Fig.~\ref{fig:Hopf}c). 

Using VIB, we learn a continuous $h_t$ which can be computed directly from samples of the state variable. 
The encoding $h_t$ learned by VIB closely approximates the angle coordinate (Fig.~\ref{fig:Hopf}e-f) and is nearly uncorrelated with $r$ (inset).
This shows that our mathematical results illustrated for exact IB in Sec.~\ref{si_sec:derivations} hold also in the approximate framework of VIB.

\subsection*{Optimal encoding for Hopf normal form dynamics}
\label{si:hopf_opt_enc}

In Fig.~\ref{fig:Hopf} we show that IB learns the angle coordinate when applied to a dynamical system above a Hopf bifurcation. Here, we show that this is expected.
We begin by deriving the adjoint transfer operator in polar coordinates, and then continue to solve for its eigenfunctions.
In Cartesian coordinates, the equations of motion for the particle are
\begin{align}
    \dot x &= (\mu - x^2 - y^2)x - \omega y
    \\
    \dot y &= (\mu - x^2 - y^2)y - \omega x
\end{align}
Expressed in polar coordinates, we have
\begin{align}
    f_x \hat{e}_x &= f_x \cos\theta\hat{e}_r - f_x \sin\theta \hat{e}_\theta
    \\
    f_y \hat{e}_y &= f_y \sin\theta\hat{e}_r + f_y \cos\theta \hat{e}_\theta
\end{align}
From this we can compute $\mathcal{L}_{\mathcal{K}}\phi=f_i \partial_i\phi$,
\begin{align}
    f_i \partial_i\phi &= (\partial_r\phi)(\mu-r^2)r + \frac{1}{r}(\partial_\theta \phi)\omega r
    \\
    &= r(\mu-r^2)(\partial_r\phi) + \omega\partial_\theta \phi
\end{align}
Because this differential equation is separable, we can find eigenfunctions by looking for eigenfunctions that are a function of either $r$ or $\theta$. Recall that for deterministic dynamics, a product of adjoint transfer operator (or \textit{Koopman} operator) eigenfunctions is again an eigenfunction \cite{ModernKoopman}.
The eigenvalue equation for the radial coordinate is solved in Ref. \cite{gaspard1995}, and in Ref. \cite{page2019} it is shown that for this system there is no globally valid Koopman decomposition.
However we are only interested in a subset of eigenfunctions, in particular those with eigenvalue with real part $\text{Re}\infev_i\approx 0$, which may not suffice to approximate arbitrary functions of the state variable.

For the angle coordinate alone the situation is much simpler,
\begin{align*}
    \omega \partial_\theta \phi(\theta) = \lambda\phi(\theta)
\end{align*}
which is solved by functions $ \phi_n(\theta) = \exp\left(\frac{\infev_n}{\omega}\theta\right)$.
Periodicity requires $\frac{\lambda}{\omega}2\pi = i2\pi n$, which leads to $\lambda=i n \omega$. 
The eigenfunctions and eigenvalues are then 
\begin{align*}
    \phi_n(\theta)=e^{in\theta},\quad\quad \infev_n=in \omega.
    \label{eq:hopf_efcts}
\end{align*}
The corresponding eigenfunctions of the Perron-Frobenius operator are given by $\rho=\ee^{-in\theta}$.
We now consider an encoding $p(h|x)$.
\begin{align}
    p(h|r,\theta)&\propto \exp\left\{
    \sum_n^\infty \ee^{\infev_n\Delta t}
    \phi_n(r,\theta)
    \int r' dr'd\theta'
    \rho_n(r',\theta')\log p(r', \theta' | h)
    \right\}
    \\
    &= \exp\left\{
    \sum_n^\infty \ee^{in\omega\Delta t}
    \int r' dr'd\theta'
    \ee^{in(\theta - \theta')}\log p(r', \theta' | h)\right\}
     + \mathcal{O}(e^{\text{Re}\infev_k \Delta t})
\end{align}

where we retain only the eigenvalues with zero real part; the $\infev_k$ in the above refer to those eigenvalues with non-zero real part. After neglecting these terms, it can be seen directly that $p(h|r,\theta)=p(h|\theta)$. 
The coordinates $(r', \theta')$ are used to denote $(r_{t+\Delta t}, \theta_{t+\Delta t})$. The expression can be further simplified by replacing the sum over $n$ with a delta function, which leads to
\begin{align}
    p(h|r,\theta)&\propto \exp\left\{
    \int r' dr'
    \log p(r', (\theta+\omega\Delta t) | h)\right\}.
    \label{eq:hopf_enc_2}
\end{align}

\begin{figure}[tp]
    \centering
    \includegraphics[width=0.7\textwidth]{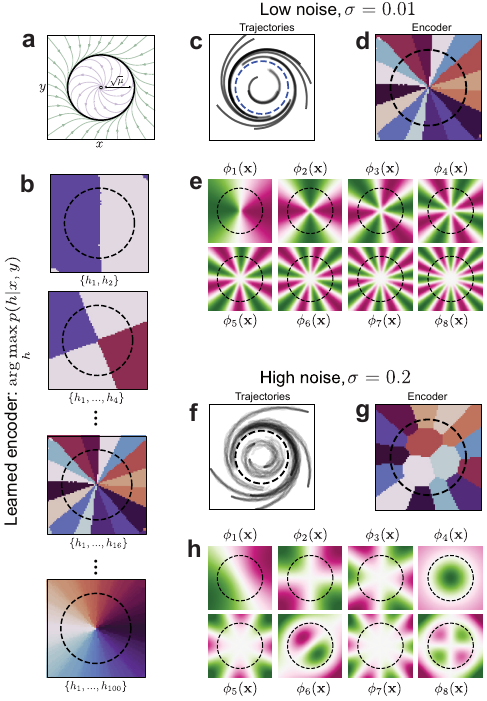}
    \caption{\textbf{Eigenfunctions and IB partitions for the Hopf Oscillator.}
    (a) Phase portrait for the deterministic Hopf oscillator.
    (b) Partitions found by IB for high beta but a restricted encoding alphabet $N_H$. 
    (c) Example trajectories of the nearly deterministic Hopf oscillator, with small noise amplitude $\sigma=0.01$. 
    (d) IB partition of the low noise dynamics for a $N_H=16$ encoding alphabet (same as appears in (b)).
    (e) For small noise, the first several eigenfunctions obtained numerically approximate the numerically expected ones of the form $\cos n\theta$.
    (f) Simulated trajectories of Hopf oscillator with higher noise amplitude $\sigma=0.2$.
    (g) IB partition of the low noise dynamics for a $N_H=16$ encoding alphabet; note the dependence on $r$.
    (h) The presence of noise changes the eigenfunctions, in particular they depend on $r$. 
    }
    \label{fig:si_hopf}
\end{figure}

We next ask whether an encoder of the form $p(h|\theta)\propto \delta(h - \theta)$ is a solution.
Recall that the probability distribution appearing in Eq.~\eqref{eq:hopf_enc_2} is a distribution over \emph{future} positions, $p(R_{t+\Delta t}=r',\Theta_{t+\Delta t}= \theta + \omega\Delta t | h)$.
This distribution can be calculated as
\begin{align*}
    p(r_{t+\Delta t},\theta_{t+\Delta t}| h) &= \int r dr d\theta
    p(r_{t+\Delta t},\theta_{t+\Delta t}| r_{t},\theta_{t})
    p(r,\theta_t | h) 
    \\
    &\propto \delta((\theta_{t+\Delta t} - \omega\Delta t) - h)
\end{align*}
where we used that the two terms in the integrand of the first equation are both delta functions.
Plugging this into Eq.~\eqref{eq:hopf_enc_2} shows that $p(h|\theta)\propto \delta(h - \theta)$ is consistent, and hence is an optimal encoding.
In the presence of noise, the eigenvectors are perturbed and may gain a dependence on $r$ (see Fig.~\ref{fig:si_hopf}).

\subsection*{VIB for deterministic dynamics}
 IB is an inherently probabilistic framework. 
To handle deterministic dynamics, we inject noise by introducing a stochastic sampling scheme. This step is necessary only for deterministically simulated dynamics; in our experience experimental data is sufficiently noisy on its own.
Concretely, rather than taking as our IB relevance variable $Y=X_{t+\Delta t}$, we take $Y=X_{t+\Delta t+\eta}$ where $\eta$ is a random uniformly-distributed time shift.
Despite using a different relevance variable, this yields essentially the same optimal encoding as Eq.~\eqref{si:encoder_mostinformativefeature}, where the eigenvalue $\ee^{\infev_n \Delta t}$ is replaced by 
$\int d\eta \,p(\eta)\exp(\infev_n(\Delta t + \eta))$.
Crucially, the encoder retains its dependence on the transfer operator eigenfunctions $\phi_n(x)$ as before. 
In general, this need not be the case: selecting a new relevance variable changes the IB objective and will generically lead to a different encoder.
Our choice of stochasticity which we introduce to the dynamics is chosen in such a way to preserve the form of the encoder.
%In this way, we can extend the IB framework to handle deterministic systems.

\begin{figure}[htp]
    \centering
    \includegraphics[width=0.8\linewidth]{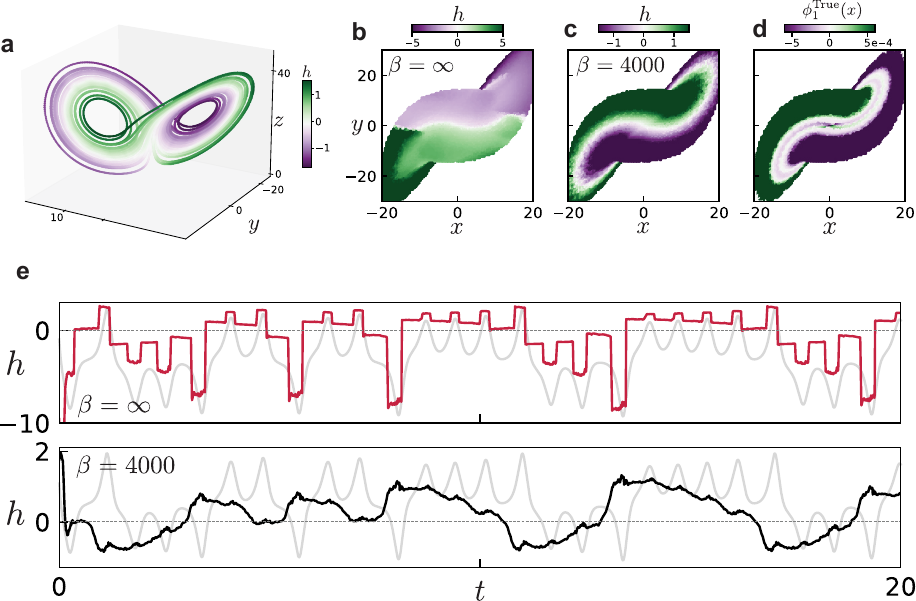}
    \caption{\response{\textbf{VIB applied to the Lorenz-63 system}
    (a) Illustration of the function found by VIB when allowed to learn a one-dimensional encoding. 
    (b-c) Learned functions shown projected onto $x-y$, left is in the limit of low compression, while right is for high compression. 
    (d) True eigenfunction, computed using the GAIO library \cite{gaio}.
    (e) Time series of $h$ in the high- and low-compression limits.
    }}
    \label{fig:lorenz}
\end{figure}

We illustrate this with a prototypical example of deterministic chaotic dynamics, the Lorenz system \cite{Lorenz1963}.
In the steady state, the state variable of the Lorenz system resides on a chaotic attractor consisting of two ``lobes'' which encircle unstable fixed points (Fig.~\ref{fig:lorenz}a).
Eigenvalues $\infev_n$ as well as right eigenvectors $\phi_n(x)$ of the Perron-Frobenius operator were computed numerically using the Ulam method. % \cite{the guy whose code I used}.
The correspondence between $\phi_1(x)$ and $h$ disappears as $\beta$ is increased (Fig.~\ref{fig:lorenz}b-d).
The dynamics of $h$ also become notably less ``slow'', and are instead nearly constant with occasional large jumps (Fig.~\ref{fig:lorenz}e). %This is reminiscent of the behavior of the encoding variable in exact IB
The encoding variable $h$ learned by VIB is the slowest varying function of the state $\vec{x}$, decaying as $\exp(\infev^{\text{Re}}_1 t)$ where $\infev^{\text{Re}}_1$ is the real part of the first subleading eigenvalue of the transfer operator (here Perron-Frobenius operator), which we show below in Fig.~\ref{fig:mixing}.
This shows that in order for $h$ to be a valid slow variable, the compression term in Eq.~\eqref{apxeq:vibloss} is crucial. 
These results are consistent with those obtained for exact IB, where we showed that the encoder incorporates the slow modes at low beta (high compression).

\section{VIB for fluid flow around a cylinder}
\label{si_sec:vib_fluid}

\subsection*{Form of the eigenfunctions}
\response{
What do the true eigenfunctions look like in this system?
Because it is well approximated by linear dynamics, eigenfunctions of the adjoint transfer operator are linear functions of the state variable, 
\begin{align}
    \phi_n[\vec{v}] = \langle \vec{v}(\vec{x}), \vec{m}^{(n)}(\vec{x})\rangle,
\end{align}
where $\vec{m}^{(n)}$ is the $n$-th mode (often referred to as a \textit{Koopman} mode \cite{ModernKoopman}) and angled brackets denote integration over space.
The true eigenfunction and corresponding modes can be computed via dynamic mode decomposition (DMD) \cite{rowley2009_dmd,schmid_2010_dmd}, as described in the SI.
The eigenfunctions for this system are in general complex, and come in conjugate pairs: $\phi_2(x) = \phi^{*}_1(x)$. In this situation any linear combination of $\phi_1$ and $\phi_2$ will decay at the same rate, and hence we expect to learn some arbitrary combination of the two dominant eigenfunctions, or equivalently a combination of the real and imaginary parts of $\phi_1$. 
We therefore take a two-dimensional encoding variable $[h_0, h_1]$, so that it can represent the full complex eigenfunction rather than only the real or imaginary part.}

\response{
Our learned latent variables are oscillatory with the correct frequency as shown in Fig. 5b-c of the main text.
%The fact that the latent variables oscillate as expected does not on its own imply that they are learning the correct eigenfunctions: there may be many such functions which exhibit the same oscillatory behavior, for example the flow velocity at one well-chosen pixel.
A more stringent test is whether we are also learning the correct mode $\vec{m}^{(1)}$. 
From the true eigenfunctions, the modes can be extracted by computing the gradient
\begin{align}
   \frac{\partial \phi_n}{\partial v_j} = m_j^{(n)}.
\end{align} 
As the learned function $h[\vec{v}]$ is a neural network, we can efficiently compute gradients of the network with respect to the input field
\begin{align}
   \frac{\partial h}{\partial v_j} = m^{(\text{IB})}_j + g_{\text{res},j}(\vec{v}(\vec{x})),
\end{align}
where we have separated the part of the gradient which is independent of $\vec{v}$ from a residual part which is dependent on $\vec{v}$.
If $h$ corresponds to the true eigenfunction, we expect that $\vec{m}^{(\text{IB})}$ is approximately equal to the Koopman mode  $\vec{m}^{(1)}$, and that $\vec{g}_{\text{res}}$ is small.
%The gradients of the VIB encoder indeed correspond closely to the subleading DMD mode $\vec{m}^{(1)}$, and the residual $\vec{g}_{\text{res}}$ is small relative to $\vec{m}^{(1)}$. 
We indeed find this to be the case; Fig.~5d of the main text shows these gradients averaged over several instantiations of the neural network, which corresponds strongly to the true mode. 
Details concerning both the averaging procedure and the residuals $\vec{g}_{\text{res}}$ can be found in the SI.
This shows that variational IB not only recovers the essential oscillatory nature of the dynamics, but does so by learning the correct slowly varying functions of the state variable given by the adjoint transfer operator eigenfunctions.}

\subsection*{Eigenfunction computation via dynamic mode decomposition}
Here we provide some additional details on the computation used to generate components of Fig.~5 of the main text.
The true Koopman modes were computed using dynamical mode decomposition (DMD) \cite{rowley2009_dmd,schmid_2010_dmd,tu2014}, also described below in Section~\ref{sct:compare:dmd}.
DMD attempts to find a finite dimensional approximation of the Koopman operator using $n$ snapshots of the system's state $x\in \mathbb{R}^d$ which are assembled into a data matrix $\mathbf{X}\in\mathbb{R}^{n\times d}$.
One then attempts to find a linear evolution operator $K$ which propagates the state forward in time
\begin{align*}
    \mathbf{X}_{t+\Delta t} = K \mathbf{X}_{t}.
\end{align*}
The approximate Koopman operator is given by the least squares solution $K=\mathbf{X}_{t+\Delta t}\mathbf{X}_{t}^{+}$ where $\mathbf{X}_{t}^{+}$ denotes the pseudo-inverse of $\mathbf{X}_{t}$. Approximate Koopman eigenfunctions are given by $\phi_n(x)= x\cdot w_n $ where $w_n$ denotes the $n$th eigenvector of the matrix $K$ and is known as the $n$th DMD mode, denoted by $m^{(n)}$ in the main text.

\subsection*{Details on gradient analysis}
The fluid flow of a von Karmen vortex street is well approximated by linear dynamics, which can be understood by recognizing that the system is poised just after a Hopf bifurcation so that there is an angle coordinate which rotates with constant angular velocity.
This constant rotation can be described by a linear dynamical system.
It follows from linearity that eigenfunctions of the adjoint transfer operator are given by linear functions of the state variable, 
\begin{align}
    \phi_n[\vec{v}] = \langle \vec{v}(\vec{x}), \vec{m}^{(n)}(\vec{x})\rangle,
    \label{fluid_phi}
\end{align}
where $\vec{m}^{(n)}$ is the $n$-th ``Koopman mode'' and angled brackets denote integration over space.
To compute these modes and the corresponding transfer operator spectrum, we use dynamic mode decomposition (DMD; see SI Section~\ref{si_sec:vib_fluid}) \cite{rowley2009_dmd,schmid_2010_dmd}.

We allow VIB to learn a two-dimensional encoding variable $[h_0, h_1]$, so that it can learn the complete first eigenfunction rather than only the real or imaginary part.
For the purposes of comparing our learned variable with $\phi_1$, we construct a complex $h=h_0 + i h_1$ out of the two learned components. 

\begin{figure}[tp]
    \centering
    \includegraphics[width=0.5\textwidth]{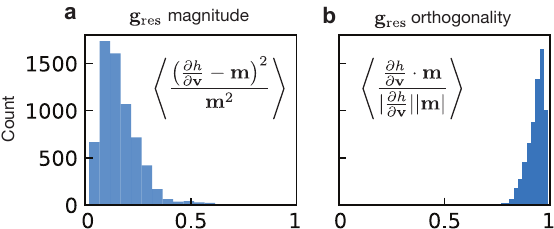}
    \caption{\textbf{VIB gradients for fluid flow are nearly linear.}
    We compare the linear part of the gradients of the VIB network to the residual which depends on the input field $\vec{v}(\vec{x})$.
    (a) The residual parts are smaller in magnitude than the linear part $\vec{m}$. Angle brackets denote average over space. 
    (b) The residual parts are nearly orthogonal to the mean $\vec{m}$, as we see that the projection of the full gradient onto the mean is nearly 1. 
    }
    \label{apxfig:gradsarelinear}
\end{figure}

To understand which function has been learned by the neural network, we check whether the learned functions $h_i[\vec{v}]$ are of the form Eq.~\eqref{fluid_phi}.
This can be done by examining gradients of the network with respect to the input field 
\begin{align}
   \frac{\partial h}{\partial v_j} = m^{(\text{IB})}_j + g_{\text{res},j}(\vec{v}(\vec{x}))
\end{align}
where we have separated the part of the gradient which is independent of $\vec{v}$ from a residual part which is dependent on $\vec{v}$.
Gradients of the true eigenfunctions are given simply by
\begin{align}
   \frac{\partial \phi_n}{\partial v_j} = m_j^{(n)}.
\end{align} 
We can then directly compare our latent variables with the true eigenfunctions by comparing their derivatives. 
If $h$ corresponds to the true eigenfunction, we expect that $\vec{m}^{(\text{IB})}$ is approximately equal to the Koopman mode  $\vec{m}^{(1)}$, and that $\vec{g}_{\text{res}}$ is small.
While it is unclear how to perform this decomposition in a general setting, we assume that the residual component $g_{\text{res},j}$ averages to zero over an oscillation period of the flow field $\vec{v}$. 
Then, $m^{(\text{IB})}_j \approx \langle\frac{\partial h}{\partial v_j} \rangle_t$ and $g_{\text{res},j}$ is given by variations about the mean. We see in Fig.~\ref{apxfig:gradsarelinear} that these variations are much smaller than the mean in magnitude, and that they are essentially orthogonal to the mean vector. From this, we conclude that the gradients are given primarily by the constant part $\vec{m}$.

The learned latent functions vary for different training instances of the neural network. 
To extract the \textit{average} gradient, we use PCA in an approach similar to that in Ref.~\cite{Goekmen2021}.
The learned functions can have arbitrary sign structure; $h=h_0 + ih_1$ is just as likely to be learned as $h=-h_0 + ih_1$, for example. While in principle the network could learn arbitrary rotations, rather than simply changes in sign, we observe this is not the case.
The distribution of gradients forms clusters in the high-dimensional gradient space corresponding to the four possible permutations of sign. PCA picks out the directions separating these clusters, as these are precisely the directions along which the data varies the most. This procedure gives an average gradient, while taking the varying sign structure into account. For reference, the gradient of a single instantiation of the VIB network can be seen in Fig.~\ref{apxfig:vae}b.

\section{Chaotic systems}
\label{si_sec:chaotic_systems}

Chaotic systems exhibit many features which make them ``difficult'' to forecast. For example, they reside on strange attractors with fractal structure; even infinitesimal shifts in initial conditions lead to large deviations in a short time; and they are ``mixing'', meaning that correlations decay to zero in time which is typically characteristic of stochastic systems.
Although the goal of IB is compression (and not forecasting) we ask how the IB-optimal encoding reflects these properties. We do this in the contexts of the Lorenz-63 system \cite{Lorenz1963}, a hyperchaotic system \cite{azar2015hyperchaotic}, and the Kuramoto-Sivashinsky system \cite{kuramoto1978KS,sivashinsky1977ks}.

\subsubsection*{Fractal attractors} 
\response{
A unique feature of strange attractors is their fractal structure. While this is an interesting theoretical property, it is difficult to estimate from data in practice. The dominant methods for estimating fractal dimension from a time-series are little changed from those introduced nearly 50 years ago \cite{theiler1990estimating}. It is not known how to estimate the Hausdorff dimension of an attractor \cite{datseris2023estimating}; one typically relies on an estimate of the correlation dimension instead, which in many cases is close to the Hausdorff dimension \cite{grassberger1981hausdorff}.}

\response{
One difficulty in the estimation of correlation dimension is that it requires the computation of distances between states of the system. In high-dimensional systems the effect of noise on measures of distance can become large: the ratio of distances between a point to its nearest and farthest neighbors approaches 1 \cite{aggarwal_highdim,beyer1999nearest}, which can lead to large corrections to the correlation dimension
\cite{nolte2001}.
We will show that the encoding learned by VIB preserves the fractal dimension of the full system. This opens the door to using it in conjunction with denoising strategies to reliably determine the fractal dimension of very high-dimensional systems.}

\response{
We consider two systems, a driven one-dimensional sine-Gordon (sG) equation and a one-dimensional Kuramoto-Sivashinsky (KS) equation. 
Both systems are described by time-evolving partial differential equations, and are hence high-dimensional.
However, they exhibit low-dimensional dynamics. 
The sine-Gordon system is quasiperiodic and can thus be described by dynamics on a torus, while the KS system is chaotic and evolves on a low-dimensional strange attractor.
We show in Fig.~\ref{fig:KS} and Fig. 6 of the main text that the fractal dimension of the respective \textit{embedded} attractors is consistent with estimates of the full attractors.
We find that the fractal dimension of attractor embedded in latent space coincides with that of the full system, which is much more computationally-expensive to estimate.
This suggests that in the case where the true fractal dimension is unknown, it can be recovered by looking only at the dynamics in the encoding space.}

\subsubsection*{Mixing dynamics} 
\response{
In terms of the transfer operator spectrum, a system is mixing if $P$ has only $\lambda=1$ in its point spectrum \cite{lasota2013chaos, colbrook2024rigorous}, and otherwise consists only of a continuous spectrum (please see our comment below on ``continuous spectra'').
A more experimentally-relevant property of mixing systems is that correlations decay to zero $\langle A(t)B(0)\rangle - \langle A\rangle\langle B\rangle \overset{t\rightarrow\infty}{\rightarrow} 0$. 
In this context, the primary quantity of interest is the subleading eigenvalue of the transfer operator spectrum, as this bounds the rate of decay \cite{froyland1997}. 
Here we ask whether the latent variables learned by VIB, which approximate the subleading eigenfunction of the transfer operator, enable us to estimate mixing rates. As a demonstration, we consider the chaotic Lorenz-63 system as well as a hyperchaotic system governed by the equations of motion
\cite{azar2015hyperchaotic}:
\begin{align*}
    \dot{x} &= \alpha(y-x)+w
    \\
    \dot{y} &= x(\beta-z)-y
    \\
    \dot{z} &= xy - \gamma z
    \\
    \dot{w} &= \delta w + \frac{1}{2}y^2-xz
\end{align*}
where $(\alpha,\beta,\gamma,\delta)=(10,28,2.7,2.2)$.}

\begin{figure}[tp]
    \centering
    \includegraphics[width=0.65\linewidth]{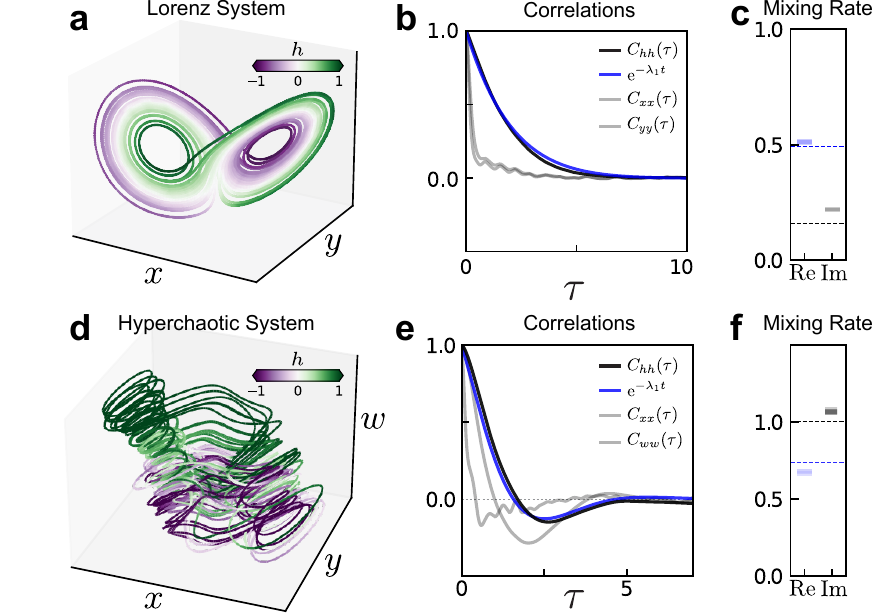}
    \caption{
    \response{
    \textbf{Mixing in chaotic systems}
    \\
    By finding the dominant eigenfunction of the transfer function, one can estimate the mixing rate of dynamical systems which we demonstrate on (a-c) the Lorenz system and (d-f) a 4-dimensional hyperchaotic system .
    (a) Attractor of the Lorenz system colored by the learned encoding value.
    (b) Time autocorrelations for the encoding $C_{hh}(\tau)=\langle h(t)h(t+\tau)\rangle_t - \langle h(t)\rangle_t^2$ (black) compared to the mixing rate computed by estimating the full transfer operator (blue; see text). The encoding variable decays over a much slower time scale than the state variable $x$ (gray).
    (c) Mixing rate estimated by fitting the blue curve in (b) to an exponential. We show the fitted real and imaginary parts of the decay rate separately. The true values are given by dashed lines of the respective color. 
    (d-f) Show the analogous results for the hyperchaotic system mentioned in the text. Note that the full state is four-dimensional, so that in (d) we are only showing a projection onto three selected components.
    }
    }
    \label{fig:mixing}
\end{figure}

\response{
To ensure that our latent variable isolates the first eigenfunction and does not include further subleading eigenfunctions, we restrict $h$ to be one-dimensional and take a small compression parameter $\beta$. The results of the learned encoding are shown in Fig.~\ref{fig:mixing}a,d.
The correlations of the latent variable decay on a slower time-scale than single components of the system Fig.~\ref{fig:mixing}b,e.
For these relatively low-dimensional systems, we can independently approximate the dominant eigenvalue $\lambda_1$ using an Ulam approach, which we do using the software package GAIO \cite{gaio}.
We find good agreement between these eigenvalues and the decay rate estimated from a fit to the correlation function $C(\tau)=\langle h(t+\tau)h(t)\rangle-\langle h(t)\rangle^2$ where averages are taken over time Fig.~\ref{fig:mixing}c,f. 
}
    
\subsubsection*{Exponential divergence of nearby trajectories} 
\response{
Exponential divergence of nearby trajectories is a fundamental issue when attempting to forecast chaotic systems \cite{mikhaeil2022difficulty}. For that reason, modern methods typically avoid training neural networks to forecast dynamics using mean-square error alone, but rather include higher-order information about the dynamics, such as Jacobians \cite{nisha_neurips2024} or place emphasis instead on preserving long-term statistics of the system \cite{schiff2024dyslim,jiang2024training}. However, the goal of our work is not to predict the full dynamics: neither at the level of the trajectories, nor at the statistical level. Instead our goal is to find an appropriate dimensionally-reduced representation of the state of the system.}

\begin{figure}[tp]
    \centering
    \includegraphics[width=0.95\linewidth]{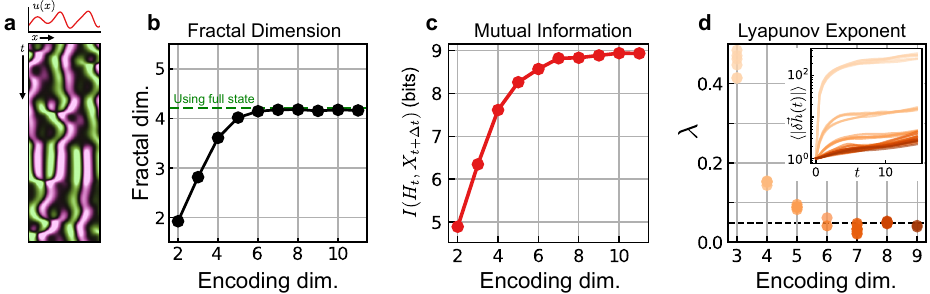}
    \caption{
    \response{
    \textbf{Fractal dimension of VIB encodings}
    \\
     (a) Fractal dimension of the embedded Kuramoto-Sivashinsky attractor with varying encoding dimension. We estimate fractal dimension here by computing the correlation dimension \cite{datseris2023estimating,Procaccia_1985}.
    %The gray dashed line shows the upper bound on the fractal dimension given by the Kaplan-Yorke conjecture \cite{kaplanyorke,yangradons2012}.
    Green dashed line denotes the dimension of the attractor approximated for the full system, which lives in an $\mathbb{R}^N$, $N=128$ discretization of the system's domain. 
    \textbf{b}: Fractal dimension of the embedded system as a function of encoding dimension. Dashed line shows the fractal dimension of the system measured using the full state. 
    \textbf{c}: Mutual information between the embedding and the future state, as a function of encoding dimension.
    \textbf{d}: At sufficiently high embedding dimension, encodings preserve the system's Lyapunov exponent.
    Dashed line shows the true exponent of $\lambda\approx 0.048$.
    Inset shows the distance $\lvert\delta h\rvert $ as a function of time for points which start close together (initial distance is normalized to one for comparison).
    }
    }
    \label{fig:KS}
\end{figure}

\response{
In Fig.~\ref{fig:KS}a we consider the Kuramoto-Sivashinsky model, which governs the chaotic evolution of a one-dimensional continuous field \cite{kuramoto1978KS,sivashinsky1977ks}. Though the system itself is high-dimensional (depending on the coarseness of discretization of the continuous field) it lives on a low-dimensional attractor (Fig.~\ref{fig:KS}b) \cite{lan2008unstable,cvitanovic1988invariant}.
We see that the top Lyapunov exponent, which characterizes the exponential divergence of nearby trajectories, is preserved by VIB when the dimensionality of the latent space is sufficient to fully encode the system's attractor Fig.~\ref{fig:KS}c. 
This fact is not necessarily surprising; as long as the encoding is a smooth transformation of points on the system's attractor, this will generally be true.
To see this, recall the definition of the top Lyapunov exponent \cite{eckmann1985ergodic}
\begin{align*}
    \lambda_x = \lim_{t\rightarrow\infty}\frac{1}{t}\log \lVert \delta x(t) \lVert
\end{align*}
where the small perturbation evolves according to $\dot{\delta x}_i=\frac{\partial f_i}{\partial x_j} \delta x_j$, with $f$ governing the system's evolution $\dot x = f(x)$.
Under a transformation $y=g(x)$, a small perturbation $\delta x$ yields the perturbation $\delta y_i = \frac{\partial g_i}{\partial x_j}\delta x_j$, or more concisely, $\delta y=D_xg \cdot\delta x$
Hence, the Lyapunov exponent estimated using the $y$ variables is given by
\begin{align*}
    \lambda_y &= \lim_{t\rightarrow\infty}\frac{1}{t}\log \lVert \delta y(t) \lVert= \lim_{t\rightarrow\infty}\frac{1}{t}\log \lVert (D_x g)\delta x(t) \lVert
\end{align*}
We assume that the attractor in $x$ is bounded, and that the transformation $g$ is sufficiently smooth and invertible so that the attractor in $y$ is also bounded and $D_x g\leq C$ for some constant $C$.
In addition, we will use the fact that $\lVert Ax\lVert \geq \sigma_{\text{min}}\lVert x\lVert$ where $\sigma_{\text{min}}$ is the minimal singular value of $A$.
The Lyapunov exponent can consequently be bounded
\begin{align*}\lim_{t\rightarrow\infty}\frac{1}{t}\log \sigma_{\text{min}} 
    \leq \lambda_y -\lambda_x\leq \lim_{t\rightarrow\infty}\frac{1}{t}\log C
\end{align*}
Assuming $\sigma_{\text{min}}$ and $C$ are both finite, in the limit this implies $\lambda_y=\lambda_x$.}

\response{
This result breaks down when the transformation is non-invertible, in which case we have $\sigma_{\text{min}}\rightarrow 0$. 
So, while we expect the Lyapunov exponents to be captured as long as the encoding recovers the full attractor, for smaller dimensions the exponents may differ. This effect is demonstrated in Fig.~\ref{fig:KS}d.
The inset shows the evolution of a small perturbation $\langle \lvert \vec{\delta h}\rvert(t)\rangle$ over time, where the average is taken over $\mathcal{O}(10^4)$ initial conditions. Different lines of the same color denote different instantiations of the same neural network (cross validation).
For small dimensions, points which are distinct in the full system get mapped close together in the embedding space. In the case that they come from different trajectories, they may move apart very quickly (and could in principle even move in the opposite direction). This explains the rapid growth observed for small dimensions.
For sufficiently high dimensional embeddings the exponent approaches the true value $\lambda\approx 0.048$, suggesting that the network has learned the full attractor.
This demonstrates that the latent representation found by VIB retains the essential dynamical features of the full system.}

\subsubsection*{Continuous spectra} 
    \response{
    In our work we take the perspective that noise is an unavoidable and essential feature of the system we aim to describe, which is reflected in our choice to invoke information theory and our focus on stochastic dynamical systems. 
    While continuous spectra are relevant for the study of \textit{deterministic} chaotic systems \cite{ModernKoopman}, in practice a system's essential spectrum will shrink due to the presence of noise \cite{froyland2013, Cvitanovic2012}.
    This does not mean that continuous spectra are meaningless, just that the accompanying singularities are ``smoothed.'' Consider for example a Brownian particle in a double-well potential, as shown in the main text. 
    In the deterministic limit, the system exhibits a continuous spectrum at the bifurcation $\mu=0$ which is associated with critical slowing down and the algebraic decay of correlations \cite{gaspard1995}. 
    However, the presence of noise renders the spectrum discrete for all values of $\mu$. While the system still ``slows down'' near the deterministic bifurcation point, the massive qualitative change in the system's dynamics is smoothed over \cite{gaspard1995}.
    }
    
\section{Cyanobacteria experiments}
\label{si_sec:bacteria}

\begin{figure}
    \centering
    \includegraphics[width=\textwidth]{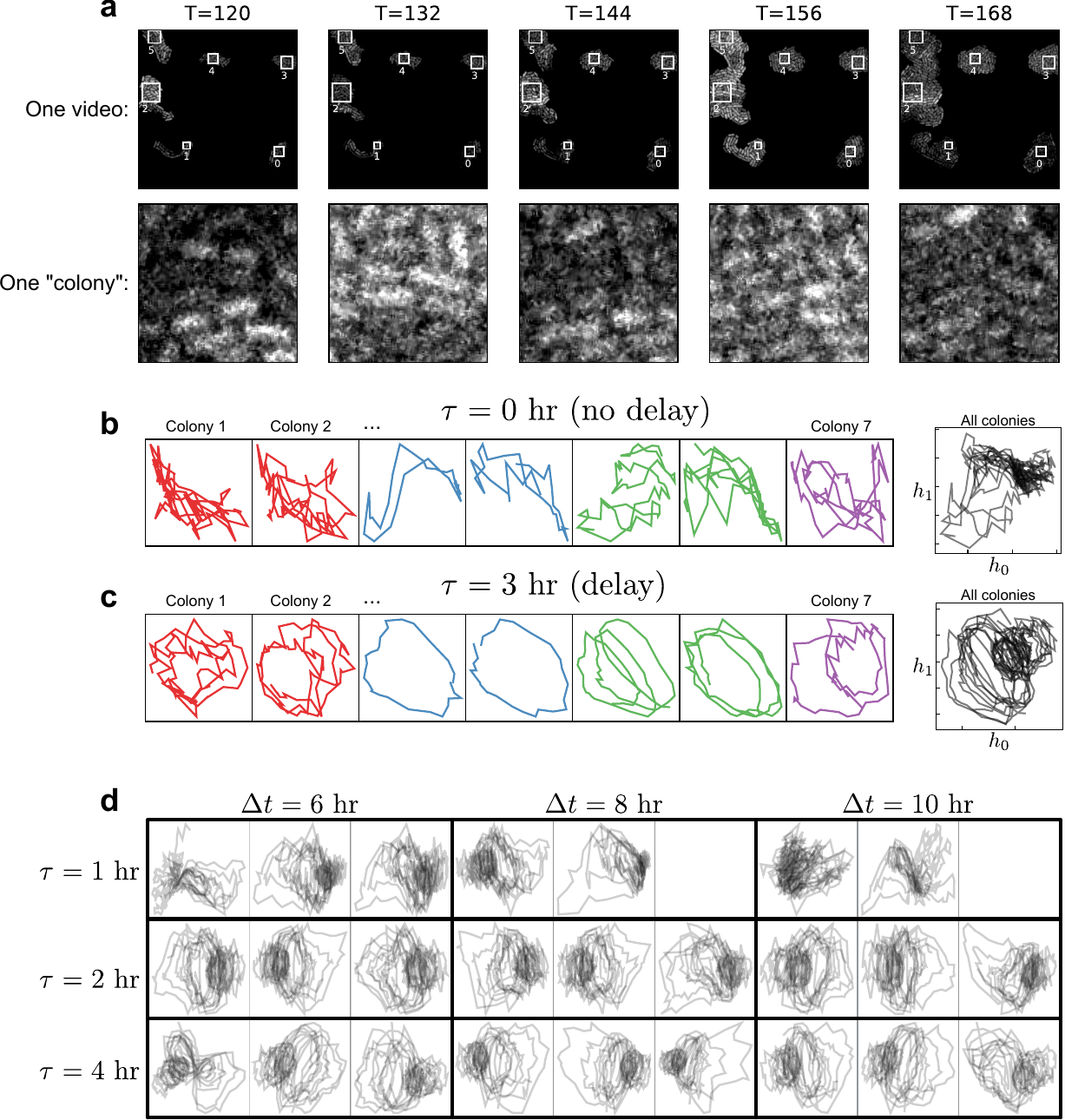}
    \caption{\textbf{Cyanobacteria data and choice of time delay}
    (a) The cyanobacteria dataset consists of 5 videos, where each has multiple (up to 3) colonies. (Top) time evolution of one video, where the interiors of the colonies are outlined with white boxes. The dataset is composed of all colonies which are greater than 64px in height and width. 
    (Bottom) evolution in time of one selected colony.
    (b) Learned VIB encodings when no time delay is used, i.e. $\tau=0$. Left shows the evolution of (a subset of) individual colonies, colored by the original experimental video which they belong to. Note that the axes scale is adjusted for each trajectory so that it fills the plot.
    Right panel shows all colonies in latent space.
    (c) Same as above, but with the 3 hr time delay used in the main text.
    (d) Hyperparameter sweep over different choices of time delay $\tau$ and prediction horizon $\Delta t$. Here each set of three plots was trained with the same parameters but different neural network instantiations to understand how robust these latent spaces are.
    }
    \label{sifig:cyanobac}
\end{figure}

The experimental parameters used in the cyanobacteria experiments are described in detail in \cite{ChewRust2018}. 
In brief, the authors in \cite{ChewRust2018} control the translation of the KaiA protein with a theophylline riboswitch, allowing them to tune the copy number of KaiA proteins in the bacteria by modulating the concentration of theophylline.
The clock state of each individual bacteria is visualized with a fluorescent marker EYFP driven by the \textit{kaiBC} promoter. Colonies are imaged once per hour.
The full dataset consists of 5 videos, such as the one shown in Fig.~\ref{sifig:cyanobac}, which each contains several colonies. For each video we isolate regions which are filled by bacteria at all times to eliminate the effect of exponential colony growth, which otherwise dominates the VIB results. 
Each latent trajectory shown in Fig.~\ref{sifig:cyanobac} as well as Fig.~6 in the main text corresponds to the trajectory of one single colony evolving under one of four theophylline conditions. There are 10 trajectories (colonies) in total, coming from 4 experimental conditions (colonies with radius smaller than 64 pixels are not considered).
Figure~\ref{sifig:cyanobac} also shows the effect of the choice of time delay, which in the main text we take as $\tau=3$ hr.

\textit{\textbf{Synchronization measurements}}
We measure the synchronization of the cyanobacterial oscillations using a metric inspired by a locally-coupled Kuramoto model which considers a spatially-varying phase field $\theta(\mathbf{x}, t)$.
In terms of this field, an order parameter can be computed as 
\begin{align}
    r(t)\ee^{i\psi(t)} = \frac{1}{V}\left|\int \dd\mathbf{x} \ee^{i\theta(\mathbf{x}, t)} \right|
\end{align}
where $V$ refers to the volume being integrated over.
The value of $r(t)$ is referred to as the synchronization order parameter, while $\psi(t)$ is the average phase.

\begin{figure}[htp]
    \centering
    \includegraphics[width=0.5\textwidth]{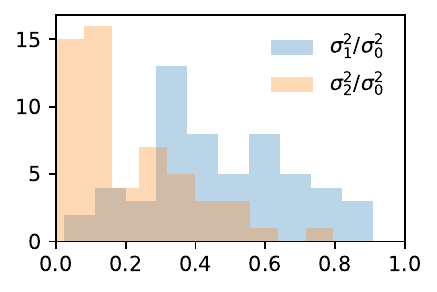}
    \caption{\textbf{Variance along principal component directions}
    For each DVIB model trained on the cyanobacteria dataset, we compute the principal component decomposition of the resulting point cloud in latent space. 
    Here we show the variance along each principal component for 60 instantiations of the DVIB model. Variances are normalized by the variance in the first principal component direction. 
    }
    \label{sifig:cyanobac_pca}
\end{figure}

To calculate the field $\theta(\mathbf{x}, t)$ from an intensity field $I(\mathbf{x}, t)$ we imagine that the intensity field represents one component of the complex phase, for example $I(\mathbf{x}, t)=\sin(\theta(\mathbf{x}, t))$.
The other component can be accessed using a time-delay, 
$\cos(\theta(\mathbf{x}, t)) = \cos(\theta(\mathbf{x}, t+\tau)) = I(\mathbf{x}, t+\tau)=$, where $\tau$ here should be chosen so that the intensity field undergoes one quarter of a full oscillation. As we know the true period of the circadian cycle is $24$ hours/frames, we choose a delay of $\tau=8$ frames.
Then, the phase can be computed as
\begin{align*}
    \theta(\mathbf{x}, t+\tau)\approx \arctan \frac{I(\mathbf{x}, t)}{I(\mathbf{x}, t+\tau)}.
\end{align*}
In Fig.~\ref{sifig:kuramoto} we show the results of VIB when applied to a locally-coupled Kuramoto model. Here we learn the same latent features which undergo oscillatory dynamics of varying radius, where the radius corresponds to the synchronization order parameter.

\section{Simulation Parameters}
\label{si_sec:simulation_parameters}

\subsubsection*{Triple well simulations}
For the dynamics of the triple well we work directly with the force $F=-\partial_x U$,
\begin{align*}
    F(x) &= \frac{-1}{200}\left( 9375 x^5 - 7500 x^3 + 1100 x - 20 \right).
    %F(x) &= \frac{-1}{200}\left( 3 (5x)^5 - 60 (5x)^3 + 220 (5x) \right) + 0.1.
\end{align*}
The evolution of the Brownian particle's position $x_t$ is given by
\begin{align}
    \dd x_t = \frac{F(x_t)}{\gamma}\dd t + \sigma\sqrt{\dd t} \eta_t
    \label{eq:sde}
\end{align}
where $\eta_t$ is white noise with unit variance. The noise magnitude $\sigma$ is related to the diffusion constant in Eq. \ref{apxeq:langevin} by $D=\frac{\sigma^2}{2}$. We use $\sigma=1.0$ and $\gamma=0.2$

Simulations were performed with $N_{\text{init}}=10^5$ initial conditions, with 300 trajectories generated from each initial condition. The state is evolved for $100$ steps at $\dd t=2\cdot 10^{-3}$.
The transfer matrix is approximated by binning the space $x\in[-1,1]$ with $N_{\text{bins}}=100$ bins.
IB is performed using a time delay $\Delta t=64$ steps.

\subsubsection*{Pitchfork bifurcation simulations}
The system was evolved according to the stochastic differential equation Eq.~\ref{eq:sde} with $F(x)=-\mu x - x^3$, $\gamma=1$ and $\sigma=0.1$.
For each value of $\mu$, $10^5$ initial conditions were simulated with 2000 trajectories starting at each initial condition. These were evolved for 100 time steps of $\dd t=2\cdot 10^{-3}$.

\subsubsection*{Hopf oscillator simulations}
The system was evolved according to Eq.~\ref{eq:sde} with the force given by Eq.\ref{eq:hopf_eom} above (transformed into cartesian coordinates), with $\gamma=1$ and $\sigma=10^{-2}$, $\omega=4.0$ and $\mu=0.25$.
For each value of $\mu$, $10^6$ initial conditions were simulated with 1000 trajectories starting at each initial condition. These were evolved for 50 time steps of $\dd t=2\cdot 10^{-3}$.

\subsubsection*{Lorenz system simulations}
The system was evolved according to 
\begin{align}
    \dot x &= \sigma (y-x)
    \\
    \dot y &= x(\rho - z) - y
    \\
    \dot z &= xy - \beta z
\end{align}
with $\rho=28$, $\beta=8/3$ and $\sigma=10$.
We took $10^3$ initial conditions from which trajectories were simulated for $10^5$ steps with $\dd t=2\cdot 10^{-3}$.
We compute the true eigenfunctions using the GAIO library \cite{gaio}.

\subsubsection*{Driven sine-Gordon simulations}
The equation of motion of the continuous field $\phi(x)$ is \cite{bishop_sinegordon}
\begin{align}
    \partial_t^2\phi - \partial_x^2\phi +\sin\phi 
    &=\varepsilon[-\alpha \partial_t\phi + \Gamma\cos(\omega t)].
\end{align}
The system is periodic in space, and was simulated with initial conditions
\begin{align}
    \phi(x,t)\big|_{t=0}&=4\arctan\left[\frac{\sqrt{1-\omega_{\text{br}}^2}}{\omega_{\text{br}}\cosh\left(\sqrt{1-\omega_{\text{br}}^2}(x-L/2)\right)}\right]
    \\
    \partial_t\phi(x,t)\big|_{t=0}&=0
\end{align}
where $\omega_{\text{br}}=0.77$ is the frequency of the breather initial condition and $L$ is the system size.
Key simulation parameters are $\text{d}t=10^{-2}$, $L=24$ (discretized onto $N_{\text{bins}}=256$ bins), $\varepsilon\alpha=0.04$, $\varepsilon\Gamma=0.105$, and $\omega=0.87$ (see Ref.~\cite{bishop_sinegordon}).

\subsubsection*{Kuramoto-Sivashinsky simulations}
The Kuramoto Sivashinsky model describes the evolution of a one-dimensional field $u(x)$ according to 
\begin{align}
    \partial_t u = -\partial_x^4u-\partial_x^2u-u\partial_xu 
\end{align}
Key simulation parameters are the system size $L=22$, time resolution $\text{d} t=2\cdot10^{-3}$, total simulation time $T=300$, and spatial resolution $N_{\text{bins}}=128$.

\subsubsection*{Hyperchaotic system simulations}
As described above, the system was evolved according to \cite{azar2015hyperchaotic}
\begin{align*}
    \dot{x} &= \alpha(y-x)+w
    \\
    \dot{y} &= x(\beta-z)-y
    \\
    \dot{z} &= xy - \gamma z
    \\
    \dot{w} &= \delta w + \frac{1}{2}y^2-xz
\end{align*}
where $(\alpha,\beta,\gamma,\delta)=(10,28,2.7,2.2)$.
We took $10^2$ initial conditions, which were allowed a burn-in time of $t_{\text{burn-in}}=1000$ steps to reach the attractor, after which they were simulated for $10^5$ steps with $\dd t=10^{-2}$.

\subsubsection*{Fluid flow simulations}
The fluid flow simulations simulations are contained in the ``Cylinder in Crossflow'' dataset, downloaded from Ref.~\cite{flowdata}. We take the dataset at Reynolds number 150, and interpolate the velocity field from the unstructured 6569-node mesh to a regular grid of $300\times150$ pixels.
The VIB networks are trained with $\beta=10^{7}$ and a time buffer $\Delta t$ chosen randomly between $[4,24]$ (see our comment on time randomization in ``VIB for deterministic dynamics'' in Section~\ref{si_sec:dvib}).

\subsubsection*{Locally-coupled Kuramoto model}
\begin{figure}[tp]
    \centering
    \includegraphics[width=\textwidth]{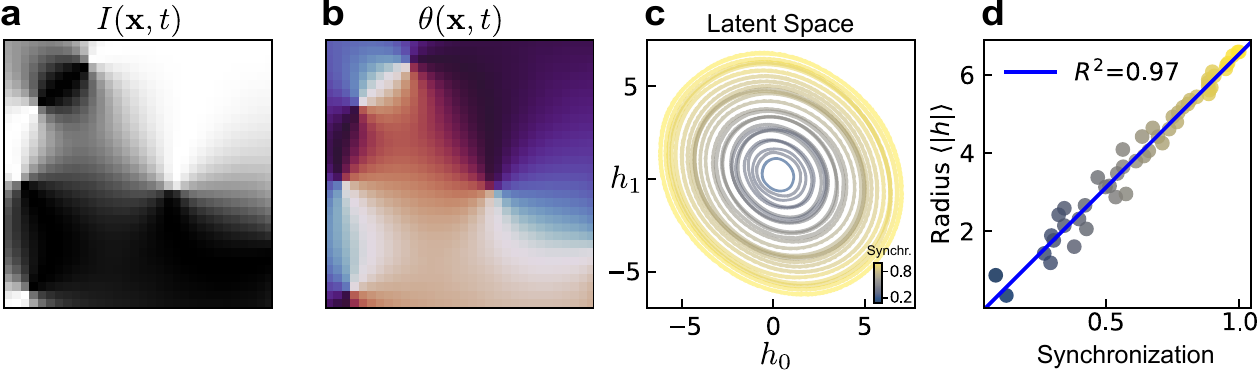}
    \caption{\textbf{VIB applied to simulated Kuramoto model} 
    Inputs are made to mimic the intensity field of the cyanobacteria data. The intensity field (a) is derived from the phase field (b) according to $I(\mathbf{x},t) = \frac{1}{2}\cos \theta(\mathbf{x},t) - 1$.
    (c) DVIB trained with two latent variables learns oscillations. Here each trajectory is colored by its synchronization order parameter. 
    (d) Latent oscillations are very highly correlated with synchronization order parameter.}
    \label{sifig:kuramoto}
\end{figure}

In SI Fig.~\ref{sifig:kuramoto} we perform variational IB on a locally-coupled Kuramoto model and find the latent variables learn a synchronization order parameter as in the cyanobacteria data. This model considers a spatially-distributed field of phases which evolves according to
\begin{align*}
    \partial_t\theta(\mathbf{x}_i,t) = \omega(\mathbf{x}_i) + J \sum_{j\in\mathcal{N}_i} \sin(\theta(\mathbf{x}_i,t) - \theta(\mathbf{x}_j,t))
\end{align*}
where $\mathbf{x}_i$ denotes the location of gridpoint $i$ and $\mathcal{N}_i$ denotes the sites neighboring $i$.
These dynamics are integrated using the Euler method, where we take $\omega(\mathbf{x}) = \text{const}=0.05$ for the natural frequencies, and $J=1.0$.

\section{Comparison of various dimensionality reduction methods}

Variational IB (VIB) is by no means the only numerical method for performing data-driven model reduction.
In this section we compare the performance of variational IB (VIB) to several common data-driven model reduction or inference methods.
%We emphasize that the goal of VIB is not to construct a neural network which generates the most accurate predictions possible, but rather one which generates well-behaved, predictive collective variables.
VIB uses a neural network to identify the relevant variables that are most predictive of the future. This construction learns \textit{dynamically} relevant variables in contrast to methods such as principal component analysis or diffusion maps (in the absence of additional stringent assumptions), and it finds potentially non-linear variables in an agnostic way without requiring a tailored library of non-linearities (such as extended dynamic mode decomposition). 
In addition to these factors, VIB produces physically well-defined relevant variables -- transfer operator eigenfunctions -- in contrast to other deep methods such as (variational) autoencoders. 
In Fig.~\ref{sifig:comparison} we compare the low-dimensional latent space trajectories of the reduced models obtained using several of these standard methods to those output by VIB. In the remainder of this section we discuss these methods in detail.

\subsection*{Principal Component Analysis/Proper Orthogonal Decomposition}

Principal component analysis, or PCA, is a \textit{linear} method that projects the data onto a subspace which accounts for the most variance in the dataset \cite{bishop2006pattern}. 
In the dynamical systems literature, PCA is also known as a proper orthogonal decomposition (POD) \cite{berkooz_POD}.
The starting point for PCA is a collection of \textit{samples} $\mathbf{x}^{(k)}$, where each sample has $N_{\text{feat}}$ different \textit{features}: $\mathbf{x}^{(k)}\in \mathbb{R}^{N_{\text{feat}}}$.
Given these samples, one computes a correlation matrix between every pair of features
\begin{align*}
    C_{ij} = \frac{1}{N_{\text{samples}}}\sum_{n=1}^{N_{\text{samples}}} (x^{(n)}_i - \bar{x}_i)(x^{(n)}_j - \bar{x}_j).
\end{align*}
The principal components of the data are eigenvectors of the covariance matrix; the dominant principal component determines the direction of most variation in the dataset. 
In practice, if many samples are present the correlation matrix is expensive to compute. Writing the dataset as a matrix, $\mathbf{X}\in\mathbb{R}^{N_{\text{samples}}\times N_{\text{feat}}}$, computing the correlation matrix requires evaluating the matrix product $\mathbf{C} \propto \mathbf{X}^T \mathbf{X}$.
Instead, a more computationally efficient way is to compute the singular value decomposition (SVD) of the data matrix, 
\begin{align*}
    \mathbf{X} = \mathbf{U}\mathbf{\Sigma}\mathbf{V}^T.
\end{align*}
The principal components are given by columns of $\mathbf{V}$, and the singular values in the matrix $\mathbf{\Sigma}$ are the square root of the variance along each principal component direction.
PCA's relation to SVD also means that one can interpret a reconstruction of $\mathbf{X}$ using only $k$ principal components as the optimal rank-$k$ approximation of the dataset, where optimality is here defined with respect to the Frobenius norm.

Compared to VIB, the linearity assumptions underlying PCA can be viewed either as a benefit or a drawback.
The benefit is that a linear projection can be intuitive to understand in terms of the original data, and that one can compute this projection quickly. 
The potential downside is that a linear projection may not be sufficient or appropriate for systems that are highly non-linear.
In addition, PCA makes no reference to the dynamics of the system, unlike VIB. An example of where PCA can fail is shown in Fig.~3 of Ref.~\cite{Klus2018}. In words, imagine a particle is hopping in a 2D double-well potential, where the wells are centered at $x\pm 1$. If the vertical extent of the wells is very large ($|y|\gg 0$), then the dominant principal component will be along the $y$-direction, even though the interesting dynamics are the hops between wells, which happens in the $x$-direction.

\subsection*{Dynamic Mode Decomposition}
\label{sct:compare:dmd}
Dynamic Mode Decomposition (DMD) starts from the assumption that the system evolves linearly \cite{rowley2009_dmd,schmid_2010_dmd}. Concretely, if $\mathbf{x}_t\in \mathbb{R}^{n_{\text{feat}}}$ denotes the observed quantity at one point in time, then one assumes dynamics given by
\begin{align*}
    \mathbf{x}_{t+1} = \mathbf{A}_{\text{DMD}}\mathbf{x}_t.
\end{align*}
The matrix $\mathbf{A}$ is a finite-dimensional approximation to the Koopman operator \cite{ModernKoopman}. 

To find the matrix $\mathbf{A}_{\text{DMD}}$, we start by assembling a collection of samples into a data matrix $\mathbf{X}_t\in\mathbb{R}^{N_{\text{samples}}\times N_{\text{feat}}}$. In addition to this, we assemble a time-shifted matrix $\mathbf{X}_{t+1}$ of the same shape as  $\mathbf{X}_{t}$.
In the scenario where we only have one trajectory of duration $T$, each sample may correspond to the system at a measured time point (excluding the final time), so that 
\begin{align*}
    \mathbf{X}_t=
    \begin{pmatrix}
        - \mathbf{x}^{(1)} -
        \\
        - \mathbf{x}^{(2)} -
        \\
         \vdots 
        \\
        - \mathbf{x}^{(T-1)} -
    \end{pmatrix},
    \quad
    \mathbf{X}_{t+1}=
    \begin{pmatrix}
        - \mathbf{x}^{(2)} -
        \\
        - \mathbf{x}^{(3)} -
        \\
         \vdots 
        \\
        - \mathbf{x}^{(T)} -
    \end{pmatrix}
\end{align*}

The matrix $\mathbf{A}_{\text{DMD}}$ is given by the least squares solution
\begin{align*}
    \mathbf{A}_{\text{DMD}}^T = \mathbf{X}_{t}^{+}\mathbf{X}_{t+1}
\end{align*}
where $\mathbf{M}^{+}$ denotes the pseudoinverse of the matrix $\mathbf{M}$. Eigenvectors $\mathbf{w}$ of the matrix $\mathbf{A}_{\text{DMD}}$ are Koopman modes, which correspond to left eigenfunctions of the transfer operator. The evolution of these Koopman modes in time is given by the time-dependent amplitude $a_i(t)=\mathbf{x}_t\cdot\mathbf{w}$, where we take $\mathbf{x}_t$ here to be a single measurement at time $t$.

Similar to PCA, DMD reduces the dimensionality of the dataset by finding a linear projection of the data onto a subspace of the full features space. However, a key difference between the two approaches is that DMD uses the system's dynamics to identify an ``optimal'' subspace, whereas PCA identifies a subspace based only on the steady-state distribution of the data.

DMD in its original formulation assumes that the Koopman operator linearly evolves the observed state variable. However, the true Koopman operator evolves arbitrary non-linear functions of the state variable forward in time. DMD has therefore been extended to account for non-linear functions through ``extended DMD'', or eDMD \cite{williams2015}. eDMD augments the state vector with non-linear transformations of the state. As an example, a state vector $[-\vec{x}-]$ may be replaced by monomials $[-\vec{x}-,-\vec{x}^2-,-\vec{x}^3-]$ (where here we understand exponentiation as element-wise). Other approaches also exist, such as augmenting the state vector by several time-delayed state vectors \cite{brunton2017}. Rather than constructing the full $d \times d$ matrix, where $d$ is the dimension of the (possibly augmented) state, one can directly construct a low-rank approximation of $K$ by computing the reduced-rank singular value decomposition (SVD) of the matrix $K$ from the SVDs of $\mathbf{X}_{t}$ and $\mathbf{X}_{t+\Delta t}$.

Similar to PCA, DMD is primarily limited in its assumption of linear dynamics. 
In some cases this can be resolved with eDMD, which requires that one identifies a suitable set of non-linear terms to account for potential non-linear eigenfunctions of the Koopman operator. It is unclear how to choose this set of functions in a generic setting, which constitutes the biggest disadvantage to VIB.
The features for eDMD must be hand-selected, which in some cases may defeat the purpose of using it as a feature-learning tool.
VIB is not subject to this restriction, and can learn relevant features directly from the data.

\subsection*{Independent Component Analysis}

Independent component analysis (ICA) was originally formulated in the context of blind source separation, where a useful picture is the ``cocktail party problem'' \cite{ica_hyvarinen2001}. Imagine you place a set of microphones in a room at a cocktail party; these microphones will record the combination of all the conversations happening at once. The goal of blind source separation is to find a way to, from the recorded signals, isolate the original conversations. Mathematically, ICA finds a solution to the equation
\begin{align}
    \mathbf{x} = \mathbf{A}_{\text{ICA}}\mathbf{s}.
    \label{eq:icadef}
\end{align}
Here, $\mathbf{x}$ denotes the recorded signal, $\mathbf{s}$ denotes the independent sources (conversations), and $\mathbf{A}_{\text{ICA}}$ is the mixing matrix. 
The only measured quantity is the vector $\mathbf{x}$; both the mixing matrix and the independent sources must be learned.

To find one independent component, we start with an initial guess $y=\mathbf{b}^T\mathbf{x}$ which, for a correct choice of $\mathbf{b}$, should be equal to some component $s_i$. 
To identify the correct $\mathbf{b}$ we will use the fact that sums of independent variables are more Gaussian than the original variables themselves, which follows from the central limit theorem.
By assumption, the observed signals are a linear combination of independent components, so $y$ is also: $y=\mathbf{q}^T\mathbf{s}$.
If multiple $q_i$ are non-zero, this will be more non-Gaussian than if only one is non-zero. 
Thus, $y=s_i$ for the choice of $\mathbf{b}$ for which $y$ is maximally non-Gaussian \cite{ica_hyvarinen2001}. 

Numerically, one optimizes the deviation from Gaussianity using an approximation of the ``negentropy'' of the distribution of $y$: $\mathcal{J}_{\text{neg}}=S(y_{\text{Gaussian}})-S(y)$, where $y_{\text{Gaussian}}$ is a normally distribution random variable and $S$ is the Shannon entropy. The intuition for this formulation is that the $\mathcal{J}_{\text{neg}}$ is minimized if $y$ has a unit-Gaussian distribution, so that it serves as a metric for non-Gaussianity. In practice one uses an approximation for $\mathcal{J}_{\text{neg}}$, see Ref.~\cite{ica_hyvarinen2001} for details. The optimal $\mathbf{b}$ can then be found by doing gradient ascent on this objective function.

On some level, independent component analysis (ICA) is similar to PCA in that it searches for a linear projection of the data onto a subspace of the feature space. 
In its basic implementation ICA, like PCA, does not incorporate dynamics in contrast to DMD or VIB.
The essential difference between PCA and ICA is the assumption of independence of the components. In PCA, unless the data distribution is actually multivariate Gaussian, the components are unlikely to be independent. 
IB makes no assumptions about the independence of learned encoding variables with each other, only about their level of independence with the original state. 
We note that ICA can be understood as minimizing the mutual information between encoding variables, $I(H_1;H_2)$, compared to IB's minimization of the mutual information with the original state \cite{ica_hyvarinen2001}.
Which objective is more desirable depends on the system at hand, but in general we do not expect these to lead to the same encodings.

\subsection*{Time-lagged independent component analysis}
The leveraging of non-Gaussianity in ICA is required due to the fact that, if using whitened data ($\mathbf{z}=\mathbf{V}\mathbf{x}$ so that $\mathbf{z}$ is zero mean ,$\mathbb{E}[\mathbf{z}]=\mathbf{0}$, and unit variance, $\mathbb{E}[\mathbf{z}\mathbf{z}^T]=\mathbf{I}$), then the mixing matrix $\mathbf{A}_{\text{ICA}}$ cannot be estimated if the components $\mathbf{s}$ are normal Gaussian variables: any orthogonal matrix will satisfy Eq.~\eqref{eq:icadef}.
For dynamical systems, however, one can get around this requirement of non-Gaussianity. In particular, one can use a time-lagged variable $\mathbf{z}_{t-\tau} = \mathbf{A}_{\text{ICA}}\mathbf{s}_{t-\tau}$ to compute the mixing matrix from the time-correlation matrix of the signal $\mathbf{z}_t$ \cite{ica_hyvarinen2001}. 
In particular, we have
\begin{align*}
    \mathbb{E}[\mathbf{z}_t\mathbf{z}_{t-\tau}^T] &= \mathbf{A}_{\text{ICA}}\mathbb{E}[\mathbf{s}_t\mathbf{s}_{t-\tau}^T]\mathbf{A}_{\text{ICA}}^T
    \\
    &= \mathbf{A}_{\text{ICA}}\mathbf{D}\mathbf{A}_{\text{ICA}}^T,
\end{align*}
where $\mathbf{D}$ is a diagonal matrix.
In going to the bottom line, we used the assumption that the components $s_i$ are independent not just instantaneously, but also for a time lag $\tau$. 
From here we can read off that the matrix $\mathbf{A}_{\text{ICA}}$ is composed of eigenvectors of the correlation matrix of the whitened signal. When written in terms of the original (unwhitened) data $\mathbf{x}_t$ and $\mathbf{x}_{t-\tau}$ it can be shown these eigenvectors are nothing other than the eigenvectors of the transfer matrix $\mathbf{A}_{\text{DMD}}$ \cite{Klus2018}. Thus, the two methods are equivalent.

\subsection*{Diffusion maps}
Diffusion maps are a technique which attempts to approximate the Perron-Frobenius operator, or rather the integral kernel $p(x_{t+\Delta t}|x_t)$ \cite{coifman2006_diffusionmaps,coifman2008}. Given this approximation, one computes eigenfunctions and uses them as a low-dimensional parameterization of the data (``diffusion coordinates'').

This method takes data pairs $\{x^{(i)}_{t}, x^{(i)}_{t+\Delta t}\}_i$ and approximates the probability of observing these two points via a \textit{kernel} $p(x^{(i)}_{t}, x^{(j)}_{t+\Delta t})\approx k(x^{(i)}_{t}, x^{(j)}_{t+\Delta t})$, where one typically takes a Gaussian Ansatz
\begin{align*}
    k_\epsilon(x,y)\propto \exp\left[\frac{-(x-y)^2}{\epsilon}\right].
\end{align*}
From this, one can assemble the conditional probability distributions into a matrix
\begin{align*}
    P_{ij} = p(x^{(j)}_{t+\Delta t}|x^{(i)}_t) = \frac{p(x^{(i)}_{t}, x^{(j)}_{t+\Delta t})}{p(x^{(i)}_{t})}=\frac{k(x^{(i)}_{t}, x^{(j)}_{t+\Delta t})}{\sum_j k(x^{(i)}_{t}, x^{(j)}_{t+\Delta t})}.
\end{align*}
This matrix describes the evolution of probability distributions on a \textit{graph} where each node is a data point.
In practice, a symmetrized version of $P$ is constructed and the learned eigenvectors are adjusted after the diagonalization \cite{coifman2006_diffusionmaps,coifman2008}.
To compute the diffusion coordinates of an arbitrary point that wasn't in the original dataset, one inverts the definition of the adjoint transfer operator eigenfunction
\begin{align*}
    \phi_i(\vec{x}_{\text{new}}) \approx \frac{1}{\lambda_i}\sum_k P^{\dagger}_{jk}\phi_i(\vec{x}_k)
\end{align*}
where $\lambda_i$ is the $i$-th eigenvalue.

Diffusion maps have the advantage, relative to DMD, that they find the full Perron-Frobenius operator and not a linear approximation to it. However, while DMD can isolate the dominant eigenvectors of the operator using reduced-rank SVD, it is less clear how they can be extracted with diffusion maps without first computing the full matrix $P$. We note that VIB, like DMD, also directly learns the dominant modes and does not require estimation of the full transfer operator.

\subsection*{Deep (Variational) Autoencoders}
Autoencoders (AEs) belong to a class of deep learning methods used for model reduction. 
Such approaches have successfully been applied in various domains for forecasting the dynamics of complex systems in terms of simpler latent dynamics \cite{peterluPRX,jcolenpnas}.
Autoencoders are composed of an encoder which compresses the observable $\vec{x}$ into a lower-dimensional \textit{latent} variable $\vec{z}$, and a decoder which attempts to reconstruct the original state $\vec{x}$.
Variational autoencoders aim to learn a probability distribution over observations $p_{\theta}(\vec{x})\approx p(\vec{x})$ from which one can directly sample \cite{kingma2022vae}. 
This is done by assuming that the latent variable is low dimensional, and optimizing the objective
\begin{align*}
    \mathcal{L}_{\beta\text{-VAE}}=\mathbb{E}_{q_{\phi}(z|x)}[-\log p_{\theta}(x|z)] - \beta D_{\text{KL}}(q(z|x)\lVert \hat{p}(z))
\end{align*}
where $q_{\phi}(z|x)$ is the posterior on the latent variables $z$ and is parameterized by a neural network (encoder) with parameters $\phi$, and $p_{\theta}(x|z)$ denotes the decoding network with parameters $\theta$. Strictly speaking, we present in the above objective function the $\beta$-VAE loss \cite{higgins2017betavae}. With $\beta=1$, which is the case for the original VAE \cite{kingma2022vae}, the objective is an upper bound on the log likelihood $-\mathbb{E}[\log p_{\theta}(x)]$ which is minimized when $p_{\theta}(x)$ is equal to the true data distribution $p(x)$. The term $\beta$ controls compression as in the IB objective, however it has the opposite effect: small $\beta_{\text{IB}}$ corresponds to high compression, while small $\beta_{\text{VAE}}$ corresponds to low compression.

In cases where one directly computes the probabilities $p_\theta(x|z)$ the first term can be evaluated directly, else it is typically replaced with an $L_2$ loss $\lVert x - g_{\theta}(z)\lVert^2$ (where $g_{\theta}$ is a deterministic neural network), which is equivalent to assuming a Gaussian Ansatz for $p_{\theta}$ with a fixed variance.

\begin{figure}[ht]
    \centering
    \includegraphics[width=0.8\textwidth]{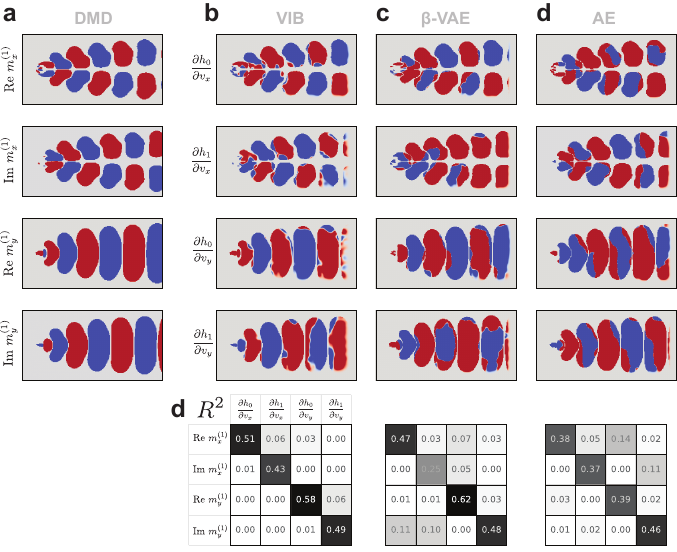}
    \caption{\textbf{Learned latent variables for fluid flows with various deep neural networks.}
    To compare latent variables we compare their gradients as in Fig.~5 of the main text, which should correspond to Koopman modes. Here we show the gradients masked by regions in the DMD mode which have large amplitude.
    (a) True Koopman modes, with real and imaginary parts shown, obtained via DMD.
    (b) Corresponding modes for a VIB network. Matrix in the bottom row shows the $R^2$ values obtained by regressing the pixel values of the VIB gradients against the different components of the Koopman modes. 
    (c) Corresponding modes for a $\beta$-variational autoencoder (VAE). The VAE has the same encoder structure as the VIB network, while the decoder has an inverted architecture to predict the future state $x_{t+\Delta t}$ from the latent variable $h_t$. Bottom row shows $R^2$ values, analogously to (b).
    (d) Corresponding modes for an autoencoder.
    }
    \label{apxfig:vae}
\end{figure}

The VIB loss function is very similar to the $\beta$-VAE loss \cite{burgess2018}. Rather than attempting to reconstruct the original state $x$, VIB replaces this term with an estimate of the mutual information between the latent variable $z$ and some other relevance variable $y$ (for us, $y=x_{t+\Delta t}$). 
In contrast to DMD and diffusion maps, neither VIB nor $\beta$-VAEs make any mention of transfer operators and are instead motivated by purely statistical considerations. 
As we show in the main text, the latent variables learned by VIB correspond to eigenfunctions of the transfer operator. While we cannot claim that that $\beta$-VAEs learn the same thing, some preliminary results in Fig.~\ref{apxfig:vae} suggest they may coincide to some degree.
In the cyanobacteria dataset, we see that the latent variables learned by a $\beta$-VAE show the same qualitative structure as in VIB, but are less smooth (Fig.~\ref{sifig:comparison}).

\subsection*{Other neural networks} 
Other deep architectures can also be used for model reduction. For example, Ref.~\cite{Vlachas2022} uses recurrent neural networks (RNNs) to learn the evolution of macroscopic variables. To ensure stability and fidelity, the macroscopic variables are periodically ``lifted'' to the full microscopic state, which is then evolved for several time steps to recalibrate the RNN's hidden state. While interpretability of the latent variables has yet to be explored in such models, we expect the addition of VIB-like objective functions may aid interpretability without harming performance.

As another example, \cite{takeishi2017_neurips} attempts to learn the full Koopman operator using neural networks. This can be thought of as an extension of eDMD, where instead of prescribing the library of nonlinear terms by hand, they can be learned by a neural network. The latent dynamics are then encouraged to be linear by minimizing the deviation between the true future (latent) state $\vec{z}_{t+\Delta t}$ and its linear approximation found by least squares. 
Similar approaches have also been explored in Refs.~\cite{bakarji2023, lusch2018}.
As an illustration, we trained one such network on the cyanobacteria which can be seen in Fig.~\ref{sifig:comparison}.
This approach was applied primarily to deterministic systems. We expect that combining their method of encouraging linear latent dynamics together with the VIB objective function may be fruitful and lead to more well-behaved latent variables.

\end{document}